\documentclass[floats,floatfix,prd,twocolumn,superscriptaddress,nofootinbib,
longbibliography]{revtex4-2} \usepackage{amsfonts,amssymb,amsmath}
\usepackage[caption=false]{subfig} %default subfig option screws up captions
\usepackage{graphicx, hyperref, tabularx, dcolumn}

\newcommand{\dual}{\,{}^*\!}

\usepackage{color, xcolor}

\usepackage{orcidlink}

\newcommand{\Cornell}{\affiliation{Cornell Center for Astrophysics and Planetary
    Science, Cornell University, Ithaca, New York 14853, USA}}
\newcommand{\Caltech}{\affiliation{Theoretical Astrophysics 350-17, California
    Institute of Technology, Pasadena, CA 91125, USA}}
\newcommand{\Fullerton}{\affiliation{Nicholas and Lee Begovich Center for
    Gravitational-Wave Physics and Astronomy, California State University
    Fullerton, Fullerton, CA 92834, USA}}
\newcommand{\Icts}{\affiliation{International Centre for Theoretical Sciences,
    Tata Institute of Fundamental Research, Bangalore 560089, India}}
\newcommand{\MaxPlanck}{\affiliation{Max Planck Institute for Gravitational
    Physics (Albert Einstein Institute), Am M{\"u}hlenberg 1, D-14476 Potsdam,
    Germany}} \newcommand{\Unh}{\affiliation{Department of Physics \& Astronomy,
    University of New Hampshire, Durham, New Hampshire 03824, USA}}
\newcommand{\Wsu}{\affiliation{Department of Physics \& Astronomy, Washington
    State University, Pullman, Washington 99164, USA}}
\newcommand{\Koln}{\affiliation{I.~Physikalisches Institut, Universit\"at zu
    K\"oln, Z\"ulpicher Stra{\ss}e 77, 50937, K\"oln, Germany}}
\newcommand{\Coimbra}{\affiliation{CFisUC, Department of Physics, University of
    Coimbra, 3004-516 Coimbra, Portugal}}

\begin{document}

\title{Simulating magnetized neutron stars with discontinuous Galerkin methods}

\author{Nils Deppe \orcidlink{0000-0003-4557-4115}} \email{ndeppe@caltech.edu}
\Caltech \author{Fran\c{c}ois H\'{e}bert \orcidlink{0000-0001-9009-6955}}
\Caltech \author{Lawrence E.~Kidder \orcidlink{0000-0001-5392-7342}} \Cornell
\author{William Throwe \orcidlink{0000-0001-5059-4378}} \Cornell \author{Isha
  Anantpurkar \orcidlink{0000-0002-5814-4109}} \Cornell \author{Crist\'obal
  Armaza \orcidlink{0000-0002-1791-0743}} \Cornell \author{Gabriel S.~Bonilla
  \orcidlink{0000-0003-4502-528X}} \Cornell \Fullerton \author{Michael Boyle
  \orcidlink{0000-0002-5075-5116}} \Cornell \author{Himanshu Chaudhary
  \orcidlink{0000-0002-4101-0534}} \Caltech \author{Matthew D.~Duez} \Wsu
\author{Nils L.~Vu \orcidlink{0000-0002-5767-3949}} \MaxPlanck
\author{Francois Foucart \orcidlink{0000-0003-4617-4738}} \Unh \author{Matthew
  Giesler \orcidlink{0000-0003-2300-893X}} \Cornell \author{Jason S.~Guo}
\Cornell \author{Yoonsoo Kim \orcidlink{0000-0002-4305-6026}} \Caltech
\author{Prayush Kumar \orcidlink{0000-0001-5523-4603}} \Cornell \Icts
\author{Isaac Legred \orcidlink{0000-0002-9523-9617}} \Caltech \author{Dongjun
  Li \orcidlink{0000-0002-1962-680X}} \Caltech \author{Geoffrey Lovelace
  \orcidlink{0000-0002-7084-1070}} \Fullerton \author{Sizheng Ma
  \orcidlink{0000-0002-4645-453X}} \Caltech \author{Alexandra Macedo
  \orcidlink{0000-0001-5139-9013}} \Fullerton \author{Denyz Melchor
  \orcidlink{0000-0002-7854-1953}} \Fullerton \author{Marlo Morales
  \orcidlink{0000-0002-0593-4318}} \Fullerton \author{Jordan Moxon
  \orcidlink{0000-0001-9891-8677}} \Caltech \author{Kyle C.~Nelli
  \orcidlink{0000-0003-2426-8768}} \Caltech \author{Eamonn O'Shea} \Cornell
\author{Harald P.~Pfeiffer \orcidlink{0000-0001-9288-519X}} \MaxPlanck
\author{Teresita Ramirez \orcidlink{0000-0003-0994-115X}} \Fullerton
\author{Hannes R.~R\"uter \orcidlink{0000-0002-3442-5360}} \Coimbra
\author{Jennifer Sanchez \orcidlink{0000-0002-5335-4924}} \Fullerton
\author{Mark A.~Scheel \orcidlink{0000-0001-6656-9134}} \Caltech \author{Sierra
  Thomas \orcidlink{0000-0003-3574-2090}} \Fullerton \author{Daniel Vieira
  \orcidlink{0000-0001-8019-0390}} \Cornell \Koln \author{Nikolas A.~Wittek
  \orcidlink{0000-0001-8575-5450}} \MaxPlanck \author{Tom Wlodarczyk} \MaxPlanck
\author{Saul A.~Teukolsky \orcidlink{0000-0001-9765-4526}} \Cornell \Caltech
\date{\today}

\begin{abstract}
  Discontinuous Galerkin methods are popular because they can achieve high order
  where the solution is smooth, because they can capture shocks while needing
  only nearest-neighbor communication, and because they are relatively easy to
  formulate on complex meshes. We perform a detailed comparison of various
  limiting strategies presented in the literature applied to the equations of
  general relativistic magnetohydrodynamics. We compare the standard
  minmod/$\Lambda\Pi^N$ limiter, the hierarchical limiter of Krivodonova, the
  simple WENO limiter, the HWENO limiter, and a discontinuous
  Galerkin-finite-difference hybrid method.  The ultimate goal is to understand
  what limiting strategies are able to robustly simulate magnetized TOV stars
  without any fine-tuning of parameters.  Among the limiters explored here, the
  only limiting strategy we can endorse is a discontinuous
  Galerkin-finite-difference hybrid method.
\end{abstract}

\maketitle

\section{Introduction}

Many of the most energetic phenomena in the universe involve matter under
extreme gravitational conditions.  These phenomena include neutron-star binary
mergers, accretion onto black holes, and supernova explosions.  For many of
these systems, the motion of this matter is expected to generate extremely
strong magnetic fields.  The matter and magnetic fields in these systems are
governed by the equations of general relativistic magnetohydrodynamics~(GRMHD).
These equations admit a rich variety of solutions, which often include
large-scale relativistic flows and small-scale phenomena such as shocks and
turbulence.

High-resolution shock capturing (HRSC) finite-difference (FD) methods are the
current standard methods of choice for numerically evolving these solutions
since they are able to robustly handle shocks. Unfortunately, HRSC FD methods
have significant computational overhead and are less efficient than
spectral-type methods like discontinuous Galerkin~(DG) where the solution is
smooth. Additionally, achieving better than second-order convergence is
generally difficult, with recent results presented in~\cite{Most:2019kfe,
  Cipolletta:2020kgq}. The common use of second-order methods means that the
current generation of GRMHD codes is not accurate enough to provide useful
predictions for many extreme systems.  Increasing simulation resolutions can
improve this, but is very computationally expensive.  More appealing is using
numerical methods with higher convergence orders, which can increase accuracy
with significantly less cost than a similar improvement from resolution.
Unfortunately, while higher-order methods handle smooth solutions very well,
they are generally poor at handling discontinuities, such as fluid shocks,
losing accuracy and sometimes failing completely.

Additionally, the time required to run simulations is already too long for most
interesting astrophysical cases.  The performance of individual computational
processors has stagnated over the past decade, so to improve computational
speed, codes must be parallelized over more processors.  New supercomputer
clusters will soon routinely have millions of cores.  Codes designed for running
on thousands of processors generally scale poorly to massively parallel setups,
however. As problems are divided up into an increasing numbers of parts, the
amount of communication required during the simulation can become prohibitive,
particularly for high-order methods.

Discontinuous Galerkin methods~\cite{Reed1973, Hesthaven2008,
  2001JCoAM.128..187C,1998JCoPh.141..199C, Cockburn1998, Cockburn2000}, together
with a task-based parallelization strategy, have the potential to deal with
these problems.  DG methods offer high-order accuracy in smooth regions, with
the potential for robust shock capturing by some non-linear stabilization
technique. The methods are also well suited for parallelization: Their
formulation in terms of local, non-overlapping elements requires only
nearest-neighbor communication regardless of the scheme's order of convergence.
Additionally, these features allow for comparatively straightforward
$hp$-adaptivity/adaptive mesh refinement and local time-stepping, enabling
better load distribution across a large number of cores.

Despite extensive success in engineering and applied mathematics communities
over the past two decades, applications of DG in
relativity~\cite{Field:2010mn,Brown:2012me,Field:2009kk, Zumbusch:2009fe,
  Dumbser:2017okk} and astrophysics~\cite{Radice:2011qr, Mocz:2013cma,
  Zanotti:2014ppa, Endeve:2014xra} have typically been exploratory or confined
to simple problems.  However, recently there have been significant advances
toward production codes for non-relativistic~\cite{Schaal:2015ila} and
relativistic~\cite{Teukolsky:2015ega,Kidder:2016hev,Bugner:2015gqa}
hydrodynamics, special relativistic magnetohydrodynamics~\cite{Zanotti:2015mia,
  Zanotti:2014ppa, axioms7030063, Fambri:2018udk, ZANOTTI2015204,
  Zanotti:2015sns}, the Einstein equations~\cite{Miller:2016vik,
  Dumbser:2017okk}, and relativistic hydrodynamics coupled to the Einstein
equations~\cite{PhysRevD.98.044041}.  Most of these codes use the MPI to
implement a data parallelism strategy,
though~\cite{Kidder:2016hev,axioms7030063} use task-based parallelism.

In this paper we present a detailed comparison of various different limiting and
shock capturing strategies for DG methods in the context of demanding GRMHD test
problems. Specifically, we compare the classical limiters $\Lambda\Pi^N$
limiter~\cite{Cockburn1999}, Krivodonova limiter~\cite{2007JCoPh.226..879K}, and
WENO-based limiters~\cite{2013JCoPh.232..397Z, 2016CCoPh..19..944Z}, as well as
a DG-FD hybrid method similar to that of~\cite{2007JCoPh.224..970C,
  2014JCoPh.278...47D}. The ultimate goal is to simulate a magnetized and
non-magnetized TOV star in the Cowling approximation. To the best of our
knowledge this is the first time a magnetized TOV star has been simulated using
DG methods. This paper presents a crucial first step to being able to apply DG
methods to simulations of binary neutron star mergers, differentially rotating
magnetized single neutron stars, and magnetized accretion disks.

While generally the classical limiters produce the best results when applied to
the characteristic variables, these are not known analytically for GRMHD.  Even
though most test cases in this paper are in special relativity, we intentionally
apply the limiters to the conserved variables to evaluate their performance in
the form they need to be used for GRMHD. Since FD methods are also known to be
less dissipative when applied to the characteristic variables, this choice does
not put any of the limiters at a disadvantage.

The paper is organized as follows.  Section~\ref{sec:equations} describes the
formulation of GRMHD used in the problems presented here.
Section~\ref{sec:methods} describes the algorithms used by
our open-source code \texttt{SpECTRE}~\cite{spectrecode} to
solve these equations.  Results of the evolutions of a variety of GRMHD problems
are presented in Sec.~\ref{sec:results} comparing the different shock
capturing strategies

\section{Equations of GRMHD}
\label{sec:equations}

We adopt the standard 3+1 form of the spacetime metric,~(see,
e.g.,~\cite{Baumgarte:2010ndz, 2013rehy.book.....R}),
\begin{align}
  \label{eq:spacetime metric}
  ds^2 &= g_{ab}dx^a dx^b \notag \\
       &=-\alpha^2 dt^2 + \gamma_{ij}
         \left(dx^i+\beta^i dt\right) \left(dx^j +\beta^j dt\right),
\end{align}
where $\alpha$ is the lapse, $\beta^i$ the shift vector, and $\gamma_{ij}$ is
the spatial metric.  We use the Einstein summation convention, summing over
repeated indices.  Latin indices from the first part of the alphabet
$a,b,c,\ldots$ denote spacetime indices ranging from $0$ to $3$, while Latin
indices $i,j,\ldots$ are purely spatial, ranging from $1$ to $3$. We work in
units where $c = G = M_{\odot} = 1$.

\texttt{SpECTRE} currently solves equations in flux-balanced and first-order
hyperbolic form. The general form of a flux-balanced conservation law in a
curved spacetime is
\begin{align}
  \label{eq:conservation law}
  \partial_t U + \partial_i F^i = S,
\end{align}
where $U$ is the state vector, $F^i$ are the components of the flux vector, and
$S$ is the source vector.

We refer the reader to the literature~\cite{2006ApJ...637..296A, Font:2008fka,
  Baumgarte:2010ndz} for a detailed description of the equations of general
relativistic magnetohydrodynamics~(GRMHD).  If we ignore self-gravity, the GRMHD
equations constitute a closed system that may be solved on a given background
metric.  We denote the rest-mass density of the fluid by $\rho$ and its
4-velocity by $u^a$, where $u^a u_a=-1$. The dual of the Faraday tensor $F^{ab}$
is
\begin{align}
  \label{eq:Faraday dual}
  \dual F^{ab} = \tfrac{1}{2}\epsilon^{abcd}F_{cd},
\end{align}
where $\epsilon^{abcd}$ is the Levi-Civita tensor. Note that the
Levi-Civita tensor is defined here with the
convention~\cite{MTW}
that in flat spacetime $\epsilon_{0123}=+1$.
The equations governing the
evolution of the GRMHD system are:
\begin{align}
  \nabla_a(\rho u^a) &= 0\quad (\text{rest-mass conservation}), \\
  \nabla_a T^{ab} &= 0\quad (\text{energy-momentum conservation}), \\
  \nabla_a  \dual F^{ab} &= 0\quad (\text{homogeneous Maxwell equation}).
\end{align}
In the ideal MHD limit the stress tensor takes the form
\begin{equation}
  T^{ab} = (\rho h)^*u^a u^b + p^* g^{ab} - b^a b^b
\end{equation}
where
\begin{equation}
  b^a = -\dual F^{ab}u_b
\end{equation}
is the magnetic field measured in the comoving frame of the fluid, and
$(\rho h)^* = \rho h + b^2$ and $p^* = p + b^2/2$ are the enthalpy density and
fluid pressure augmented by contributions of magnetic pressure
$p_{\mathrm{mag}} = b^2/2$, respectively.

We denote the unit normal vector to the spatial hypersurfaces as $n^a$, which is
given by
\begin{align}
  \label{eq:spacetime unit normal vector}
  n^a &= \left(1/\alpha, -\beta^i/\alpha\right)^T, \\
  \label{eq:spacetime unit normal covector}
  n_a &= (-\alpha, 0, 0, 0).
\end{align}
The spatial velocity of the fluid as measured by an observer at rest in the
spatial hypersurfaces (``Eulerian observer'') is
\begin{equation}
  \label{eq:spatial velocity}
  v^i = \frac{1}{\alpha}\left(\frac{u^i}{u^0} + \beta^i\right),
\end{equation}
with a corresponding Lorentz factor $W$ given by
\begin{align}
  \label{eq:Lorentz factor}
  W &= - u^a n_a = \alpha u^0 = \frac{1}{\sqrt{1 - \gamma_{ij}v^i v^j}} \\
  \label{eq:Lorentz factor four velocity}
    &=\sqrt{1+\gamma^{ij}u_iu_j}=\sqrt{1+\gamma^{ij}W^2v_iv_j}.
\end{align}
The electric and magnetic fields as measured by an Eulerian observer are given
by
\begin{align}
  \label{eq:Eulerian electric field}
  E^i &= F^{ia}n_a = \alpha F^{0i},\\
  \label{eq:Eulerian magnetic field}
  B^i &= -\dual F^{ia}n_a = -\alpha \dual F^{0i}.
\end{align}
Finally, the comoving magnetic field $b^a$ in terms of $B^i$ is
\begin{align}
  \label{eq:b^0}
  b^0 &= \frac{W}{\alpha}B^i v_i, \\
  \label{eq:b^i}
  b^i &= \frac{B^i + \alpha b^0 u^i}{W},
\end{align}
while $b^2=b^a b_a$ is given by
\begin{equation}
  \label{eq:b^i b_i}
  b^2 = \frac{B^2}{W^2} + (B^i v_i)^2.
\end{equation}

We now recast the GRMHD equations in a 3+1 split by projecting them along and
perpendicular to $n^a$~\cite{2006ApJ...637..296A}.  One of the main
complications when solving the GRMHD equations numerically is preserving the
constraint
\begin{align}
  \label{eq:monopole constraint}
  \partial_i (\sqrt{\gamma} B^i)=0,
\end{align}
where $\gamma=\det(\gamma_{ij})$ is the determinant of the spatial metric.
Analytically, initial data evolved using the dynamical Maxwell equations are
guaranteed to preserve the constraint. However, numerical errors generate
constraint violations that need to be controlled. We opt to use the Generalized
Lagrange Multiplier (GLM) or divergence cleaning
method~\cite{2002JCoPh.175..645D} where an additional field $\Phi$ is evolved in
order to propagate constraint violations out of the domain.
Our version is very close to the one in Ref.~\cite{2014CQGra..31a5005M}.
The augmented system
can still be written in flux-balanced form, where the conserved variables are
\begin{align}
  U
  &=\sqrt{\gamma}\begin{pmatrix}
    D \\
    S_j \\
    \tau \\
    B^j \\
    \Phi
  \end{pmatrix}
  =\begin{pmatrix}
    \tilde{D} \\
    \tilde{S}_j \\
    \tilde{\tau} \\
    \tilde{B}^j \\
    \tilde{\Phi}
  \end{pmatrix}
  \notag \\
  &= \sqrt{\gamma}
    \begin{pmatrix}
      \rho W \\
      (\rho h)^* W^2 v_j - \alpha b^0 b_j \\
      (\rho h)^* W^2 - p^* - \left(\alpha b^0\right)^2 - \rho W \\
      B^j \\
      \Phi
    \end{pmatrix},
  \label{eq:conserved variables}
\end{align}
with corresponding fluxes
\begin{align}
  \label{eq:conserved fluxes}
  F^i =
  \begin{pmatrix}
    \tilde{D} v^i_\text{tr} \\
    \tilde{S}_j v^i_\text{tr} + \alpha\sqrt{\gamma} p^* \delta^i_j - \alpha
    b_j \tilde{B}^i/W\\
    \tilde{\tau} v^i_\text{tr} + \alpha\sqrt\gamma p^* v^i - \alpha^2 b^0
    \tilde{B}^i / W\\
    \tilde{B}^j v^i_\text{tr} - \alpha v^j\tilde{B}^i + \alpha
    \gamma^{ij}\tilde{\Phi}\\
    \alpha \tilde{B}^i - \tilde{\Phi} \beta^i
  \end{pmatrix},
\end{align}
and corresponding sources
\begin{align}
  \label{eq:conserved sources}
  S =
  \begin{pmatrix}
    0\\
    (\alpha/2)\tilde{S}^{kl} \partial_j\gamma_{kl}
    + \tilde{S}_k \partial_j \beta^k - \tilde{E}\partial_j \alpha\\
    \alpha \tilde{S}^{kl}K_{kl} - \tilde{S}^k\partial_k \alpha\\
    -\tilde{B}^k \partial_k \beta^j +
    \Phi\partial_k(\alpha\sqrt\gamma\gamma^{jk})\\
    \alpha\tilde{B}^k\partial_k\ln \alpha - \alpha K\tilde{\Phi} -
    \alpha\kappa\tilde{\Phi}
  \end{pmatrix}.
\end{align}
The transport velocity is defined as $v_\text{tr}^i = \alpha v^i - \beta^i$ and
the generalized energy $\tilde{E}$ and source $\tilde{S}^{ij}$ are given by
\begin{align}
  \label{eq:generalized energy}
  \tilde{E} &= \tilde{\tau} + \tilde{D},\\
  \label{eq:generalized source}
  \tilde{S}^{ij} &= \sqrt\gamma\left[(\rho h)^*W^2 v^i v^j +
                   p^*\gamma^{ij} -
                   \gamma^{ik}\gamma^{jl}b_kb_l\right].
\end{align}

The 3+1 GRMHD divergence cleaning evolution equations analytically preserve the
constraint \eqref{eq:monopole constraint}, while numerically
constraint-violating modes will be damped at a rate $\kappa$.  We typically
choose $\kappa \in [0, 10]$, but will specify the exact value used for each test
problem.  We note that the divergence cleaning method was shown to be strongly
hyperbolic in Ref.~\cite{Hilditch:2018jvf}, a necessary condition for a
well-posed evolution problem.  The primitive variables of the GRMHD system are
$\rho$, $v_i$, $B^i$, $\Phi$, and the specific internal energy $\epsilon$.

Approximate Riemann solvers use the characteristic speeds, which in the GRMHD
case require solving a nontrivial quartic equation for the fast and slow
modes. Instead, we use the approximation~\cite{2003ApJ...589..444G}:
\begin{align}
  \label{eq:grmhd char speed 1}
  \lambda_1 &= -\alpha-\beta_n, \\
  \lambda_2 &= \alpha\Lambda^--\beta_n, \\
  \lambda_{3,4,5,6,7} &= \alpha v_n - \beta_n, \\
  \lambda_8 &= \alpha \Lambda^+-\beta_n, \\
  \lambda_9 &= \alpha - \beta_n,
               \label{eq:grmhd char speed 9}
\end{align}
where $\beta_n$ and $v_n$ are the shift and spatial velocity projected along the
normal vector in the direction that we want to compute the characteristic speeds
along, and
\begin{align}
  \label{eq:grmhd lambda pm}
  \Lambda^{\pm}
  &= \dfrac{1}{1 - v^2 c_s^2}\left[v_n (1- c_s^2) \phantom{\frac{1}{2}}
    \right. \notag \\
  &\left.
    \pm c_s\sqrt{\left(1 - v^2\right)\left(1 - v^2 c_s^2 - v_n^2(1 -
    c_s^2)\right)}\right],
\end{align}
where $c_s$ is the sound speed given by
\begin{align}
  \label{eq:sound speed}
  c_s^2=\frac{1}{h} \left[ \left( \frac{\partial p}{\partial \rho}
  \right)_\epsilon +
  \frac{p}{\rho^2} \left(\frac{\partial p}{\partial \epsilon}
  \right)_\rho \right].
\end{align}

\section{Methods}
\label{sec:methods}

\subsection{The discontinuous Galerkin method}

We briefly summarize the nodal discontinuous Galerkin (DG) method for curved
spacetimes~\cite{Teukolsky:2015ega} in $d$ spatial dimensions. We decompose the
computational domain into $k$ elements, each with a reference coordinate system
$\{\xi,\eta,\zeta\}\in[-1,1]$. We denote the $i$th element by $\Omega_i$, so
our computational domain $\Omega=\cup_{i=1...k}\Omega_i$.  In this work we
consider only dimension-by-dimension affine maps. We expand the solution in each
element over a tensor product basis $\phi_{\breve{s}}$ of 1d Lagrange
polynomials $\ell_{\breve{i}}$,
\begin{align}
  \label{eqn:dgNodalInterpolation}
  U(\boldsymbol\xi)
  &=\sum_{\breve{s}}U_{\breve{s}}(t)\phi_{\breve{s}}(\boldsymbol\xi) \notag \\
  &=\sum_{\breve{\imath}}\sum_{\breve{\jmath}}\sum_{\breve{k}}
    U_{\breve{\imath}\breve{\jmath}\breve{k}}(t)
    \ell_{\breve{\imath}}\left(\xi\right)
    \ell_{\breve{\jmath}}\left(\eta\right)
    \ell_{\breve{k}}\left(\zeta\right)\,,
\end{align}
where $\xi,\eta,$ and $\zeta$ are the logical (or reference) coordinates. We use
Legendre-Gauss-Lobatto collocation points, though \texttt{SpECTRE} also supports
Legendre-Gauss points. We denote a DG scheme with 1d basis functions of degree
$N$ by $P_N$. A $P_N$ scheme is expected to converge at order
$\mathcal{O}(\Delta x^{N+1})$ for smooth solutions~\cite{Hesthaven2008}, where
$\Delta x$ is the 1d size of an element.

A spatial discretization is obtained by integrating the evolution
equations~\eqref{eq:conservation law} against the basis functions
$\phi_{\breve{s}}$
\begin{align}
  0&=\int_{\Omega_i}\left[
     \partial_tU + \partial_i F^i-S \right]
     \phi_{\breve{s}}(\mathbf{x})\,d^3x \notag \\
   &=\int_{\Omega_i}\left[
     \partial_tU + \partial_i F^i-S \right]
     \phi_{\breve{s}}(\mathbf{\xi})J\,d^3\xi,
\end{align}
where $J$ is the Jacobian determinant of the map from the reference coordinates
$\mathbf{\xi}$ to the coordinates $\mathbf{x}$.  Denoting the normal covector to
the spatial boundary of the element as $n_i$, integrating the flux divergence
term by-parts, replacing $F^in_i$ with a boundary correction/numerical flux $G$,
and undoing the integration by-parts, we obtain
\begin{align}
  &\int_{\Omega_i}
    \partial_t\left( U\right)Jd^3\xi\notag \\
  &=\oint_{\partial\Omega_i} (G-F^i n_i)\phi_{\breve{s}}(\xi)\,d^2\Sigma \notag
  \\
  &+\int_{\Omega_i} \left[\partial_iF^i+
    S\right]\phi_{\breve{s}}(\mathbf{\xi})J\,d^3\xi,
\end{align}
where $d^2\Sigma$ is the area element on the surface of the element. The area
element in the $+\zeta$ direction is given by~\cite{Teukolsky:2015ega}
\begin{align}
  \label{eq:surface element}
  d^2\Sigma
  &=\frac{\sqrt{{}^{(2)}\gamma}}{\sqrt{\gamma}}\,d\xi^1d\xi^2 \notag \\
  &=J\sqrt{\delta_{\hat{3}\hat{\imath}}
    \frac{\partial\xi^{\hat{\imath}}}{\partial x^i} \gamma^{ij}
    \frac{\partial\xi^{\hat{\jmath}}}{\partial x^j}
    \delta_{\hat{3}\hat{\jmath}}}\,d\xi^1d\xi^2.
\end{align}
Note that the normalization of the normal vectors in the $G-F^in_i$ term do not
cancel out with the term in~\eqref{eq:surface element}, as stated
in~\cite{Teukolsky:2015ega}. This is because both the inverse spatial metric and
the Jacobian may be different on each side of the boundary. Specifically, when
the spacetime is evolved, each element normalizes the normal vector using its
local inverse spatial metric.

Finally, the semi-discrete evolution equations are obtained by expanding
$U,F^i,$ and $S$ in terms of the basis
functions and evaluating the integrals by Gaussian quadrature. Our nodal DG code
uses the mass lumping approximation\footnote{``Mass lumping'' is the term that
  describes using the diagonal approximation for the mass
  matrix. See~\cite{2015JCoPh.283..408T} for more details.} when Gauss-Lobatto
points are employed.

\subsection{Numerical fluxes\label{sec:numerical fluxes}}

One of the key ingredients in conservative numerical schemes is the approximate
solution to the Riemann problem on the interface. We use the Rusanov
solver~\cite{Rusanov:1961} (also known as the local Lax-Friedrichs flux), and
the solver of Harten, Lax, and
van~Leer~(HLL)~\cite{Harten:1983a,toro2009riemann}. While both the Rusanov and
the HLL solver are quite simple, their use is standard in numerical
relativity. The Rusanov solver is given by
\begin{align}
  \label{eq:Rusanov flux}
  G^{\mathrm{Rusanov}}&=\frac{1}{2}\left(F^{k,+}n_k^{+}+F^{k,-}n_k^{-}\right)
                         \notag \\
                      &-\frac{C}{2}\left(U^+-U^-\right),
\end{align}
where $C=\max(\lvert\lambda_i(U^+)\rvert, \lvert\lambda_i(U^-)\rvert)$, and
$\lambda_i(U)$ is the set of characteristic speeds. Quantities superscripted
with a plus sign are on the exterior side of the boundary between an element and
its neighbor, while quantities superscripted with a minus sign are on the
interior side. In this section $n_k$ is the outward pointing unit normal to the
element.

The HLL solver is given by
\begin{align}
  \label{eq:HLL flux}
  G^{\mathrm{HLL}}
  &=\frac{\lambda_{\min}F^{k,+}n_k^{+} + \lambda_{\max}F^{k,-}n_k^{-}}
    {\lambda_{\max} - \lambda_{\min}} \notag\\
  &-\frac{\lambda_{\max}\lambda_{\min}}
    {\lambda_{\max}-\lambda_{\min}}\left(U^+-U^-\right),
\end{align}
where $\lambda_{\min}$ and $\lambda_{\max}$ are estimates for the fastest left-
and right-moving signal speeds, respectively. We compute the approximate signal
speeds pointwise using the scheme presented in
Ref.~\cite{Davis:1988}. Specifically,
\begin{align}
  \label{eq:approximate signal speeds}
  \lambda_{\min}&=\min(\lambda_i(U^+),\lambda_i(U^-),0),\notag \\
  \lambda_{\max}&=\max(\lambda_i(U^+),\lambda_i(U^-),0).
\end{align}

\subsection{Time stepping}
\label{sec:timestepping}

\texttt{SpECTRE} supports time integration using explicit multistep and substep
integrators.  The results presented here were obtained using either a strong
stability-preserving third-order Runge-Kutta method~\cite{Hesthaven2008} or a
self-starting Adams-Bashforth method.  \texttt{SpECTRE} additionally supports
local time-stepping when using Adams-Bashforth
schemes~\cite{2018arXiv181102499T}, but that feature was not used for any of
these problems. The maximum admissible time step size for a $P_{N}$ scheme
is~\cite{1990MaCom..54..545C}
\begin{align}
  \label{eq:DG time step size restriction}
  \Delta t\le\frac{c}{d(2N+1)}\frac{\Delta x}{\lambda_{\max}},
\end{align}
where $c$ is a time-stepper-dependent constant, $d$ is the number of spatial
dimensions, $\Delta x$ is the minimum 1d size (along each Cartesian axis) of the
element, and $\lambda_{\max}$ is the maximum characteristic speed in the
element.

\subsection{Limiting}
\label{sec:limiting}

Near shocks, discontinuities, and stellar surfaces, the DG solution may exhibit
spurious oscillations (i.e., Gibbs phenomenon) and overshoots. These
oscillations can lead to a non-physical fluid state (e.g., negative densities)
at individual grid points and prevent stable evolution of the system.  To
maintain a stable scheme, some nonlinear limiting procedure is necessary. In
general, we identify elements where the solution contains spurious oscillations
(we label these elements as ``troubled cells'') and we modify the solution on
these elements to reduce the amount of oscillation.

In this work we consider limiters that preserve the order of the DG solution
while maintaining a compact (nearest-neighbor) stencil. The compact stencil
greatly simplifies communication patterns, but, in order to provide the limiter
with sufficient information to preserve the order of the scheme, it becomes
necessary to send larger amounts of data from each element for each limiting
step. We specifically consider
\begin{itemize}
\item the $\Lambda\Pi^N$ limiter of~\cite{Cockburn1999}
\item the hierarchical limiter of Krivodonova~\cite{2007JCoPh.226..879K}
\item the simple WENO limiter of~\cite{2013JCoPh.232..397Z} (based on weighted
  essentially non-oscillatory, often abbreviated as WENO, finite volume methods)
\item the Hermite WENO (HWENO) limiter of~\cite{2016CCoPh..19..944Z}
\item a DG-finite-difference hybrid scheme similar to that
  of~\cite{2007JCoPh.224..970C, 2014JCoPh.278...47D}
\end{itemize}
Note that we do not use the limiter of Moe, Rossmanith, and
Seal~\cite{2015arXiv150703024M} because our experiments show that it is not very
robust for the kinds of problems we study here.

Below we summarize the action of these limiters. Note that because computing the
characteristic variables of the GRMHD system is complicated, we apply the
limiters to the evolved (i.e., conserved) variables.  However, we do not limit
the divergence-cleaning field $\Phi$, as it is not expected to form any
shocks. The limiters are applied at the end of each time step when using an
Adams-Bashforth method, and at the end of each substep when using a Runge-Kutta
method.

\subsubsection{$\Lambda\Pi^N$\label{sec:LambdaPiN}}

The $\Lambda\Pi^N$ limiter~\cite{Cockburn1999, 1989JCoPh..84...90C,
  1990MaCom..54..545C, 1998JCoPh.141..199C} works by reducing the spatial slope
of each variable $U$ if the data look like they may contain oscillations.
Specifically, if the slope exceeds a simple estimate based on differencing the
cell-average of $U$ vs the neighbor elements' cell-averages of $U$, then the
limiter will linearize the solution and reduce its slope in a conservative
manner. We use the total variation bounding (TVB) version of this limiter, which
only activates if the slope is above $mh^2$, where $m$ is the so-called TVB
constant and $h$ is the size of the DG element.  This procedure is repeated
independently for each variable component $U$ being limited. While quite simple
and robust, this limiter is very aggressive and can cause significant smearing
of shocks and flattening of smooth extrema.

\subsubsection{Krivodonova limiter\label{sec:Krivodonova}}

The Krivodonova limiter~\cite{2007JCoPh.226..879K} works by limiting the
coefficients of the solution's modal representation, starting with the highest
coefficient then decreasing in order until no more limiting is necessary. This
procedure is repeated independently for each variable component $U$ being
limited. Although the algorithm is only described in one or two dimensions, the
limiting algorithm is straightforwardly generalized to our 3d application. We
expand $U$ over a basis of Legendre polynomials $P_i$,
\begin{equation}
  U^{l,m,n} = \sum_{i,j,k=0,0,0}^{N,N,N} c^{l,m,n}_{i,j,k}
  P_{i}(\xi)P_{j}(\eta)P_{k}(\zeta),
\end{equation}
where the $c^{l,m,n}_{i,j,k}$ are the modal coefficients, with the superscript
$\{l,m,n\}$ representing the element indexed by $l, m, n$, and the upper bound
$N$ is the number of collocation points minus one in each of the
$\xi, \eta, \zeta$ directions.

Each coefficient is limited by comparison with the coefficients of $U$ in
neighboring elements. The new value $\tilde{c}^{l,m,n}_{i,j,k}$ of
$c^{l,m,n}_{i,j,k}$ is computed according to
\begin{align}
  \tilde{c}^{l,m,n}_{i,j,k} &= \mathrm{minmod} \Bigl( c_{i,j,k}^{l,m,n},\notag\\
                            &\alpha_i\left(c^{l+1,m,n}_{i-1,j,k}-c^{l,m,n}_{i-1,j,k}\right),
                              \alpha_i\left(c^{l,m,n}_{i-1,j,k}-c^{l-1,m,n}_{i-1,j,k}\right),\notag\\
                            &\alpha_j\left(c^{l,m+1,n}_{i,j-1,k}-c^{l,m,n}_{i,j-1,k}\right),
                              \alpha_j\left(c^{l,m,n}_{i,j-1,k}-c^{l,m-1,n}_{i,j-1,k}\right),\notag\\
                            &\alpha_k\left(c^{l,m,n+1}_{i,j,k-1}-c^{l,m,n}_{i,j,k-1}\right),
                              \alpha_k\left(c^{l,m,n}_{i,j,k-1}-c^{l,m,n-1}_{i,j,k-1}\right)\Bigr),
\end{align}
where $\mathrm{minmod}$ is the minmod function defined as
\begin{align}
  &\mathrm{minmod}(a,b,\ldots) = \notag\\
  &\left\{
    \begin{array}{ll}
      \mathrm{sgn}(a)\min(\lvert a\rvert, \lvert b\rvert, \ldots),
      & \mathrm{if} \; \mathrm{sgn}(a)=\mathrm{sgn}(b)=\mathrm{sgn}(\ldots) \\
      0, & \mathrm{otherwise},
    \end{array}\right.
\end{align}
and the $\alpha_i, \alpha_j, \alpha_k$ set the strength of the limiter. In all
cases shown in this paper, we set $\alpha_i = 1$, at the least dissipative end
of the range for these parameters%
\footnote{Whereas Krivodonova~\cite{2007JCoPh.226..879K} changes normalization
  convention for the Legendre polynomials in going from one to two dimensions,
  our convention matches their 1d convention in all cases, so that the range of
  the $\alpha_i$ parameters is given by Eq.~(14) in the reference.}.

The algorithm for limiting from highest to lowest modal coefficient is as
follows. We first compute $\tilde{c}_{N,N,N}$ (we drop the element superscripts
here). If this is equal to $c_{N,N,N}$, no limiting is done. Otherwise, we
update $c_{N,N,N} = \tilde{c}_{N,N,N}$, and compute the trio of coefficients
$\tilde{c}_{N,N,N-1}, \tilde{c}_{N,N-1,N}, \tilde{c}_{N-1,N,N}$. If \emph{all}
of these are unchanged, the limiting stops. Otherwise, we update each
coefficient and proceed to limiting all coefficients given by index permutations
such that $i+j+k=3N-2$, then $i+j+k=3N-3$, etc.~up to the three index
permutations of $c_{1,0,0}$. Finally, the limited modal coefficients are used to
recover the limited nodal values of the function $U$. Note that by not modifying
$c_{0,0,0}$ the cell-average is maintained.

\subsubsection{Simple WENO\label{sec:simple WENO}}

For the two WENO limiters, we use a troubled-cell indicator based on the TVB
minmod limiter~\cite{1989JCoPh..84...90C,1990MaCom..54..545C,
  1998JCoPh.141..199C} to determine whether limiting is needed. When needed,
each limiter uses a standard WENO procedure to reconstruct the local solution
from several different estimated solutions.

In the simple WENO limiter~\cite{2013JCoPh.232..397Z}, each variable component
$U$ being limited is checked independently: if it is flagged for slope reduction
by the minmod limiter, then this component is reconstructed.  This limiter uses
several different estimated solutions for $U$ on the troubled element labeled by
$k$. The first estimate is the unlimited local data $U^k$. Each neighbor $n$ of
$k$ also provides a ``modified'' solution estimate $U^{k_n}$; in the case of the
simple WENO limiter, this estimate is simply obtained by evaluating the
neighbor's solution $U^n$ on the grid points of the element $k$. We follow the
standard WENO algorithm of reconstructing the solution from a weighted sum of
these estimates,
\begin{equation}
  U^k_{\text{new}} = \omega_k U^k + \sum_n \omega_n U^{k_n},
\end{equation}
where the $\omega_i$ are the weights associated with each solution estimate, and
satisfy the normalization $\sum_i \omega_i = 1$.

The weights are obtained by first computing an oscillation indicator (also
called a smoothness indicator) $\sigma_i$ for each $U^i = \{U^k,U^{k_n}\}$,
which measures the amount of oscillation in the data. We use an indicator based
on Eq.~(23) of~\cite{2007JCoPh.221..693D}, but adapted for use on square or
cubical grids,
\begin{equation}
  \begin{aligned}
    \label{eq:weno oscillation indicator}
    \sigma_i = \sum_{\alpha=0}^N \sum_{\beta=0}^N
    \sum_{\substack{\gamma=0\\
        \alpha+\beta+\gamma > 0}}^N \int
    & 2^{2(\alpha+\beta+\gamma) - 1} \times \\
    & \left( \frac{\partial^{\alpha+\beta+\gamma}} {\partial \xi^{\alpha}
        \partial \eta^{\beta} \partial \zeta^{\gamma}} U^i \right)^2 d\xi\,
    d\eta\, d\zeta.
  \end{aligned}
\end{equation}
Here the restriction on the sum avoids the term that has no derivatives of
$U^i$, and the powers of two come from the interval width in the reference
coordinates. From the oscillation indicators, we compute the non-linear weights
\begin{equation}
  \bar{\omega}_i = \frac{\gamma_i}{(\epsilon + \sigma_i)^2}.
\end{equation}
Here the $\gamma_i$ are the linear weights that give the relative weight of the
local and neighbor contributions before accounting for oscillation in the data,
and $\epsilon$ is a small number to avoid the denominator vanishing. We use
standard values from the literature for both --- we take $\gamma_{k_n} = 0.001$
for the neighbor contributions (then $\gamma_k = 0.994$ for an element with six
neighbors; in general $\gamma_k$ is set by the requirement that all the
$\gamma_i$ sum to unity), and $\epsilon = 10^{-6}$.  Finally, the normalized
non-linear weights that go into the WENO reconstruction are given by
\begin{equation}
  \omega_i = \frac{\bar{\omega}_i}{\sum_i \bar{\omega}_i}.
\end{equation}
Note that the simple WENO limiter is not conservative since the neighboring
elements' polynomials do not have the same element-average as the element being
limited.

\subsubsection{HWENO\label{sec:HWENO}}

Our implementation of the HWENO limiter~\cite{2016CCoPh..19..944Z} follows
similar steps. Note that we again use the TVB minmod limiter as troubled-cell
indicator, whereas the reference uses the troubled-cell indicator
of~\cite{Krivodonova2004}.  But, in keeping with the HWENO algorithm, we check
the minmod indicator on all components of $U$ being limited, and if any
component is flagged for slope reduction, then the element is labeled as
troubled and every variable being limited is reconstructed using the WENO
procedure.

The HWENO modified solution estimates from the neighboring elements are computed
as a least-squared fit to $U$ across several elements. This broader fitting
reduces oscillations as compared to the polynomial extrapolation used in the
simple WENO estimates, and this improves robustness near shocks.  The HWENO
reconstruction uses a differently-weighted oscillation indicator, computed
similarly to Eq.~\eqref{eq:weno oscillation indicator} but with the prefactor in
the integral being instead
$(2^{2(\alpha+\beta+\gamma) - 1})/((\alpha+\beta+\gamma)!)^2$. The HWENO
algorithm explicitly guarantees conservation by constraining the reconstructed
polynomials to have the same element-average value.

\subsubsection{DG-finite-difference hybrid method\label{sec:DG-FD hybrid}}

To the best of our knowledge the idea of hybridizing efficient spectral-type
methods with robust high-resolution shock-capturing finite difference (FD) or
finite volume (FV) schemes was first presented
in~\cite{2007JCoPh.224..970C}. However, our implementation is more similar to
that of~\cite{2014JCoPh.278...47D}. The basic idea is that after a time step or
substep we check that the unlimited DG solution is satisfactory.  If it is not,
we mark the cell as troubled and retake the time step using standard FD
methods. In this paper we use monotized-central reconstruction and the same
numerical flux/boundary correction as the DG scheme uses. Our DG-FD hybrid
method is also similar to that used
in~\cite{Bugner:2015gqa}. However,~\cite{Bugner:2015gqa} did not attempt to run
the method in 3d because of memory overhead. We have not done a detailed
comparison of memory overhead between different limiting strategies, but have
not noticed any significant barriers with the DG-FD hybrid scheme. We present a
detailed description of our DG-FD hybrid method in a companion
paper~\cite{Deppe:2021ada}. Our DG-FD hybrid method is not strictly
conservative at boundaries where one element uses DG and another uses FD. This
is because on the DG element we use the boundary correction of the reconstructed
FD data, rather than the reconstructed boundary correction computed on the FD
grid. In practice we have not found any negative impact from this choice.

\subsection{Primitive recovery\label{sec:primitive recovery}}

One of the most difficult and expensive aspects of evolving the GRMHD equations
is recovering the primitive variables from the conserved variables. Several
different primitive recovery schemes are compared in~\cite{Siegel:2017sav}. We
use the recently proposed scheme of Kastaun \textit{et
  al.}\cite{Kastaun:2020uxr}. If this scheme fails to recover the primitives, we
try the Newman-Hamlin scheme~\cite{Newman2014}. If the Newman-Hamlin scheme
fails, we use the scheme of Palenzuela \textit{et
  al.}~\cite{Palenzuela:2015dqa}, and if that fails we terminate the
simulation. Note that we have not yet incorporated all the fixing procedures to
avoid recovery failure that are presented in~\cite{Kastaun:2020uxr}.

\subsection{Variable fixing\label{sec:variable fixing}}

During the evolution the conserved and primitive variables can become
non-physical or enter regimes where the evolution is no longer stable (e.g.,
zero density). When limiting the solution does not remove these unphysical or
bad values, a pointwise fixing procedure is used --- at any grid points where
the chosen conditions are not satisfied, the conserved variables are
adjusted. The fixing procedures are generally not conservative and are used only
as a fallback to ensure a stable evolution. In \texttt{SpECTRE} we currently use
two fixing algorithms: The first applies an ``atmosphere'' in low-density
regions, while the second adjusts the conserved variables in an attempt to
guarantee primitive recovery.

Our ``atmosphere'' treatment is similar to that
of~\cite{2013PhRvD..87h4006F,Galeazzi:2013mia, Muhlberger:2014pja}. We define
values $\rho_{\mathrm{atm}}$ and $\rho_{\mathrm{cutoff}}$, where
$\rho_{\mathrm{atm}}\le\rho_{\mathrm{cutoff}}$. For any point where
$\rho<\rho_{\mathrm{cutoff}}$ we set
\begin{equation}
  \label{eq:atmosphere values}
  \begin{split}
    \rho &= \rho_{\mathrm{atm}}, \\
    v^i &= 0, \\
    W &= 1.
  \end{split}
\end{equation}
When $\rho_{\mathrm{cutoff}}<\rho\le10\rho_{\mathrm{atm}}$ we require that
$v^iv_i<10^{-4}$. After the primitive variables are set to the atmosphere we
recompute the conserved variables from the primitive ones.

Our fixing of the conserved variables is based on that of
Refs.~\cite{Foucart2011, Muhlberger:2014pja}. We define $D_{\mathrm{min}}$
and $D_{\mathrm{cutoff}}$ and adjust $\tilde{D}$ if
$D<D_{\mathrm{cutoff}}$. Specifically, we set
$\tilde{D}=\sqrt{\gamma}D_{\mathrm{min}}$. We adjust $\tilde{\tau}$ such that
$\tilde{B}^2\le2\sqrt{\gamma}\left(1-\epsilon_B\right)\tilde{\tau}$, where
$\epsilon_B$ is a small number typically set to $10^{-12}$.

Finally, we adjust $\tilde{S}_i$ such that $\tilde{S}^2\le\tilde{S}^2_{\max}$,
where $\tilde{S}^2_{\max}$ is defined below. We define variables
\begin{align}
  \label{eq:tau hat}
  \hat{\tau}&=\frac{\tilde{\tau}}{\tilde{D}}, \\
  \label{eq:B hat}
  \hat{B}^2&=\frac{\tilde{B}^2}{\sqrt{\gamma}\tilde{D}}, \\
  \label{eq:S hat}
  \hat{\mu}
  &=\begin{cases}
    \frac{\tilde{S}_i\tilde{B}^i}{\sqrt{\tilde{B}^2\tilde{S}^2}}, &
    \tilde{B}^2>\tilde{D}\times10^{-16}\ \text{and}\
    \tilde{S}^2>\tilde{D}^2\times10^{-16}
    \\
    0, & \text{otherwise}
  \end{cases}
\end{align}
The Lorentz factor is bounded by
\begin{align}
  \label{eq:Lorentz factor bounds}
  \max(1,1+\hat{\tau}-\hat{B}^2)\le W\le1+\hat{\tau},
\end{align}
and is determined by finding the root of
\begin{align}
  \label{eq:fix const Lorentz factor function}
  g(W)&=\left(W+ \hat{B}^2 - \hat{\tau} - 1\right)
        \left[W^2 + \hat{B}^2 \hat{\mu}^2 (\hat{B}^2 + 2 W)\right] \notag \\
      &-\frac{\hat{B}^2}{2}\left[1+\hat{\mu}^2\left(W^2 +
        2 W \hat{B}^2 + \hat{B}^4 - 1\right)\right].
\end{align}
Using the Lorentz factor $W$ obtained by solving \eqref{eq:fix const Lorentz
  factor function} we define $\tilde{S}_{\max}$ as
\begin{multline}
  \label{eq:tilde S_max fix conservatives}
  \tilde{S}_{\max}= \tilde{S}\, \min\left(1,
    \phantom{\sqrt{\frac{a}{\left(\hat{B}^2\right)}}}
  \right. \\
  \left.\sqrt{ \frac{\left(1-\epsilon_S\right)
        \left(W+\hat{B}^2\right)^2\left(W^2-1\right)\tilde{D}^2}
      {\left(\tilde{S}^2 + \tilde{D}^2\times 10^{-16}\right) \left[W^2 +
          \hat{\mu}^2\hat{B}^2 \left(\hat{B}^2+2W\right)\right]}}\right),
\end{multline}
where $\epsilon_S$ is a small number typically set to $10^{-12}$.  We apply the
check on the conserved variables after each time or sub step before a primitive
recovery is done.

Implementing root finding for Eq.~\eqref{eq:fix const Lorentz factor function}
in a manner that is well-behaved for floating point arithmetic is
important. Specifically, we solve
\begin{align}
  \label{eq:fix const Lorentz factor 0 lower bound}
  g(W) &= \left(\cfrac{1}{2}\hat{B}^2 - \hat{\tau}\right) \left(1 + 2 \hat{B}^2
         \hat{\mu}^2 + \hat{B}^4 \hat{\mu}^2\right) \notag \\
       &+ \left(W-1\right) \left[2 \left(\hat{B}^2 - \hat{\tau}\right) \left(1
         + \hat{B}^2 \hat{\mu}^2\right) + \hat{B}^2 \hat{\mu}^2 + 1\right]
         \notag \\
       &+ \left(W-1\right)^2 \left(\hat{B}^2 - \hat{\tau} + \cfrac{3}{2}
         \hat{B}^2 \hat{\mu}^2 + 2\right) \notag \\
       &+ \left(W-1\right)^3
\end{align}
for $W-1$ when the lower bound for $W$ is 1 and
\begin{align}
  \label{eq:fix const Lorentz factor non-zero lower bound}
  g(W) &= - \cfrac{1}{2} \hat{B}^2 \left[1 + \hat{\mu}^2 \hat{\tau}
         \left(\hat{\tau} + 2\right)\right] \notag \\
       &+ \left[W-\left(1 + \hat{\tau}-\hat{B}^2\right)\right]
         \left[1 + \hat{B}^2 \hat{\mu}^2 \right. \notag \\
       &\left.+ \left(\hat{\tau}-\hat{B}^2\right)
         \left(\hat{B}^2 \hat{\mu}^2 + \hat{\tau}-\hat{B}^2 +
         2\right)\right] \notag \\
       &+ \left[W-\left(1 + \hat{\tau}-\hat{B}^2\right)\right]^2 \notag \\
       & \times \left[2 \left(\hat{\tau}-\hat{B}^2\right) + \cfrac{3}{2}
         \hat{B}^2
         \hat{\mu}^2 + 2\right]
         \notag \\
       &+ \left[W-\left(1 + \hat{\tau}-\hat{B}^2\right)\right]^3
\end{align}
for $W-\left(1 + \hat{\tau}-\hat{B}^2\right)$ when the lower bound for $W$ is
$1 + \hat{\tau}-\hat{B}^2$.

We also have a flattening algorithm inspired by~\cite{BALSARA20127504} that
reduces oscillations of the conserved variables if the solution is
unphysical. Unlike the pointwise fixing, the flattening algorithm is
conservative. In particular, we reduce the oscillations in $\tilde{D}$ if it is
negative an at any point in the cell, and we rescale $\tilde{\tau}$ to satisfy
$\tilde{B}^2\le2\sqrt{\gamma}\tilde{\tau}$. Finally, if the primitive variables
cannot be recovered we reset the conserved variables to their mean values.

\section{Numerical results}
\label{sec:results}

For all test problems we use the less dissipative HLL boundary correction. In
many cases one of the limiting strategies fails. This failure usually occurs
during the primitive recovery. However, this is a symptom of the DG and limiting
procedure producing a bad state rather than a poor primitive recovery
algorithm. All simulations are performed using \texttt{SpECTRE}
v2022.04.04~\cite{spectrecode} and the input files used are provided
alongside the arXiv version.

\subsection{1d smooth flow\label{sec:Smooth Flow}}

We consider a simple 1d smooth flow problem to test which of the limiters and
troubled-cell indicators are able to solve a smooth problem without degrading
the order of accuracy. A smooth density perturbation is advected across the
domain with a velocity $v^i$. The analytic solution is given by
\begin{align}
  \rho&=1 + 0.7 \sin[k^i (x^i-v^it)], \\
  v^i&=(0.8,0,0),\\
  k^i&=(1,0,0),\\
  p&=1,\\
  B^i&=(0,0,0),
\end{align}
and we close the system with an adiabatic equation of state,
\begin{align}
  \label{eq:ideal fluid eos}
  p = \rho\epsilon\left(\Gamma-1\right),
\end{align}
where $\Gamma$ is the adiabatic index, which we set to 1.4. We use a domain
given by $[0,2\pi]^3$ and apply periodic boundary conditions in all directions.
The time step size is $\Delta t = 2\pi/ 5120$ so that the spatial discretization
error is larger than the time stepping error for all resolutions we use.

\begin{table}
  \centering
  \caption{\label{tab:Smooth flow errors}The errors and local convergence order
    for the smooth flow problem using different limiting strategies. Note that
    the limiter is not applied if the troubled-cell indicator determines the DG
    solution to be valid. Except for the Krivodonova limiter, which is
    non-convergent, we observe the expected convergence order except when the
    solution is underresolved because too few elements are used.}
  \begin{tabularx}{\columnwidth}{@{\extracolsep{\stretch{1}}}*{4}{lccc}@{}}
    \hline\hline
    Limiter & $N_x$ & $L_2(\mathcal{E}(\rho))$ & $L_2$ order\\ \hline
    $\Lambda\Pi^N$ & \phantom{0}2 & $2.22282\times10^{-3}$ & \\
                   & \phantom{0}4 & $2.23822\times10^{-5}$ & \phantom{-0}6.63\\
                   & \phantom{0}8 & $3.18504\times10^{-7}$ & \phantom{-0}6.13\\
                   & 16 & $5.08821\times10^{-9}$ & \phantom{-0}5.97\\
    \hline
    HWENO & \phantom{0}2 & $2.22282\times10^{-3}$ & \\
                   & \phantom{0}4 & $2.23822\times10^{-5}$ & \phantom{-0}6.63\\
                   & \phantom{0}8 & $3.18504\times10^{-7}$ & \phantom{-0}6.13\\
                   & 16 & $5.08821\times10^{-9}$ & \phantom{-0}5.97\\
    \hline
    Simple WENO & \phantom{0}2 & $2.22282\times10^{-3}$ & \\
                   & \phantom{0}4 & $2.23822\times10^{-5}$ & \phantom{-0}6.63\\
                   & \phantom{0}8 & $3.18504\times10^{-7}$ & \phantom{-0}6.13\\
                   & 16 & $5.08821\times10^{-9}$ & \phantom{-0}5.97\\
    \hline
    Krivodonova & \phantom{0}2 & $3.92346\times10^{-1}$ & \\
                   & \phantom{0}4 & $4.94975\times10^{-1}$ & \phantom{-0}-0.34\\
                   & \phantom{0}8 & $4.94975\times10^{-1}$ & \phantom{-0}0.00\\
                   & 16 & $4.73294\times10^{-1}$ & \phantom{-0}0.06\\
    \hline
    DG-FD P$_5$ & \phantom{0}2 & $3.45679\times10^{-1}$ & \\
                   & \phantom{0}4 & $2.23822\times10^{-5}$ & \phantom{-}13.91\\
                   & \phantom{0}8 & $3.18504\times10^{-7}$ & \phantom{-0}6.13\\
                   & 16 & $5.08821\times10^{-9}$ & \phantom{-0}5.97\\
    \hline
    \hline
  \end{tabularx}
\end{table}

We perform a convergence test using the different limiting strategies and
present the results in Table~\ref{tab:Smooth flow errors}. We show both the
$L_2$ norm of the error and the convergence order. The $L_2$ norm is defined as
\begin{align}
  \label{eq:L2 norm}
  L_2(u)=\sqrt{\frac{1}{M}\sum_{i=0}^{M-1}u_i^2},
\end{align}
where $M$ is the total number of grid points and $u_i$ is the value of $u$ at
grid point $i$ and the convergence order is given by
\begin{align}
  \label{eq:convergence order}
  L_2\;\mathrm{order} =
  \log_2\left[\frac{L_2(\mathcal{E}_{N_x/2})}{L_2(\mathcal{E}_{N_x})}\right]
\end{align}
We see that the troubled-cell indicator for the $\Lambda\Pi^N$, HWENO, and
simple WENO limiters does not flag any cells as troubled and the full order of
accuracy of the DG scheme is preserved. For these simulations we used a TVB
constant of 1. The Krivodonova limiter completely flattens the solution and
shows no convergence. The reason is that the Krivodonova limiter is unable to
preserve a smooth solution if the flow is constant in an orthogonal
direction. This can be understood from the minmod algorithm being applied to the
neighboring coefficients. The smooth flow solution is constant in the $y$- and
$z$-directions, and so the Krivodonova limiter effectively zeros all higher
moments. The DG-FD P$_5$ scheme switches to FD when we use only two elements,
but from four to 16 elements it uses DG. The order of convergence is so large
for the $N_x=4$ case because in addition to doubling the resolution, the code
also switches from using second-order FD to sixth-order DG, causing a very large
decrease in the errors. Using a higher-order or adaptive-order FD scheme is
expected to preserve the accuracy much better when the hybrid scheme is using
FD, while still being able to capture shocks robustly and accurately.

\subsection{1d Riemann problems}

\begin{figure}
  \begin{tabular}{c}
    \subfloat[Riemann Problem 1 comparison]
    {
    \includegraphics[width=0.45\textwidth]{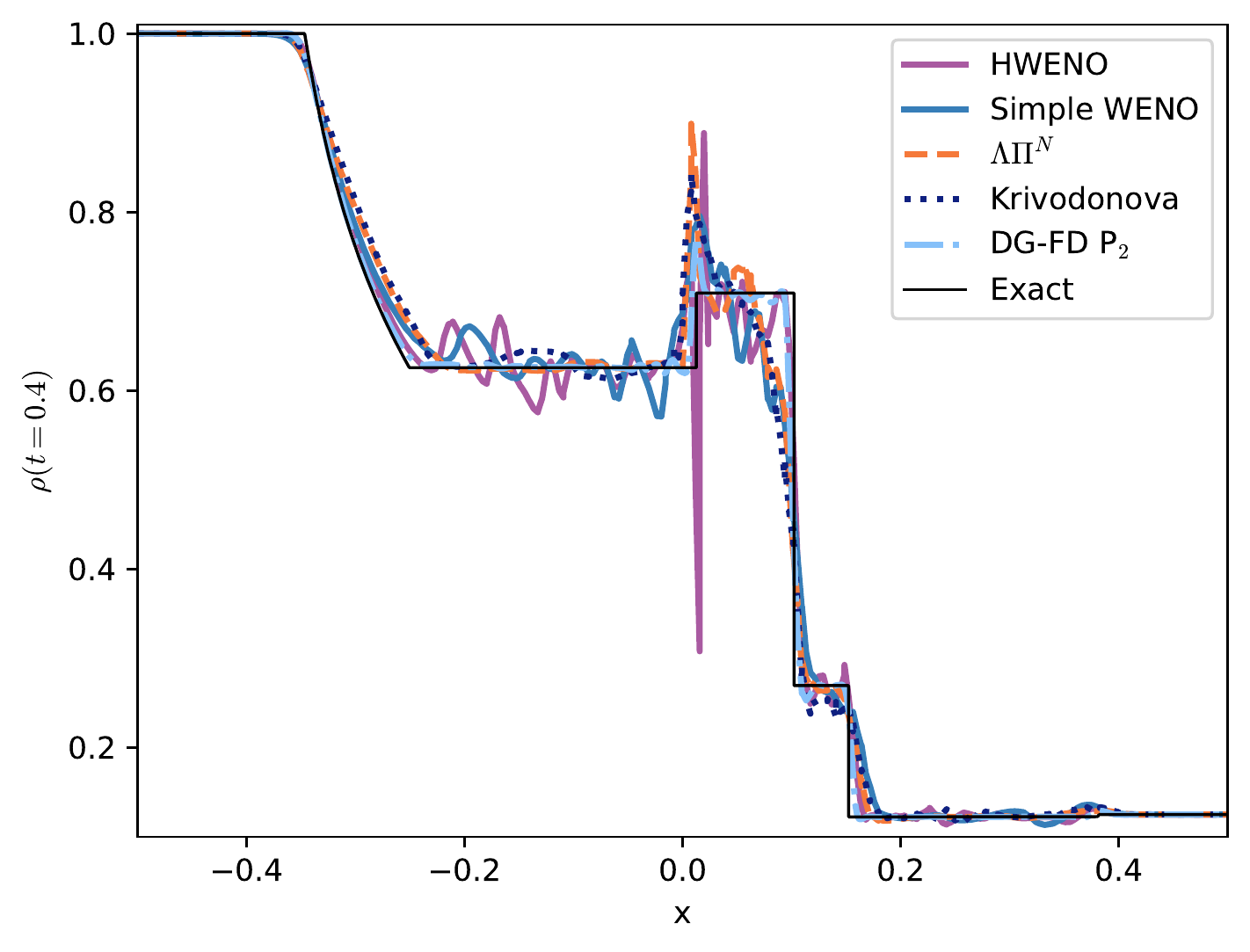}}
    \\
    \subfloat[Zoom in of Riemann Problem 1]
    {\includegraphics[width=0.45\textwidth]{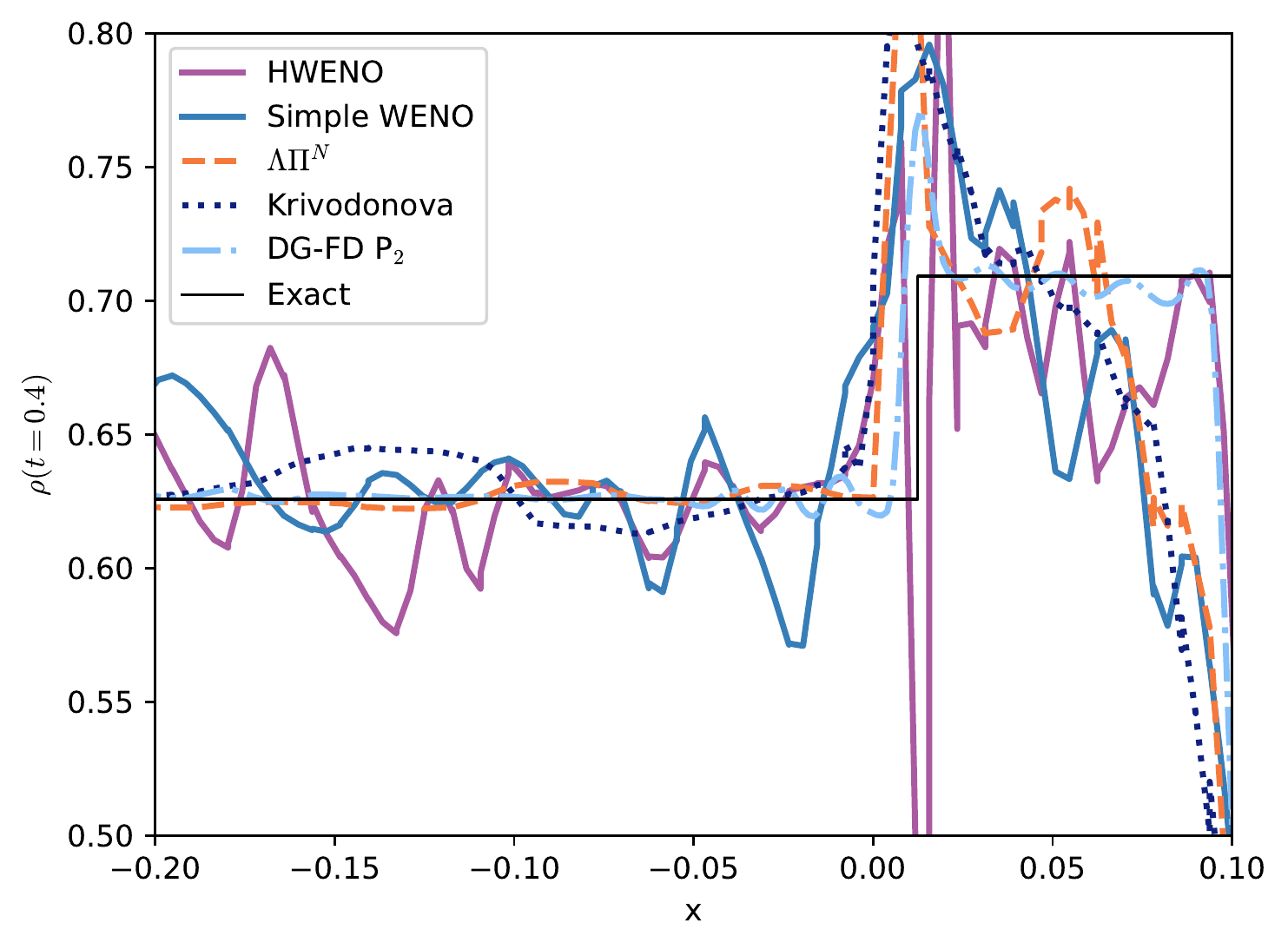}}
  \end{tabular}
  \caption{\label{fig:Rp1Comparison}A comparison of different limiters used to
    stabilize the evolution of the Riemann Problem 1 of~\cite{Balsara2001}. The
    problem is solved using 128 third-order (P$_2$) elements. The DG-FD hybrid
    scheme significantly outperforms the other limiters both in robustness and
    accuracy.}
\end{figure}

One-dimensional Riemann problems are a standard test for any scheme that must be
able to handle shocks. We will focus on the first Riemann problem (RP1)
of~\cite{Balsara2001}. The setup is given in Table~\ref{tab:Rp1 conditions}.
While not
the most challenging Riemann problem, it gives a good baseline for different
limiting strategies. We perform simulations using an SSP-RK3 method with
$\Delta t=5\times10^{-4}$. In Fig.~\ref{fig:Rp1Comparison} we show the rest mass
density $\rho$ at $t_f=0.4$ for simulations using the simple WENO, HWENO,
$\Lambda\Pi^N$, and Krivodonova limiters, as well as a run using the DG-FD
hybrid scheme. The thin black curve is the analytic solution obtained using the
Riemann solver of~\cite{Giacomazzo2006}. All simulations use 128 elements
in the $x$
direction with a P$_2$ (third-order) DG scheme, and an ideal fluid equation of
state, Eq.~\ref{eq:ideal fluid eos}.

\begin{table}
  \centering
  \caption{\label{tab:Rp1 conditions}The initial conditions for Riemann Problem
    1 of~\cite{Balsara2001}. The domain is $x\in[-0.5,0.5]$, the final time is
    $t_f=0.4$, and an ideal fluid equation of state is used with an adiabatic
    index of 2.}
  \begin{tabularx}{\columnwidth}{@{\extracolsep{\stretch{1}}}*{5}{c}@{}}
    \hline\hline
    & $\rho$ & $p$ & $v^i$ & $B^i$ \\ \hline
    $x < 0$ & 1.0\phantom{00} & 1.0 & $(0,0,0)$ & $(0.5,\phantom{-}1,0)$ \\
    $x \ge 0$ & 0.125 & 0.1 & $(0,0,0)$ & $(0.5,-1,0)$ \\
    \hline\hline
  \end{tabularx}
\end{table}

\begin{figure}
  \includegraphics[width=0.45\textwidth]{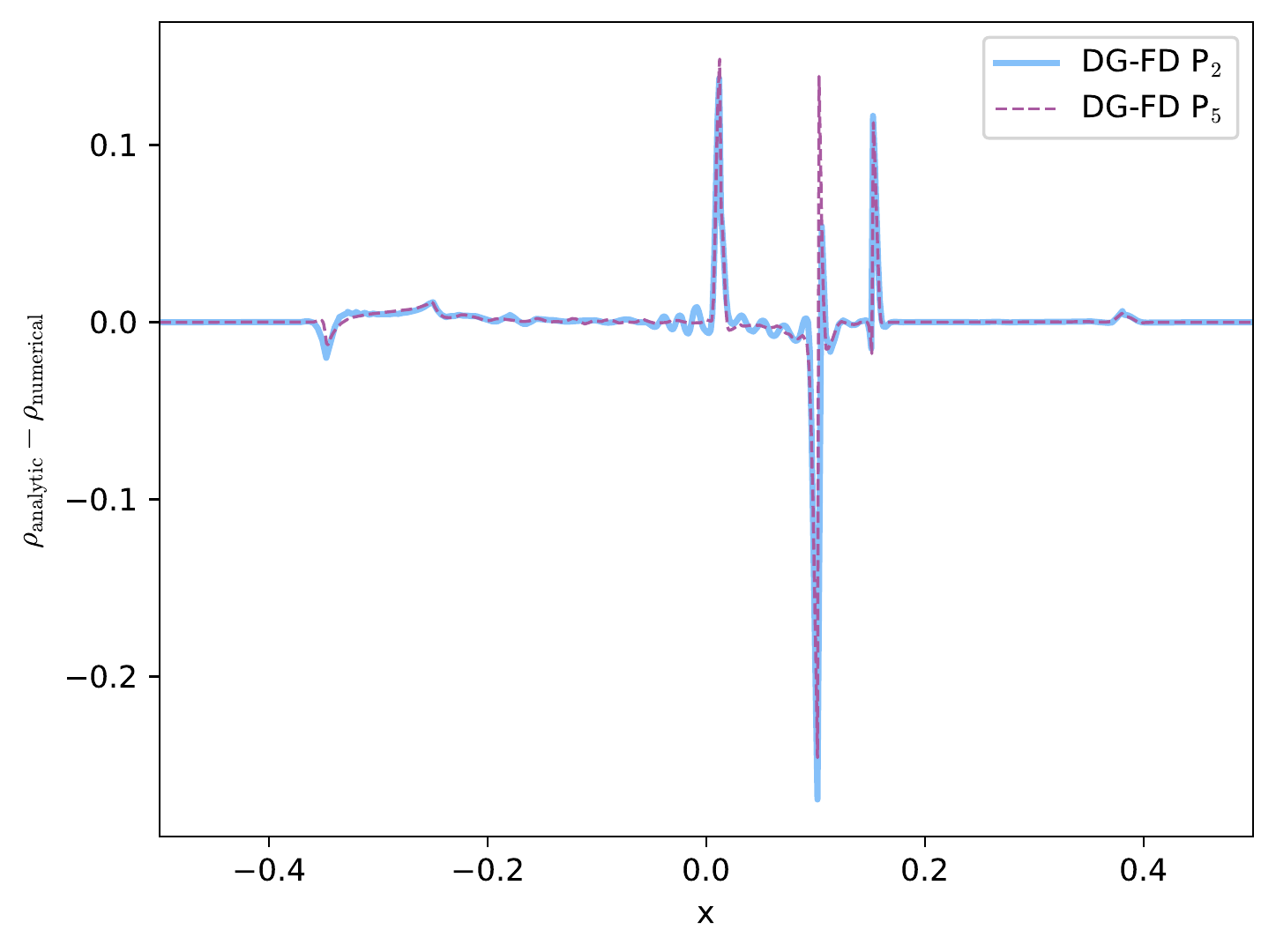}
  \caption{\label{fig:Rp1ErrorsSubcell}The difference between the analytic and
    numerical solution of the Riemann Problem 1 of~\cite{Balsara2001} at $t=0.4$
    for the DG-FD P$_2$ scheme (solid light blue curve) and the DG-FD P$_5$
    scheme (dashed purple curve). The P$_5$ scheme is able to resolve the
    discontinuities just as well as the P$_2$ scheme, while also admitting fewer
    unphysical oscillations away from the discontinuities.}
\end{figure}

While all five limiting strategies evolve to the final time, the DG-FD scheme is
the least oscillatory and is also able to resolve the discontinuities much more
accurately. The HWENO scheme is slightly less oscillatory if linear neighbor
weights of $\gamma_k=0.01$ are used instead of $\gamma_k=0.001$. However, the
simple WENO limiter fails to evolve the solution with $\gamma_k=0.01$ and such
sensitivity to parameters in the algorithm is not desirable when solving
realistic problems. Going to higher order has proven to be especially
challenging. While both the $\Lambda\Pi^N$ and the Krivodonova complete the
evolution when using a P$_5$ scheme with 64 elements (simple WENO and HWENO
fail), additional spurious oscillations are present. In comparison, the DG-FD
hybrid scheme actually has fewer oscillations when going to higher order. In
Fig.~\ref{fig:Rp1ErrorsSubcell} we plot the error of the numerical solution
using a P$_2$ DG-FD scheme with 128 elements and a P$_5$ DG-FD scheme with 64
elements. We see that the P$_5$ hybrid scheme actually has fewer oscillations
than the P$_2$ scheme, while resolving the discontinuities equally well. We
attribute this to the troubled-cell indicators actually triggering earlier
when a higher polynomial degree is used since discontinuities entering an
element rapidly dump energy into the high modes.  While we will compare the
different limiting strategies for 2d and 3d problems below, it is already quite
apparent that the DG-FD hybrid scheme is by far the most robust and accurate
method.

\subsection{2d cylindrical blast wave}

A standard test problem for GRMHD codes is the cylindrical blast
wave~\cite{2005A&A...436..503L,DelZanna:2007pk}, where a magnetized fluid
initially at rest in a constant magnetic field along the $x$-axis is
evolved. The fluid obeys the ideal fluid equation of state \eqref{eq:ideal
  fluid eos} with $\gamma=4/3$.  The fluid begins in a cylindrically symmetric
configuration, with hot, dense fluid in the region with cylindrical radius
$r < 0.8$ surrounded by a cooler, less dense fluid in the region $r > 1$. The
initial density $\rho$ and pressure $p$ of the fluid are
\begin{equation}
\begin{split}
  \rho(r < 0.8) & = 10^{-2}, \\
  \rho(r > 1.0) & = 10^{-4}, \\
  p(r < 0.8) & = 1, \\
  p(r > 1.0) & = 5 \times 10^{-4}.
\end{split}
\end{equation}
In the region $0.8 \leq r \leq 1$, the solution transitions continuously and
exponentially (i.e., transitions such that the logarithms of the pressure and
density are linear functions of $r$).  The fluid begins threaded with a uniform
magnetic field with Cartesian components
\begin{equation}
  (B^x, B^y, B^z) = (0.1, 0, 0).
\end{equation}
The magnetic field causes the blast wave to expand non-axisymmetrically.  For
all simulations we use a time step size $\Delta t=10^{-2}$ and an SSP RK3 time
integrator.

\begin{figure*}
  \begin{tabular}{ccc}
    \subfloat[DG-FD, P$_2$, $64^2$ elements]
    {\label{fig:BlastWaveSubcellP2}
    \includegraphics[width=0.3\textwidth]{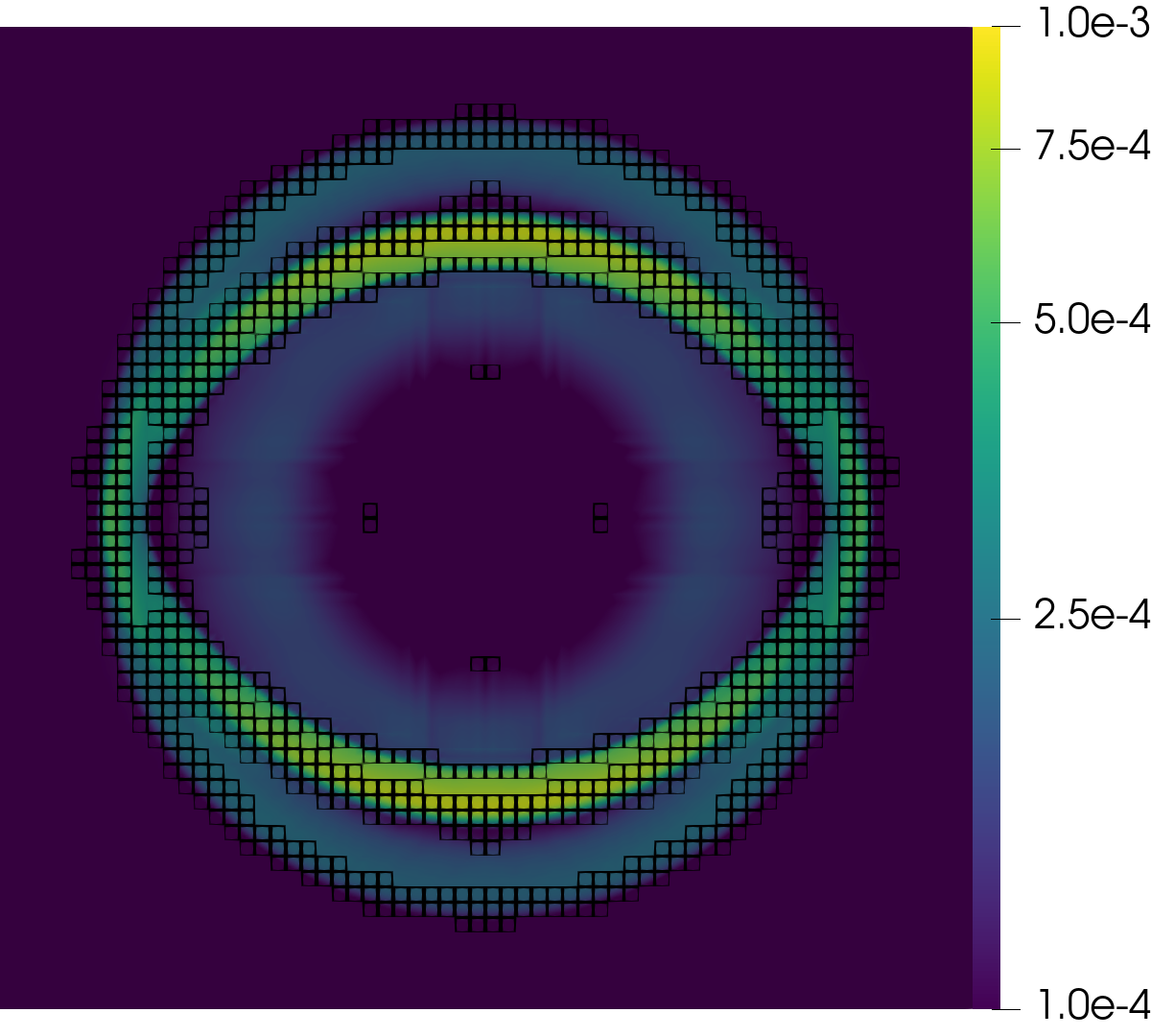}}
    &
      \subfloat[DG-FD, P$_5$, $32^2$ elements]
      {\label{fig:BlastWaveSubcellP5}
      \includegraphics[width=0.3\textwidth]{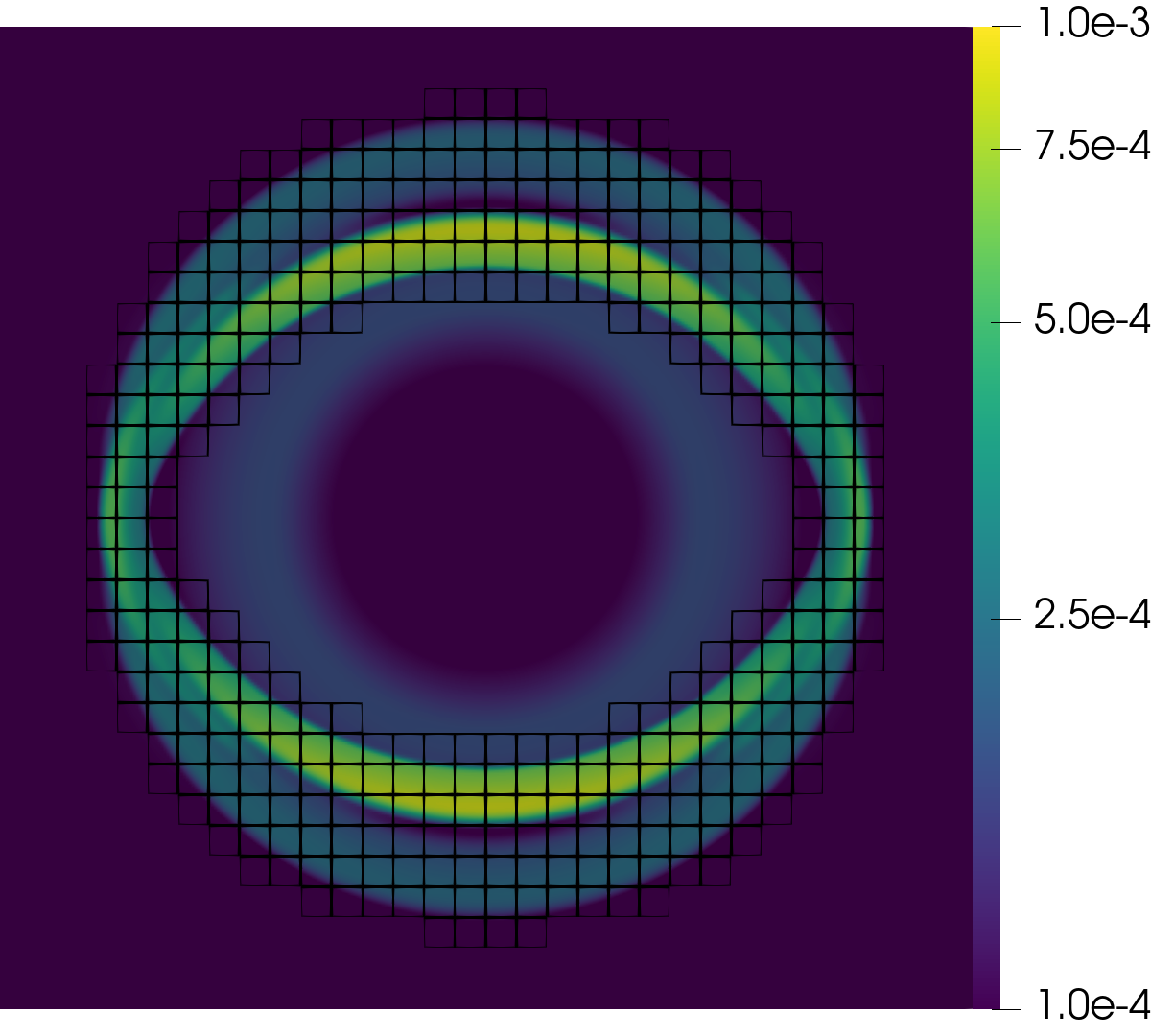}}
    &
      \subfloat[$\Lambda\Pi^N$, P$_2$, $64^2$ elements]
      {\label{fig:BlastWaveLambdaPiN}
      \includegraphics[width=0.3\textwidth]{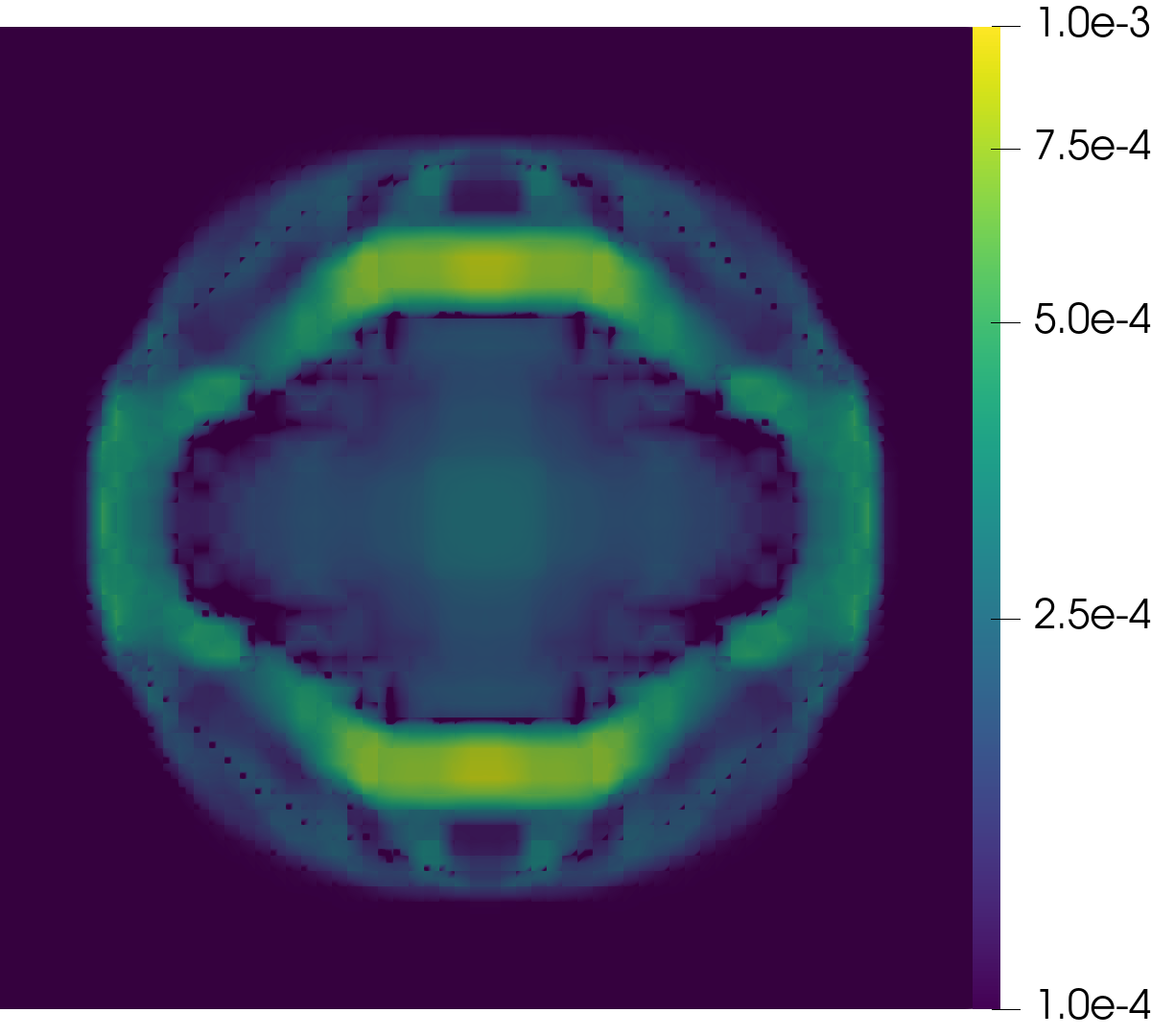}}
    \\
    \subfloat[Krivodonova, P$_2$, $64^2$ elements]
    {\label{fig:BlastWaveKrivodonova}
    \includegraphics[width=0.3\textwidth]{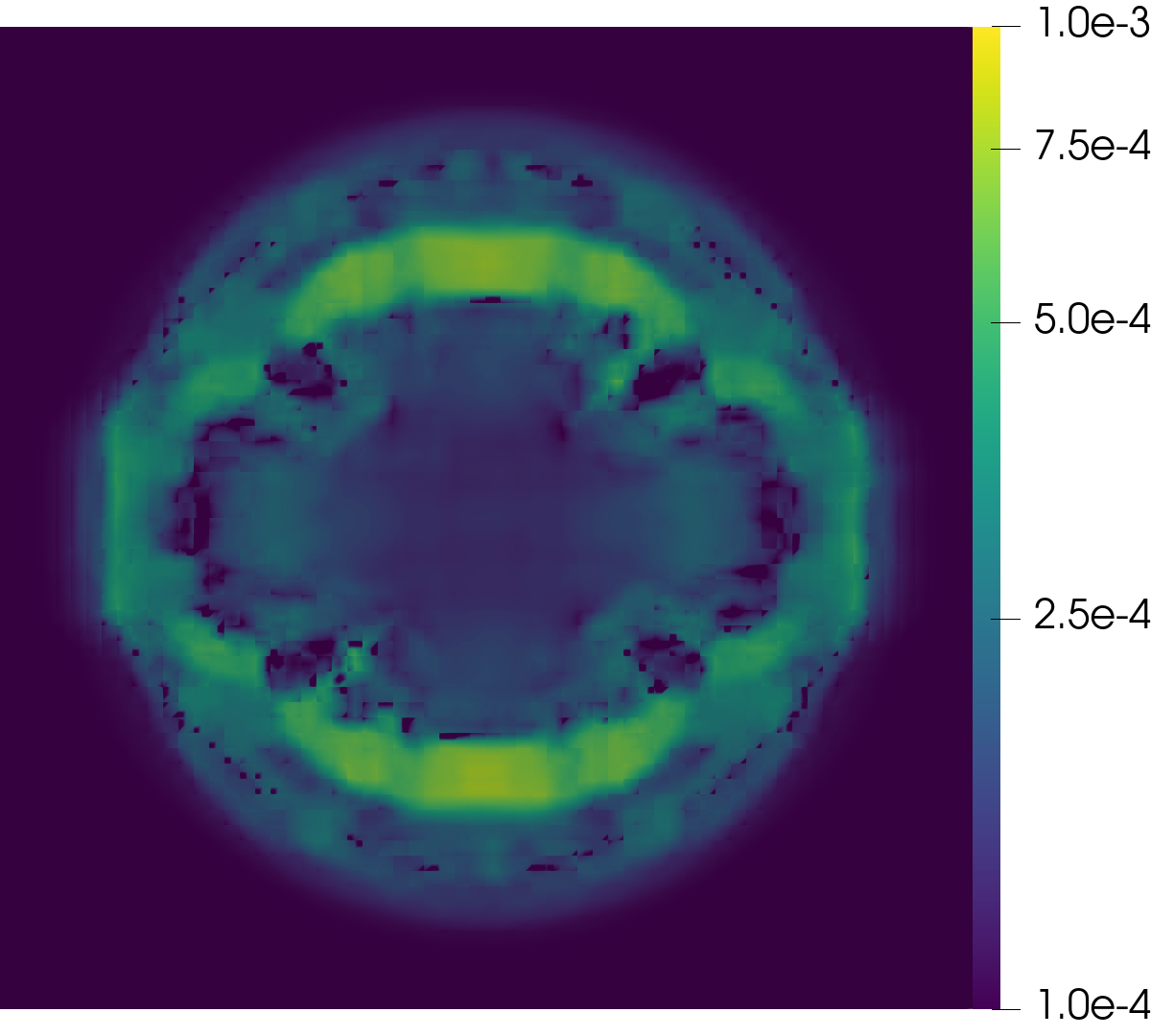}}
    &
      \subfloat[Simple WENO, P$_2$, $64^2$ elements]
      {\label{fig:BlastWaveSimpleWeno}
      \includegraphics[width=0.3\textwidth]{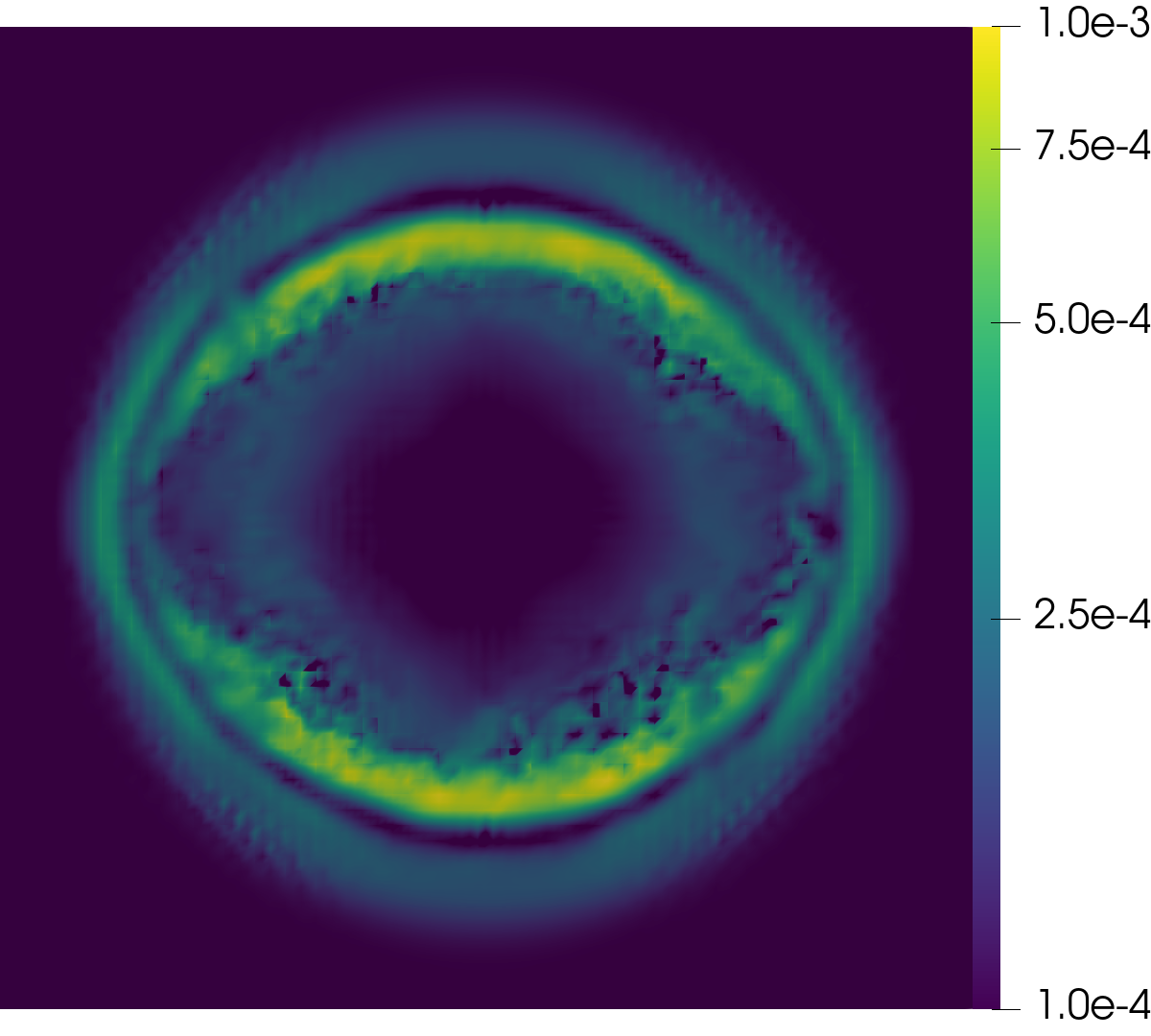}}
    &
      \subfloat[HWENO, P$_2$, $64^2$ elements]
      {\label{fig:BlastWaveHweno}
      \includegraphics[width=0.3\textwidth]{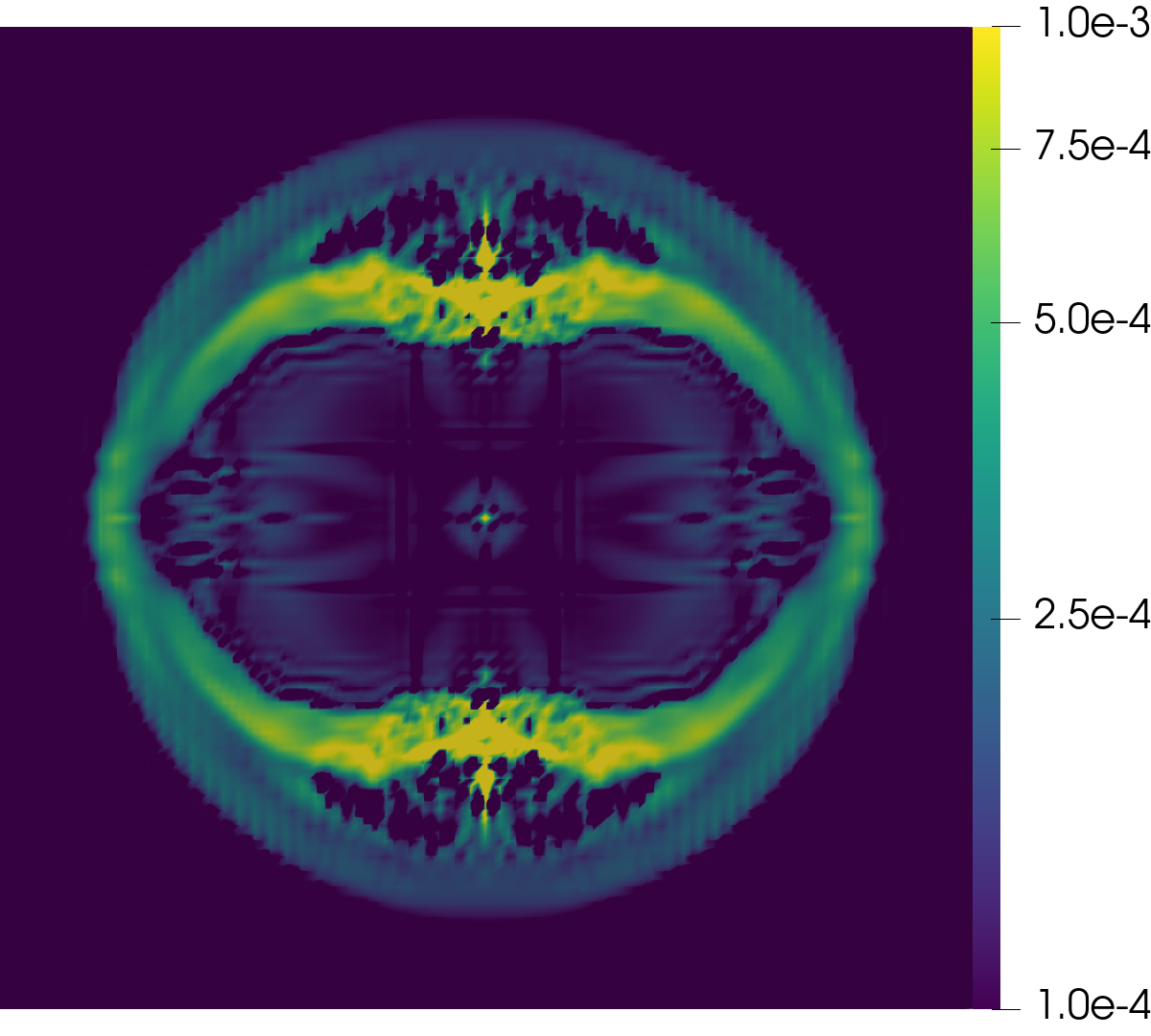}}
    \\
  \end{tabular}
  \caption{\label{fig:BlastWave}Cylindrical blast wave $\rho$ at $t=4$ comparing
    the DG-FD hybrid scheme, the $\Lambda\Pi^N$, Krivodonova, simple WENO, and
    HWENO limiters using P$_2$ DG, as well as the DG-FD scheme using P$_5$
    DG. There are 192 degrees of freedom per dimension, comparable to what is
    used when testing FD schemes. We see that only the DG-FD hybrid scheme
    really resolves the features to an acceptable level, and the $\Lambda\Pi^N$
    and Krivodonova smear out the solution almost completely. In the plots of
    the DG-FD hybrid scheme the regions surrounded by black squares have
    switched from DG to FD at the final time.}
\end{figure*}

We evolve the blast wave to time $t=4.0$ on a grid of $64\times 64 \times 1$
elements covering a cube of extent $[-6, 6]^3$ using a DG P$_2$ scheme, a
comparable resolution to what FD code tests use.  We apply periodic boundary
conditions in all directions, since the explosion does not reach the outer
boundary by $t=4.0$. Figure~\ref{fig:BlastWave} shows the logarithm of the
rest-mass density at time $t=4.0$, at the end of evolutions using the different
limiting strategies. We see from Fig.~\ref{fig:BlastWaveLambdaPiN} and
Fig.~\ref{fig:BlastWaveKrivodonova} that the $\Lambda\Pi^N$ and Krivodonova
limiters result in a very poorly resolved solution. The simple WENO evolution,
Fig.~\ref{fig:BlastWaveSimpleWeno} is much better but still not nearly as good
as a FD method with the same number of degrees of freedom. The HWENO limiter,
Fig.~\ref{fig:BlastWaveHweno}, suffers from various spurious artifacts. The
DG-FD hybrid scheme, however, again demonstrates its ability to robustly handle
discontinuities, while also resolving smooth features with very high
order. Figure~\ref{fig:BlastWaveSubcellP2} shows the result of a simulation
using a P$_2$ DG-FD scheme and Fig.~\ref{fig:BlastWaveSubcellP5} using a P$_5$
DG-FD scheme with half the number of elements. The increased resolution of a
high-order scheme is clear when comparing the P$_2$ and P$_5$ solutions in the
interior region of the blast wave. We conclude that the DG-FD hybrid scheme is
the most robust and accurate method/limiting strategy for solving the
cylindrical blast wave problem.

\subsection{2d magnetic rotor}

The second 2-dimensional test problem we study is the magnetic rotor problem
originally proposed for non-relativistic MHD~\cite{1999JCoPh.149..270B,
  2000JCoPh.161..605T} and later generalized to the relativistic
case~\cite{2010PhRvD..82h4031E, 2003A&A...400..397D}. A rapidly rotating dense
fluid cylinder is inside a lower density fluid, with a uniform pressure and
magnetic field everywhere. The magnetic braking will slow down the rotor over
time, with an approximately 90 degree rotation by the final time $t=0.4$. We use
a domain of $[-0.5,0.5]^3$ and a time step size $\Delta t=10^{-3}$ and an SSP
RK3 time integrator. An ideal fluid equation of state with $\Gamma=5/3$ is used,
and the following initial conditions are imposed:
\begin{equation}
\begin{split}
  p&=1 \\
  B^i&=(1,0,0) \\
  v^i&=\left\{
       \begin{array}{ll}
         (-y\Omega, x\Omega, 0),
         & \mathrm{if} \; r \le R_{\mathrm{rotor}}=0.1 \\
         (0,0,0), & \mathrm{otherwise},
       \end{array}\right. \\
  \rho&=\left\{
        \begin{array}{ll}
          10,
          & \mathrm{if} \; r \le R_{\mathrm{rotor}}=0.1 \\
          1, & \mathrm{otherwise},
        \end{array}\right.
\end{split}
\end{equation}
with angular velocity $\Omega = 9.95$. The choice of $\Omega$ and
$R_{\mathrm{rotor}}=0.1$ guarantees that the maximum velocity of the fluid
(0.995) is less than the speed of light.

\begin{figure*}
  \begin{tabular}{ccc}
    \subfloat[DG-FD, P$_2$, $64^2$ elements]
    {\label{fig:MagneticRotorSubcellP2}
    \includegraphics[width=0.3\textwidth]{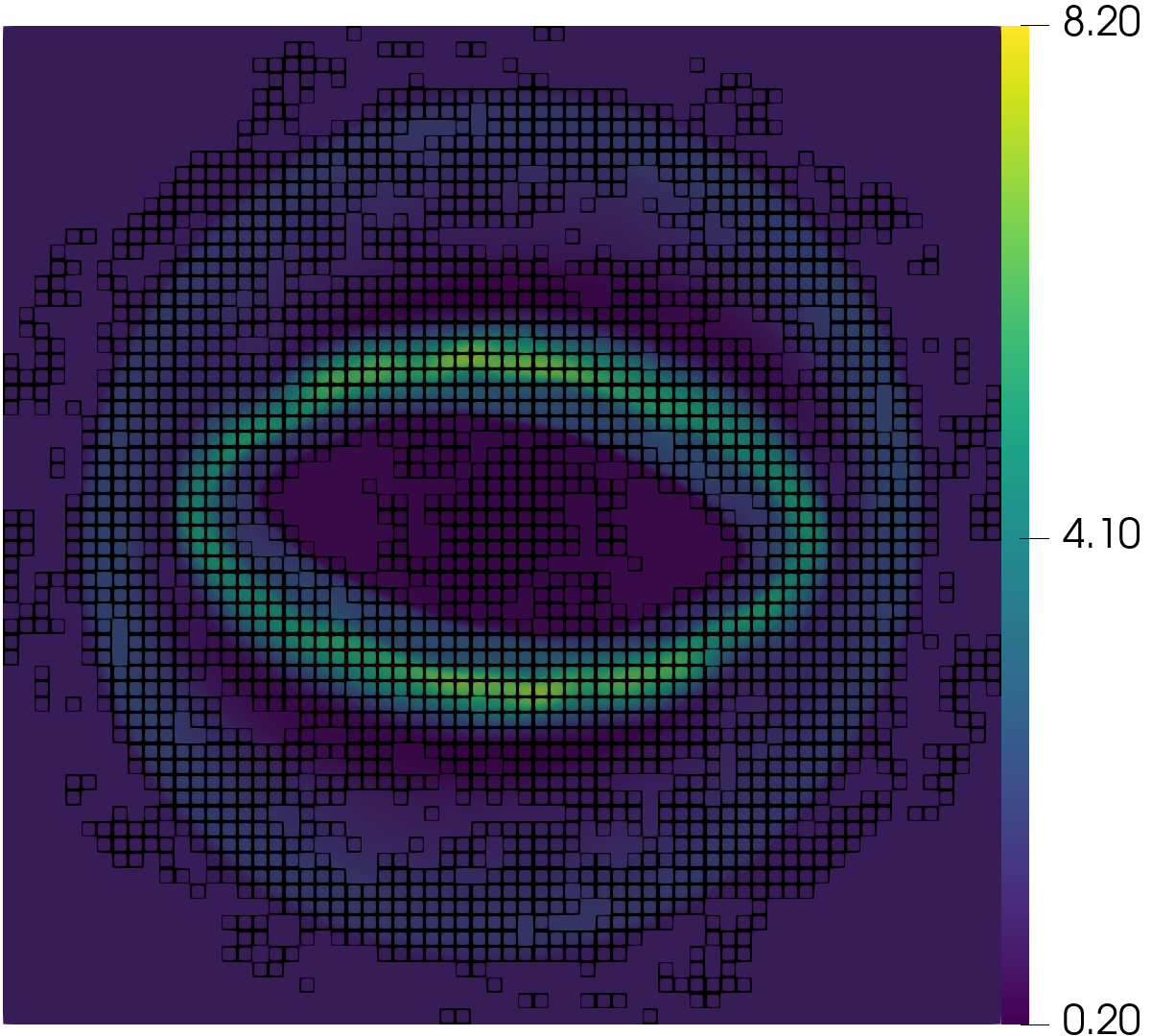}}
    &
      \subfloat[DG-FD, P$_5$, $32^2$ elements]
      {\label{fig:MagneticRotorSubcellP5}
      \includegraphics[width=0.3\textwidth]{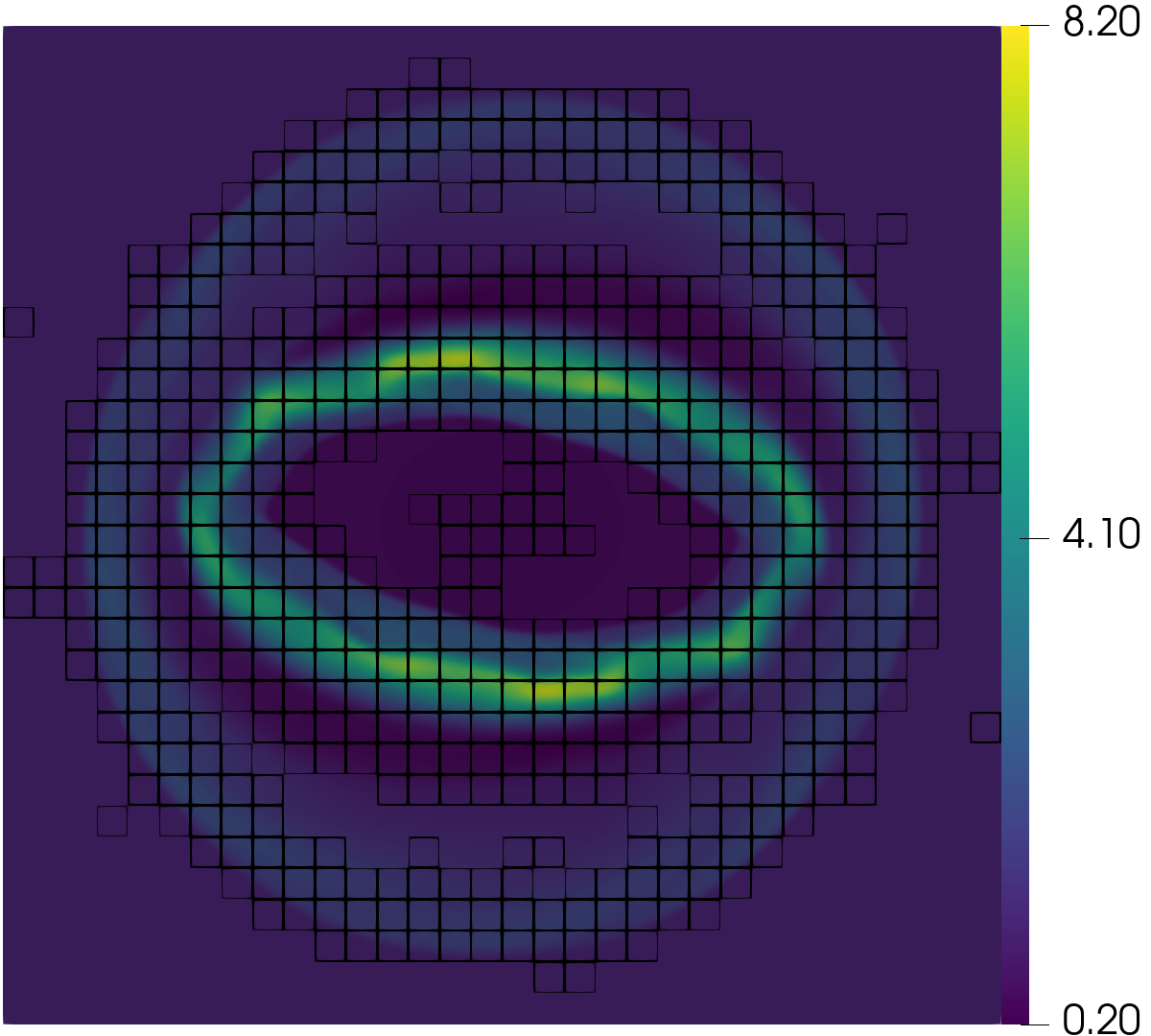}}
    &
      \subfloat[$\Lambda\Pi^N$, P$_2$, $64^2$ elements]
      {\label{fig:MagneticRotorLambdaPiN}
      \includegraphics[width=0.3\textwidth]{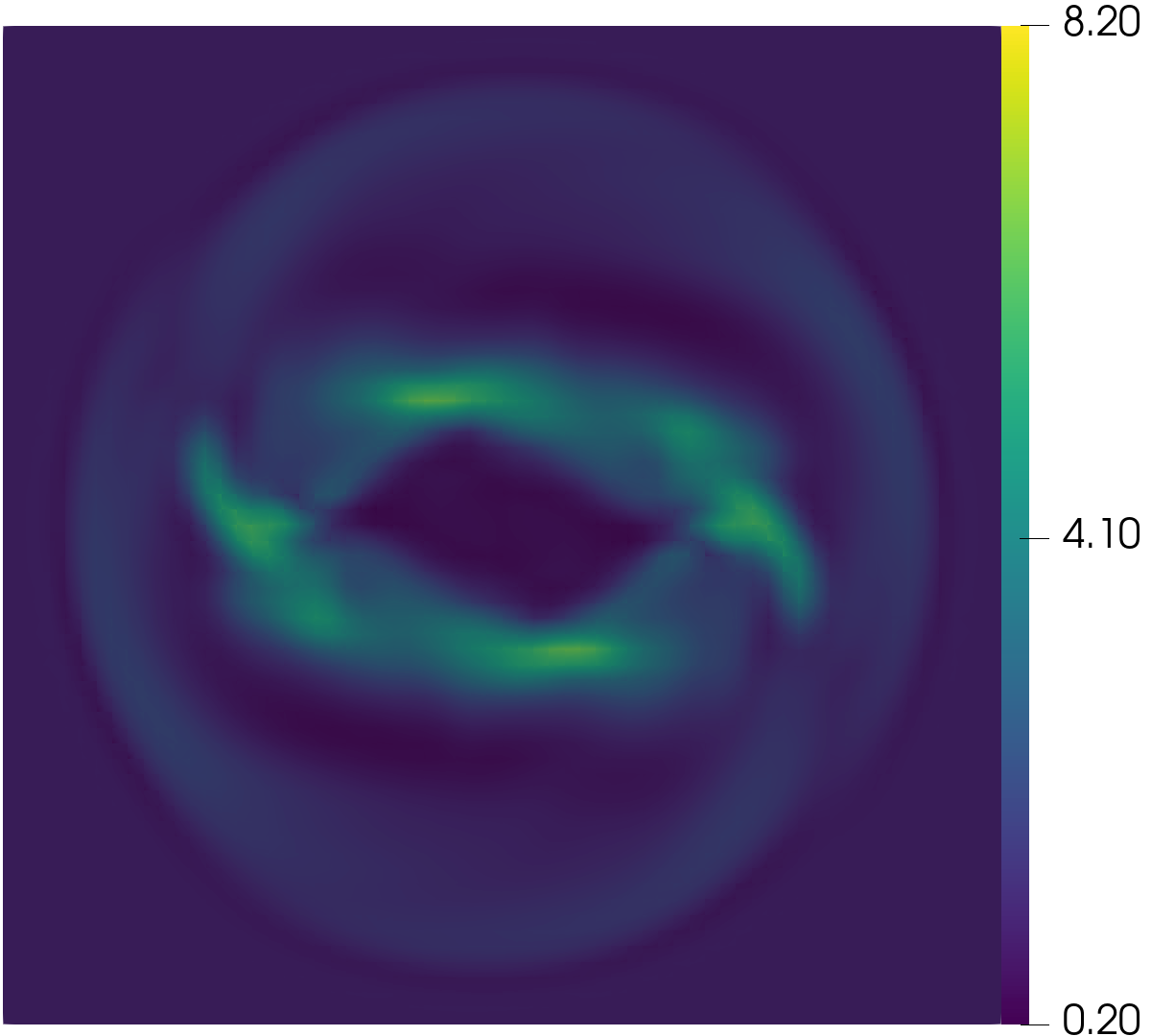}}
    \\
    \subfloat[Krivodonova, P$_2$, $64^2$ elements]
    {\label{fig:MagneticRotorKrivodonova}
    \includegraphics[width=0.3\textwidth]{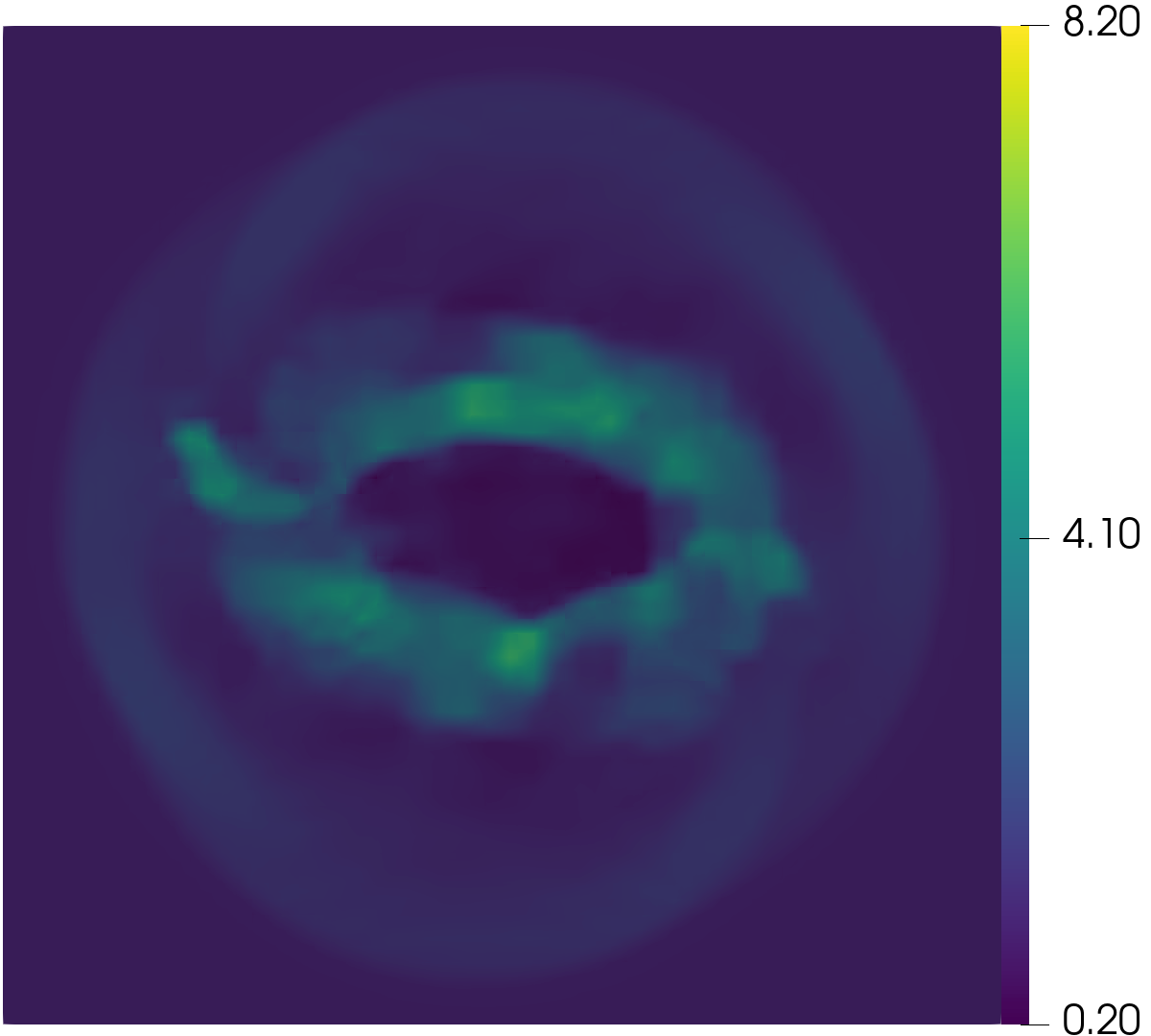}}
    &
      \subfloat[Simple WENO, P$_2$, $64^2$ elements]
      {\label{fig:MagneticRotorSimpleWeno}
      \includegraphics[width=0.3\textwidth]{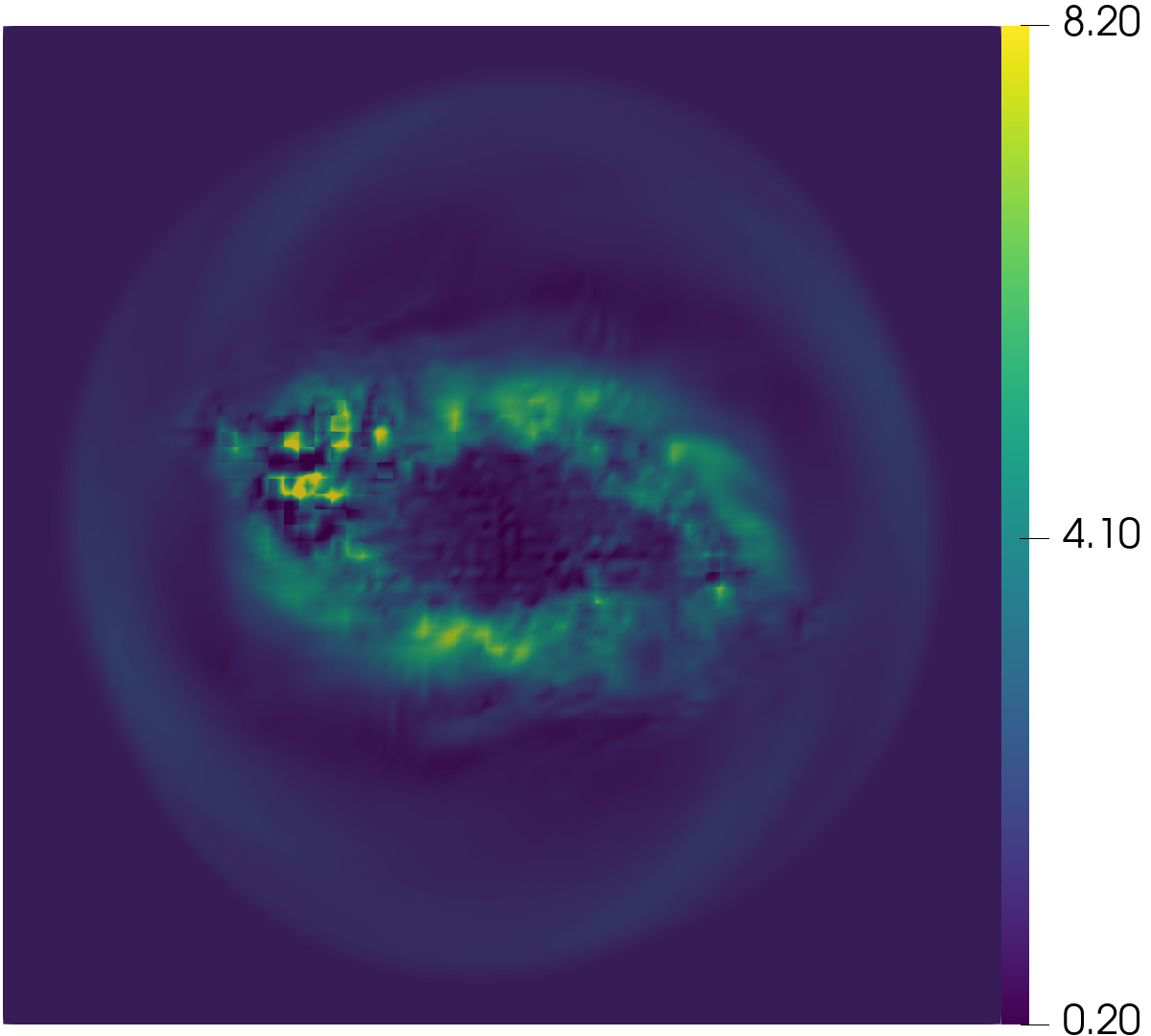}}
    &
      \subfloat[HWENO, P$_2$, $64^2$ elements]
      {\label{fig:MagneticRotorHweno}
      \includegraphics[width=0.3\textwidth]{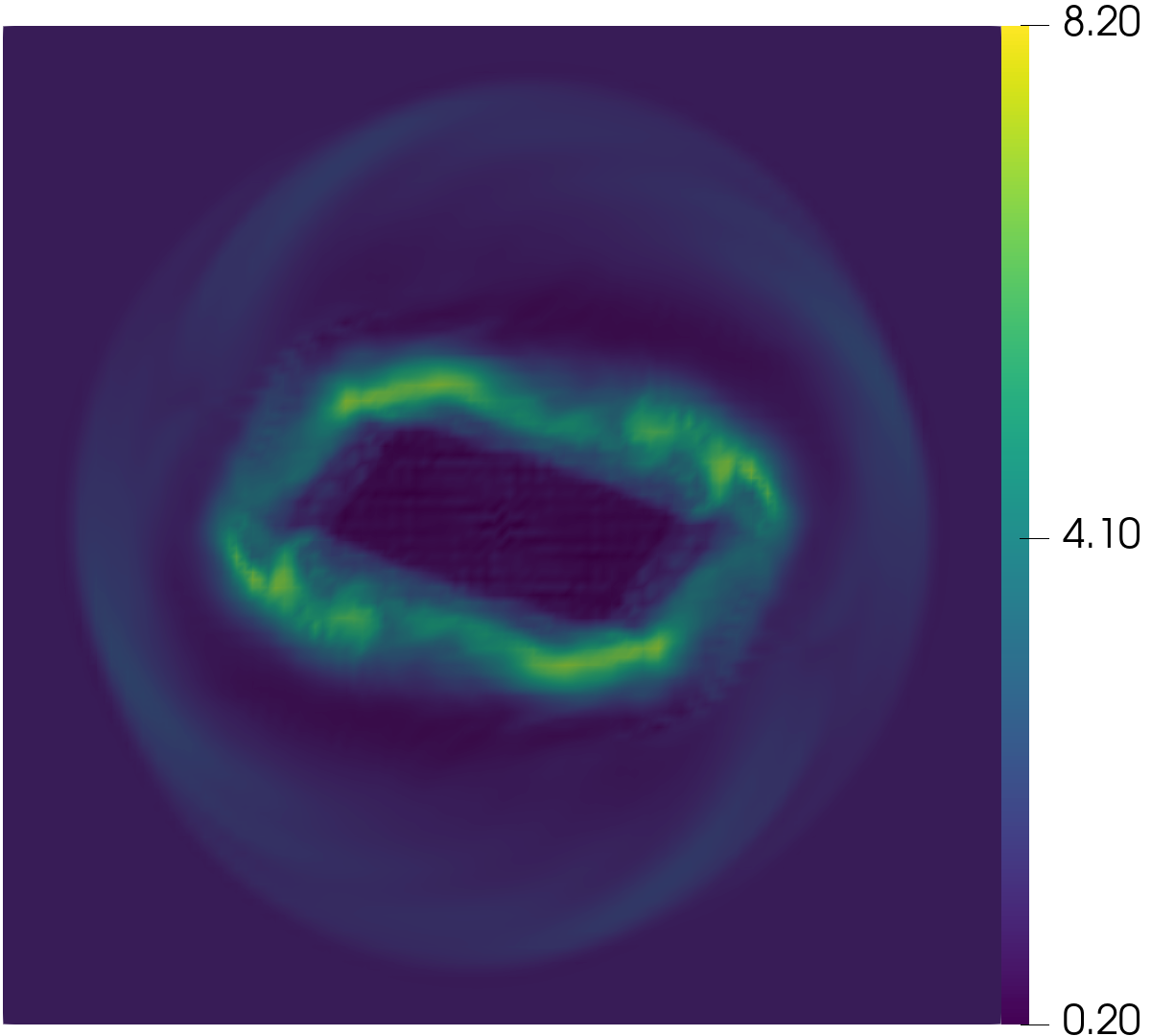}}
    \\
  \end{tabular}
  \caption{\label{fig:MagneticRotor}Magnetic rotor $\rho$ at $t=0.4$ comparing
    the DG-FD hybrid scheme, the $\Lambda\Pi^N$, Krivodonova, simple WENO, and
    HWENO limiters using P$_2$ DG, as well as the DG-FD scheme using P$_5$
    DG. There are 192 degrees of freedom per dimension, comparable to what is
    used when testing FD schemes. We see that only the DG-FD hybrid scheme
    really resolves the features to an acceptable level, and the $\Lambda\Pi^N$
    and Krivodonova smear out the solution almost completely. The simple WENO
    limiter fails to solve the problem. In the plots of the DG-FD hybrid scheme
    the regions surrounded by black squares have switched from DG to FD at
      the final time.}
\end{figure*}

We show the results of our evolutions in Fig.~\ref{fig:MagneticRotor}, which are
all done with 192 grid points and periodic boundary
conditions. Figures~\ref{fig:MagneticRotorLambdaPiN}
and~\ref{fig:MagneticRotorKrivodonova} show results using the $\Lambda\Pi^N$ and
Krivodonova limiter which both severely smear out the solution. The simple WENO
limiter suffers from spurious artifacts
(Fig.~\ref{fig:MagneticRotorSimpleWeno}), while the HWENO limiter does a
reasonable job (Fig.~\ref{fig:MagneticRotorHweno}). The DG-FD hybrid scheme is
most robust and accurate, but a fairly large number of cells end up being marked
as troubled in this problem and switched to FD. While ideally fewer cells would
be switched to FD, it is better to have a scheme that is capable of solving a
large array of problems without fine-tuning than to have a slightly different
fine-tuned scheme for each test problem.

\subsection{2d magnetic loop advection}

The third 2-dimensional test problem we study is magnetic loop advection problem
\cite{1991JCoPh..92..142D}. A magnetic loop is advected through the domain until
it returns to its starting position. We use an initial configuration very
similar to~\cite{2014CQGra..31a5005M, 2011ApJS..193....6B, 2005JCoPh.205..509G,
  2008ApJS..178..137S}, where
\begin{equation}
\begin{split}
  \rho&=1 \\
  p&=3 \\
  v^i &= (1/1.2, 1/2.4, 0) \\
  B^x &= \left\{
        \begin{array}{ll}
          -A_{\mathrm{loop}}y/R_{\mathrm{in}},
          & \mathrm{if} \; r \le R_{\mathrm{in}} \\
          -A_{\mathrm{loop}}y/r,
          & \mathrm{if} \; R_{\mathrm{in}}<r<R_{\mathrm{loop}} \\
          0, & \mathrm{otherwise},
        \end{array}\right. \\
  B^y &= \left\{
        \begin{array}{ll}
          A_{\mathrm{loop}}x / R_{\mathrm{in}},
          & \mathrm{if} \; r \le R_{\mathrm{in}} \\
          A_{\mathrm{loop}}x/r,
          & \mathrm{if} \; R_{\mathrm{in}}<r<R_{\mathrm{loop}} \\
          0, & \mathrm{otherwise},
        \end{array}\right.
\end{split}
\end{equation}
with $R_{\mathrm{loop}}=0.3$, $R_{\mathrm{in}}=0.001$, and an ideal gas equation
of state with $\Gamma=5/3$. The computational domain is $[-0.5,0.5]^3$
with $64\times64\times1$ elements and periodic boundary conditions being applied
everywhere. The final time for one period is $t=2.4$. For all simulations we use
a time step size $\Delta t=10^{-3}$ and an SSP RK3 time integrator.

\begin{figure*}
  \begin{tabular}{ccc}
    \subfloat[DG-FD, P$_2$, $64^2$ elements]
    {\label{fig:MagneticLoopBxSubcellP2}
    \includegraphics[width=0.3\textwidth]{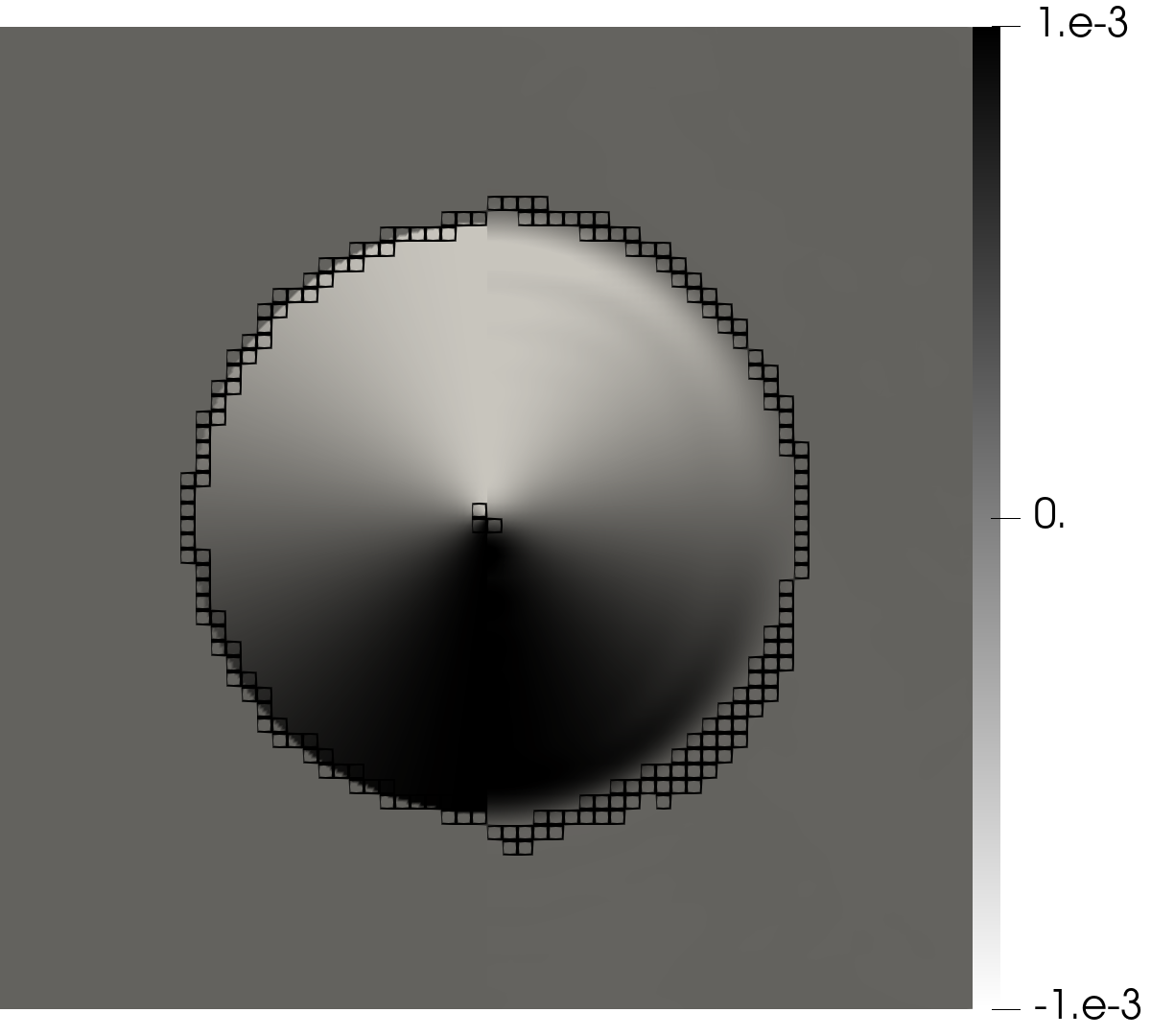}}
    &
      \subfloat[DG-FD, P$_5$, $32^2$ elements]
      {\label{fig:MagneticLoopBxSubcellP5}
      \includegraphics[width=0.3\textwidth]{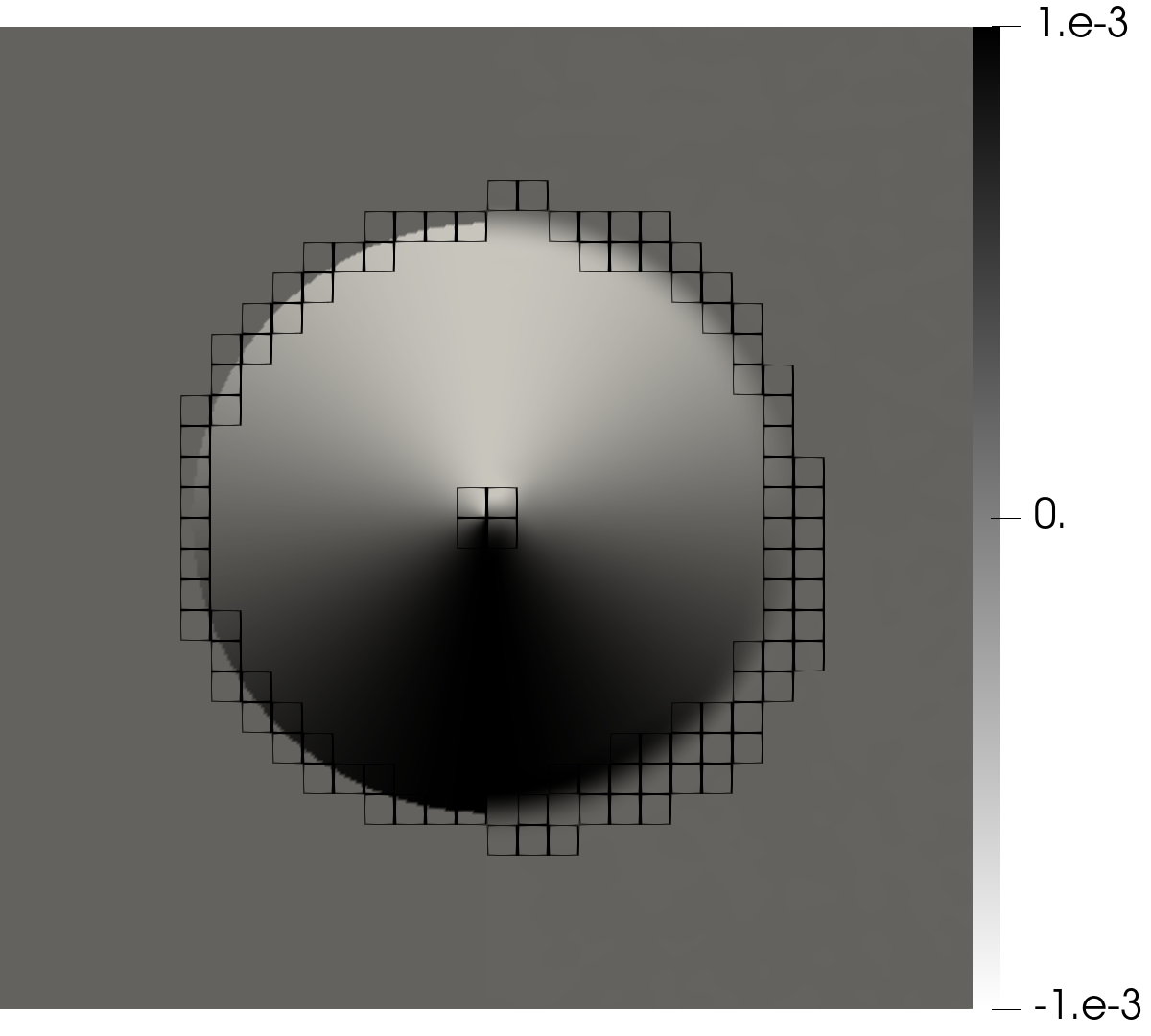}}
    &
      \subfloat[$\Lambda\Pi^N$, P$_2$, $64^2$ elements]
      {\label{fig:MagneticLoopBxLambdaPiN}
      \includegraphics[width=0.3\textwidth]{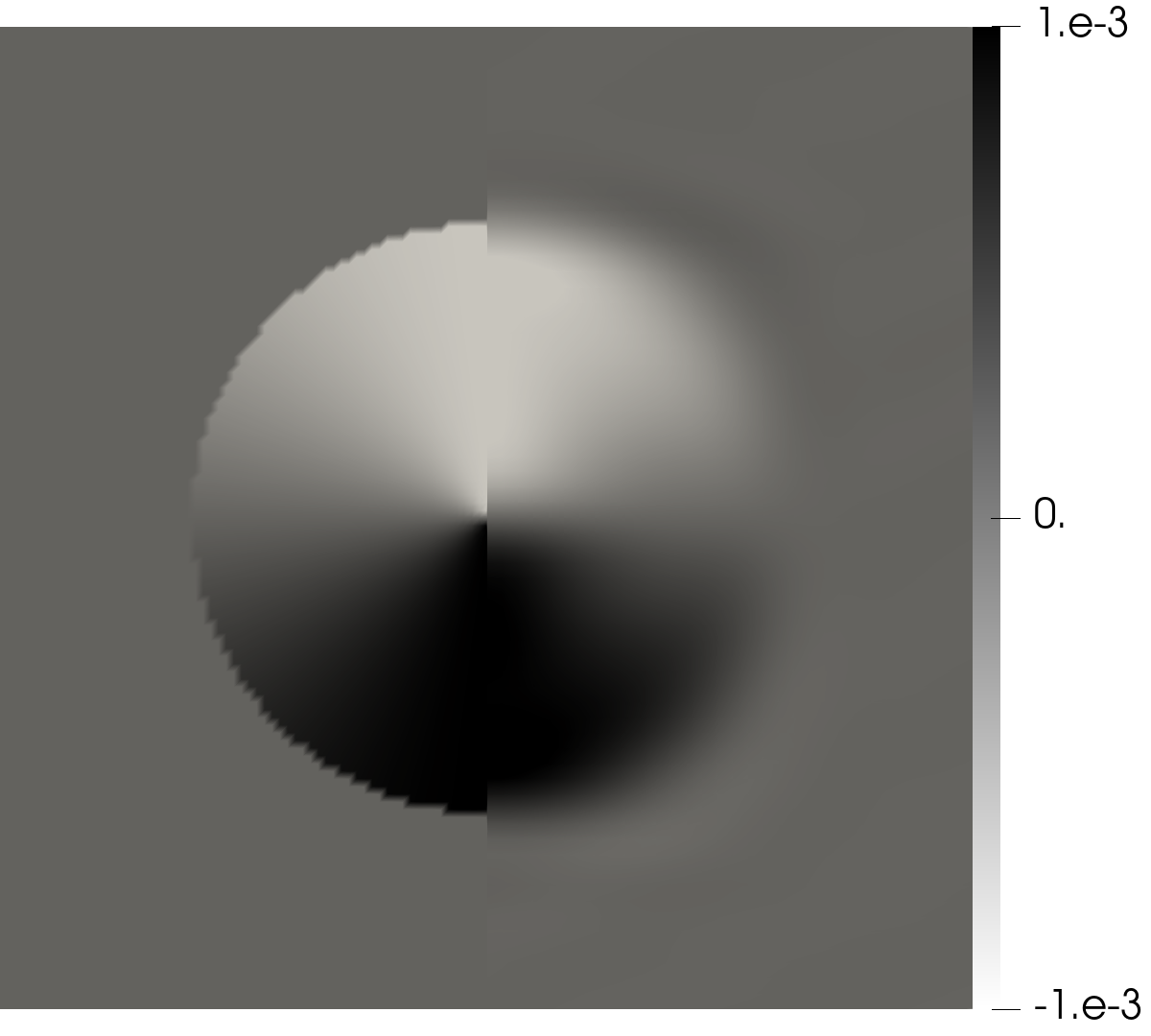}}
    \\
    \subfloat[Krivodonova, P$_2$, $64^2$ elements]
      {\label{fig:MagneticLoopBxKrivodonova}
      \includegraphics[width=0.3\textwidth]{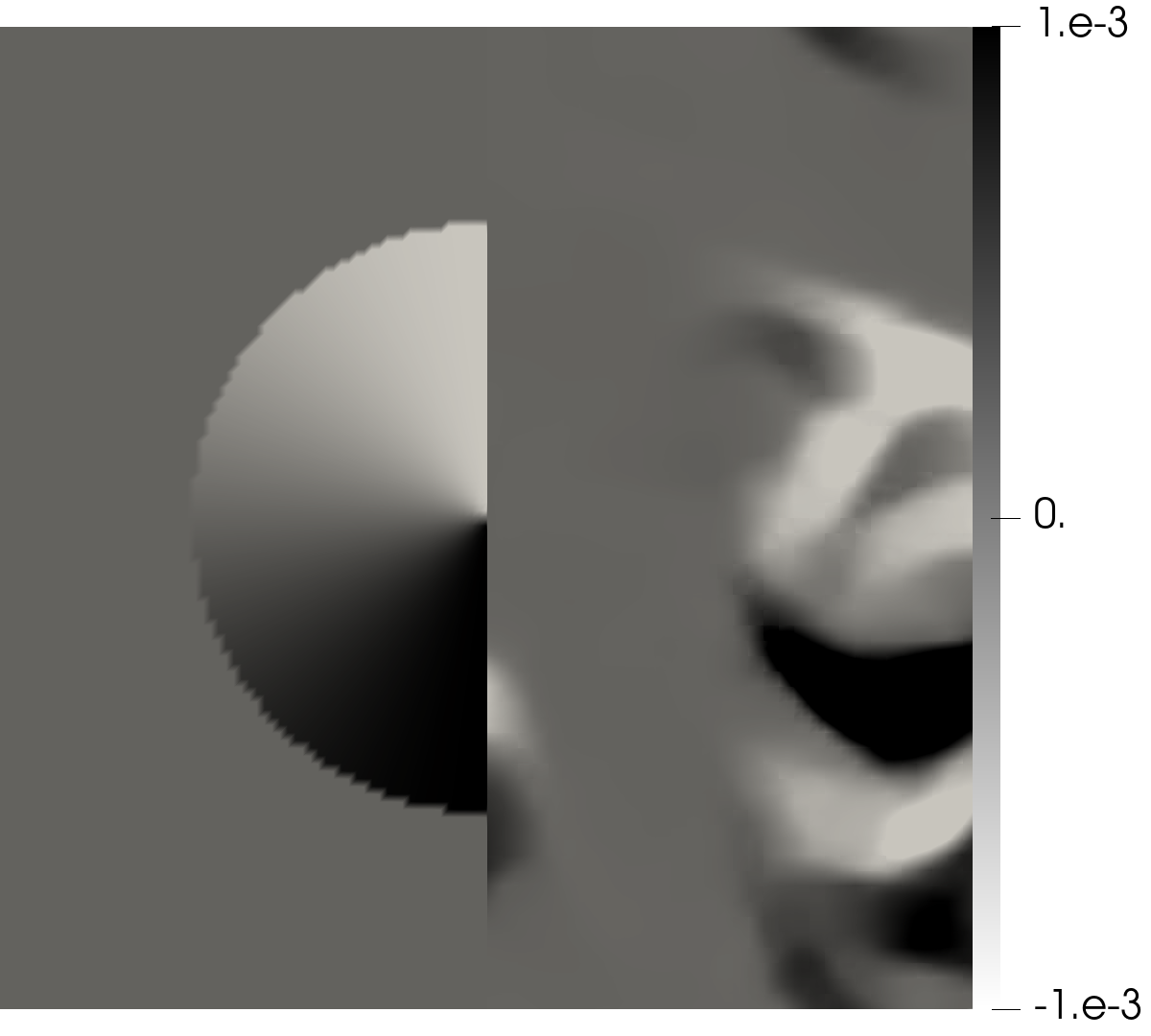}}
    &
      \subfloat[Simple WENO, P$_2$, $64^2$ elements]
      {\label{fig:MagneticLoopBxSimpleWeno}
      \includegraphics[width=0.3\textwidth]{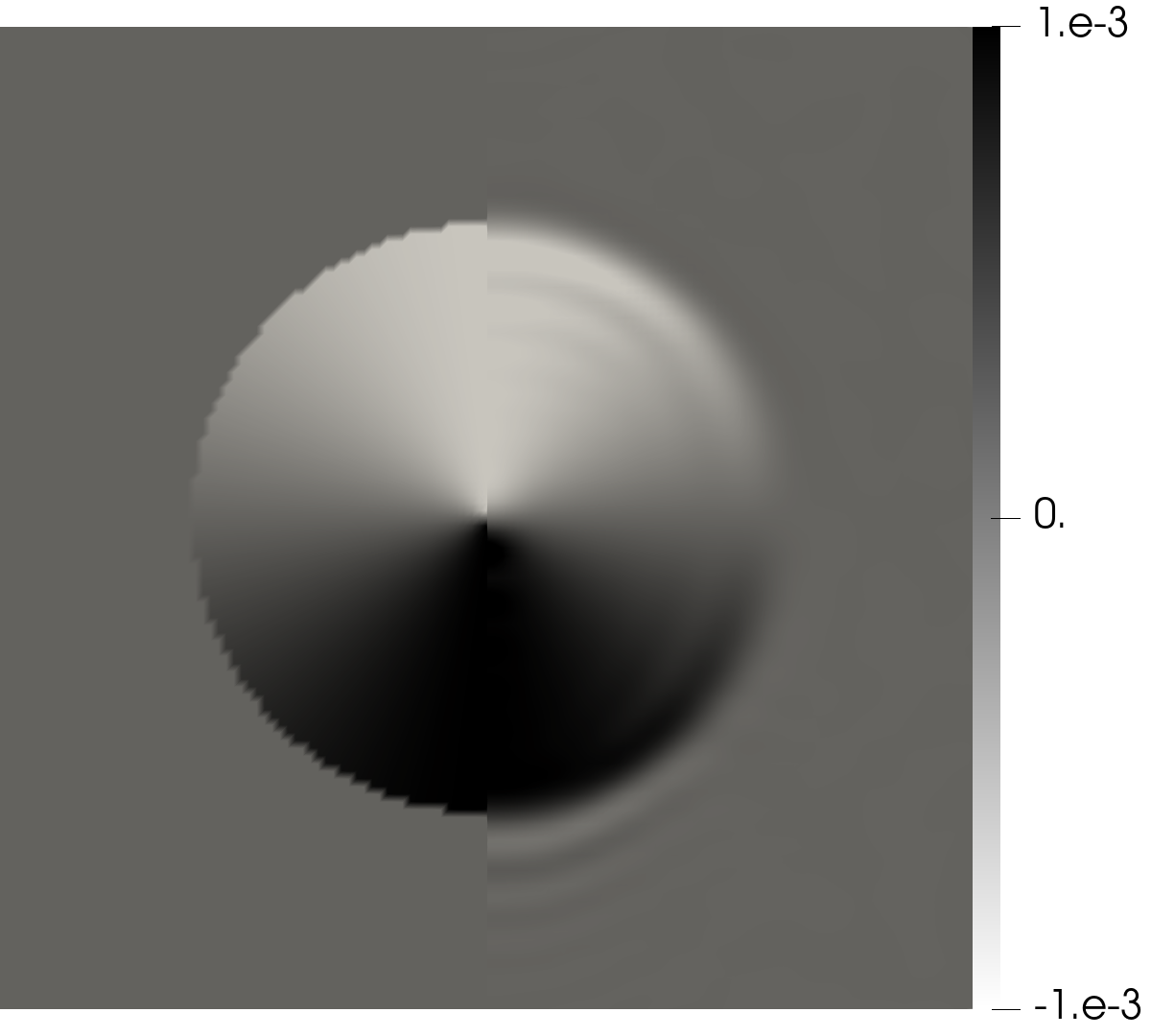}}
    &
      \subfloat[HWENO, P$_2$, $64^2$ elements]
      {\label{fig:MagneticLoopBxHweno}
      \includegraphics[width=0.3\textwidth]{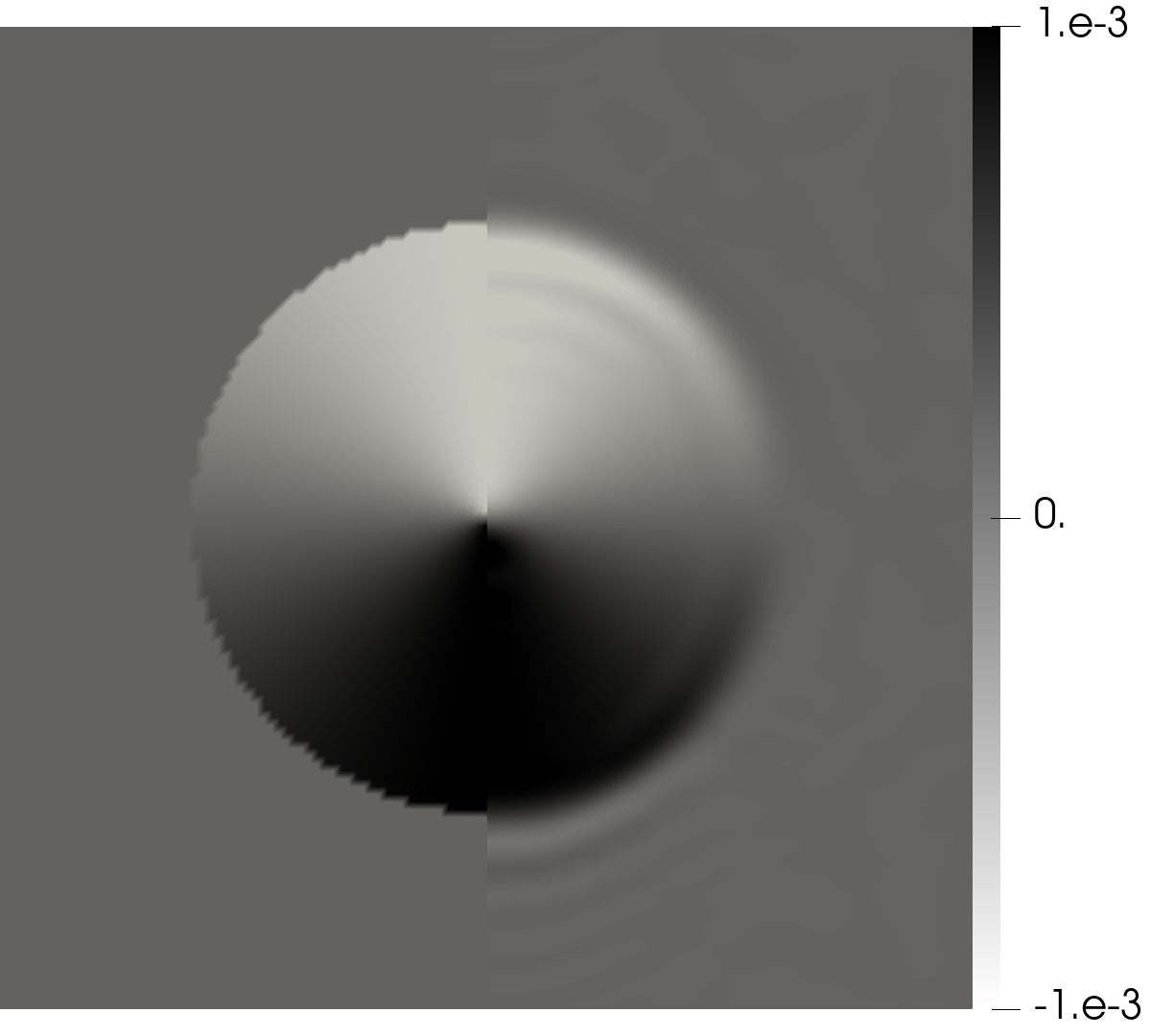}}
    \\
  \end{tabular}
  \caption{\label{fig:MagneticLoopBx}$B^x$ for the magnetic loop advection
    problem. The left half of each plot is at the initial time, while the right
    half is after one period ($t=2.4$). We compare the DG-FD hybrid scheme, the
    $\Lambda\Pi^N$, Krivodonova, simple WENO, and HWENO limiters using P$_2$ DG,
    as well as the DG-FD scheme using P$_5$ DG. There are 192 degrees of freedom
    per dimension, comparable to what is used when testing FD schemes. In the
    plots of the DG-FD hybrid scheme the regions surrounded by black squares
    have switched from DG to FD at the final time.}
\end{figure*}

In Fig.~\ref{fig:MagneticLoopBx} we plot the magnetic field component $B^x$ at
$t=0$ on the left half of each plot and after one period $t=2.4$ on the right
half of each plot for results using various limiting strategies. We use a TVB
constant of 5 for the $\Lambda\Pi^N$, simple WENO, and HWENO limiters, and use
neighbor weights $\gamma_k=0.001$ for the simple WENO and HWENO limiters. The
Krivodonova limiter completely destroys the solution and only remains stable
because of our conservative variable fixing scheme. Both WENO limiters
work quite well, maintaining the shape of the loop with only some oscillations
being generated. The DG-FD hybrid scheme again performs the best. In
Fig.~\ref{fig:MagneticLoopBxSubcellP2} we show the result using a P$_2$ DG-FD
scheme and in Fig.~\ref{fig:MagneticLoopBxSubcellP5} using a P$_5$ DG-FD
scheme. The P$_5$ scheme resolves the smooth parts of the solution more
accurately than the P$_2$ scheme, as is to be expected. The DG-FD hybrid scheme
also does not generate the spurious oscillations that are present when using the
WENO limiters. While the spurious oscillations may be reduced by fine-tuning the
TVB constant and the neighbor weights, this type of fine-tuning is not possible
for complex physics simulations and so we do not spend time searching for the
``optimal'' parameters.

\begin{figure*}
  \begin{tabular}{ccc}
    \subfloat[DG-FD, P$_2$, $64^2$ elements]
    {\label{fig:MagneticLoopPhiSubcellP2}
    \includegraphics[width=0.3\textwidth]{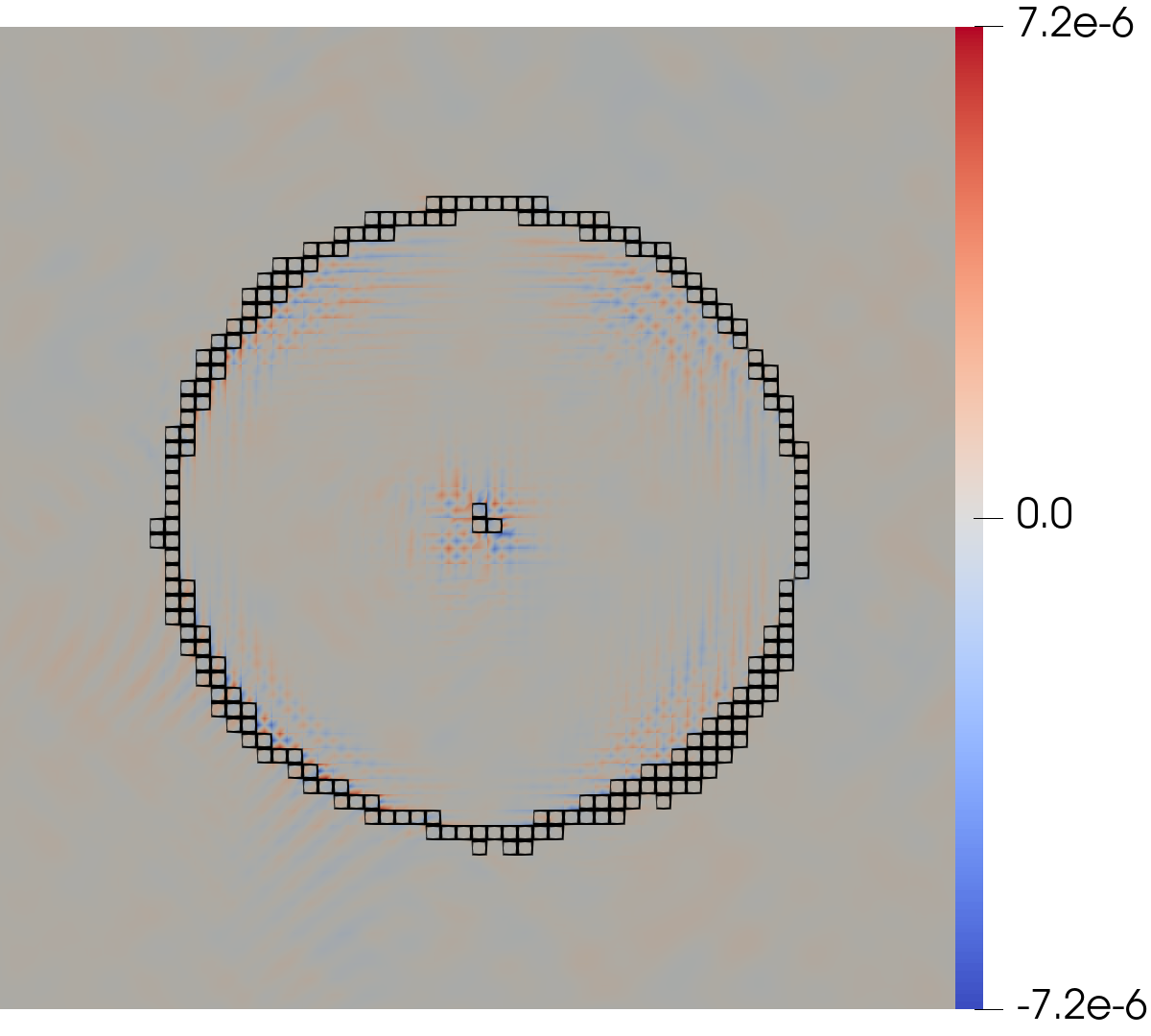}}
    &
      \subfloat[DG-FD, P$_5$, $32^2$ elements]
      {\label{fig:MagneticLoopPhiSubcellP5}
      \includegraphics[width=0.3\textwidth]{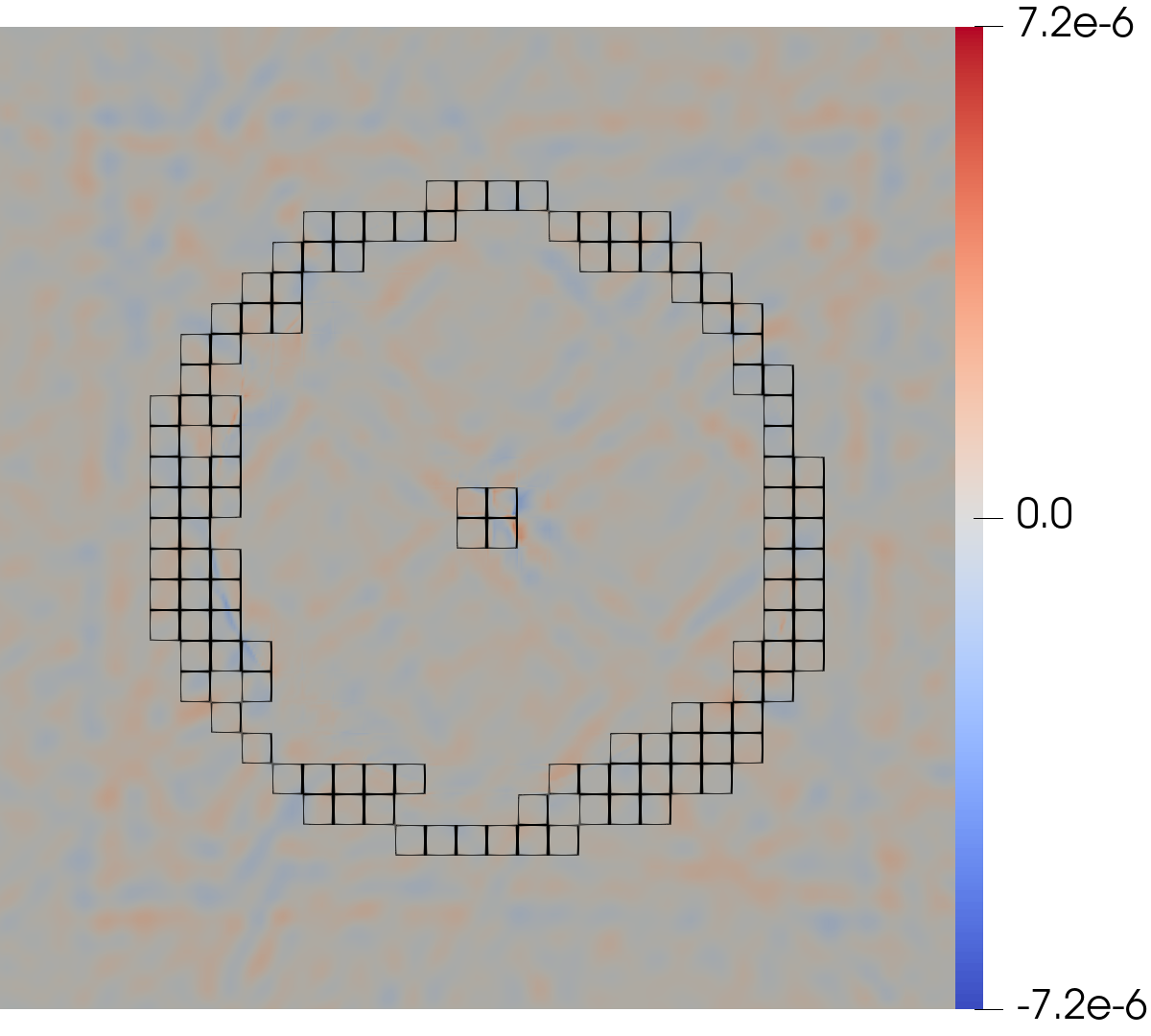}}
    &
      \subfloat[$\Lambda\Pi^N$, P$_2$, $64^2$ elements]
      {\label{fig:MagneticLoopPhiLambdaPiN}
      \includegraphics[width=0.3\textwidth]{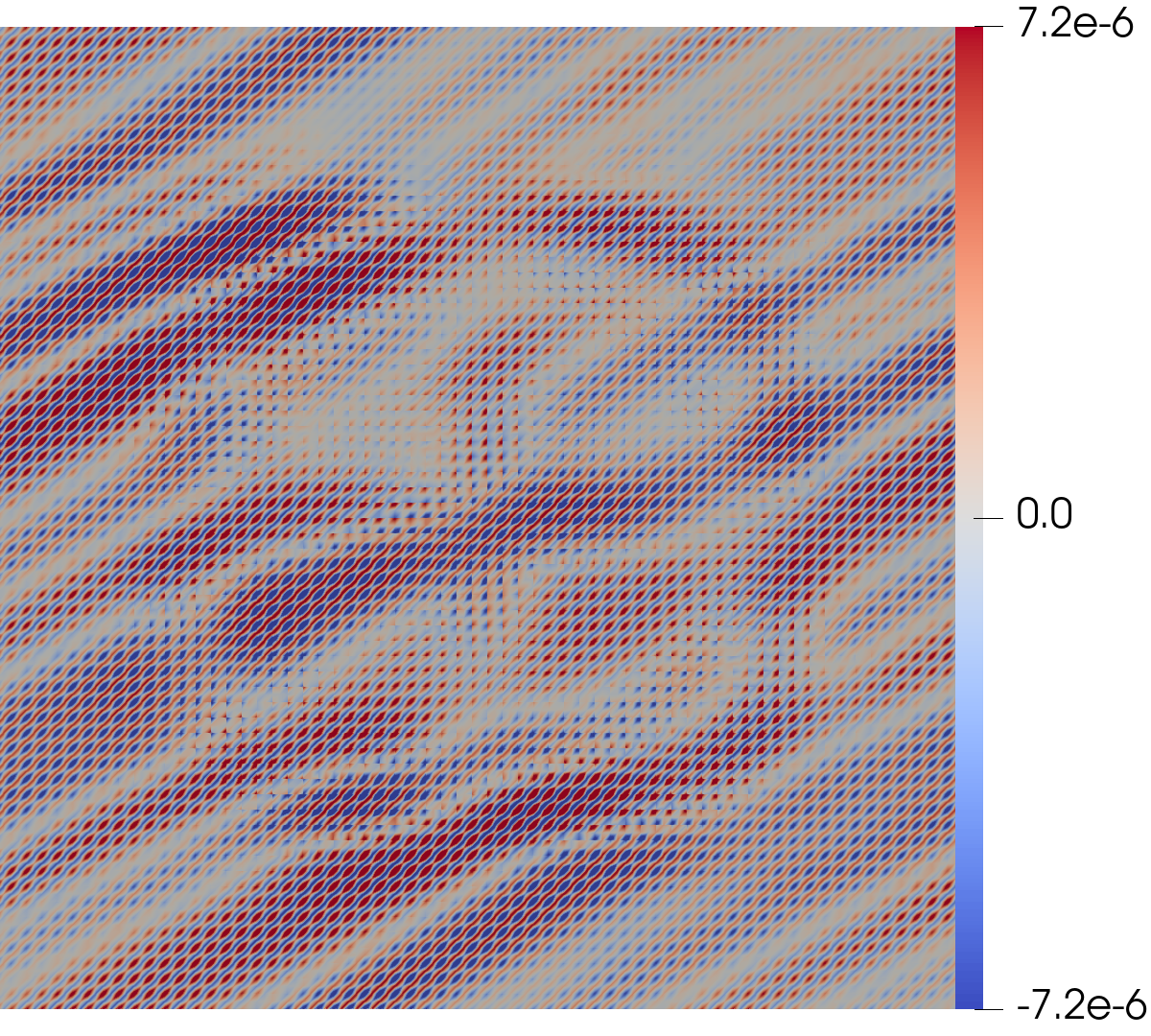}}
    \\
    \subfloat[Krivodonova, P$_2$, $64^2$ elements]
    {\label{fig:MagneticLoopPhiKrivodonova}
      \includegraphics[width=0.3\textwidth]{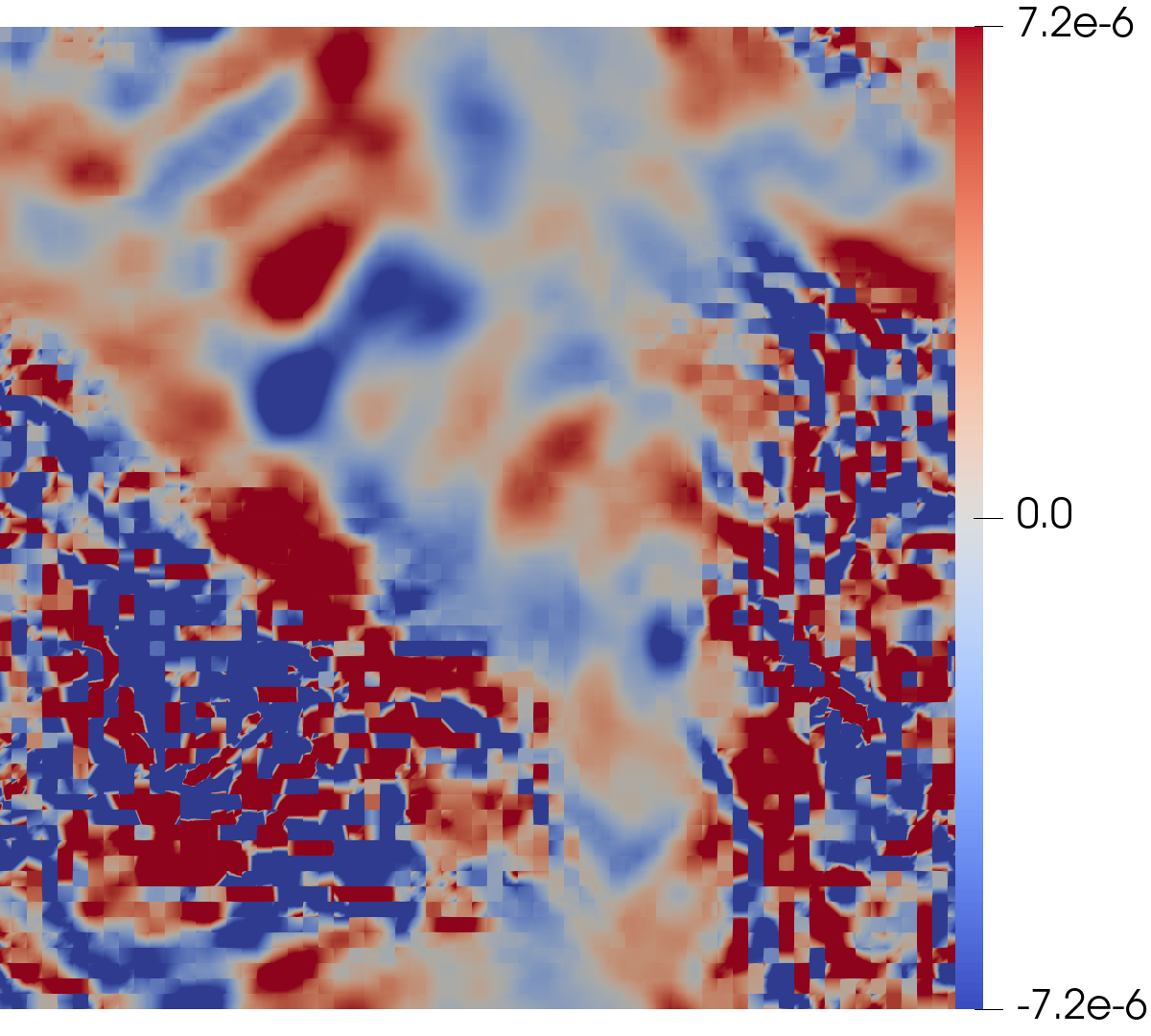}}
    &
      \subfloat[Simple WENO, P$_2$, $64^2$ elements]
      {\label{fig:MagneticLoopPhiSimpleWeno}
      \includegraphics[width=0.3\textwidth]{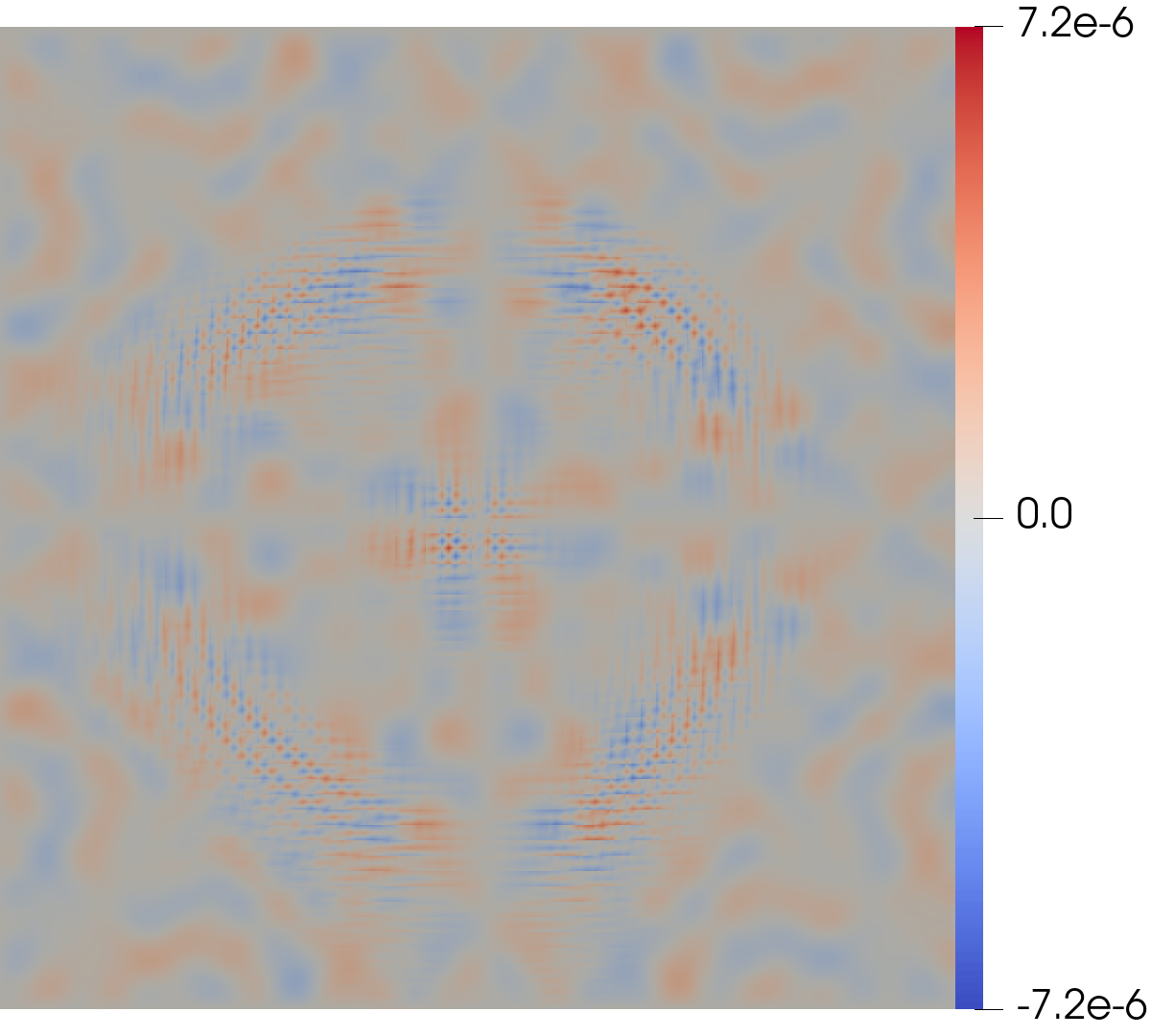}}
    &
      \subfloat[HWENO, P$_2$, $64^2$ elements]
      {\label{fig:MagneticLoopPhiHweno}
      \includegraphics[width=0.3\textwidth]{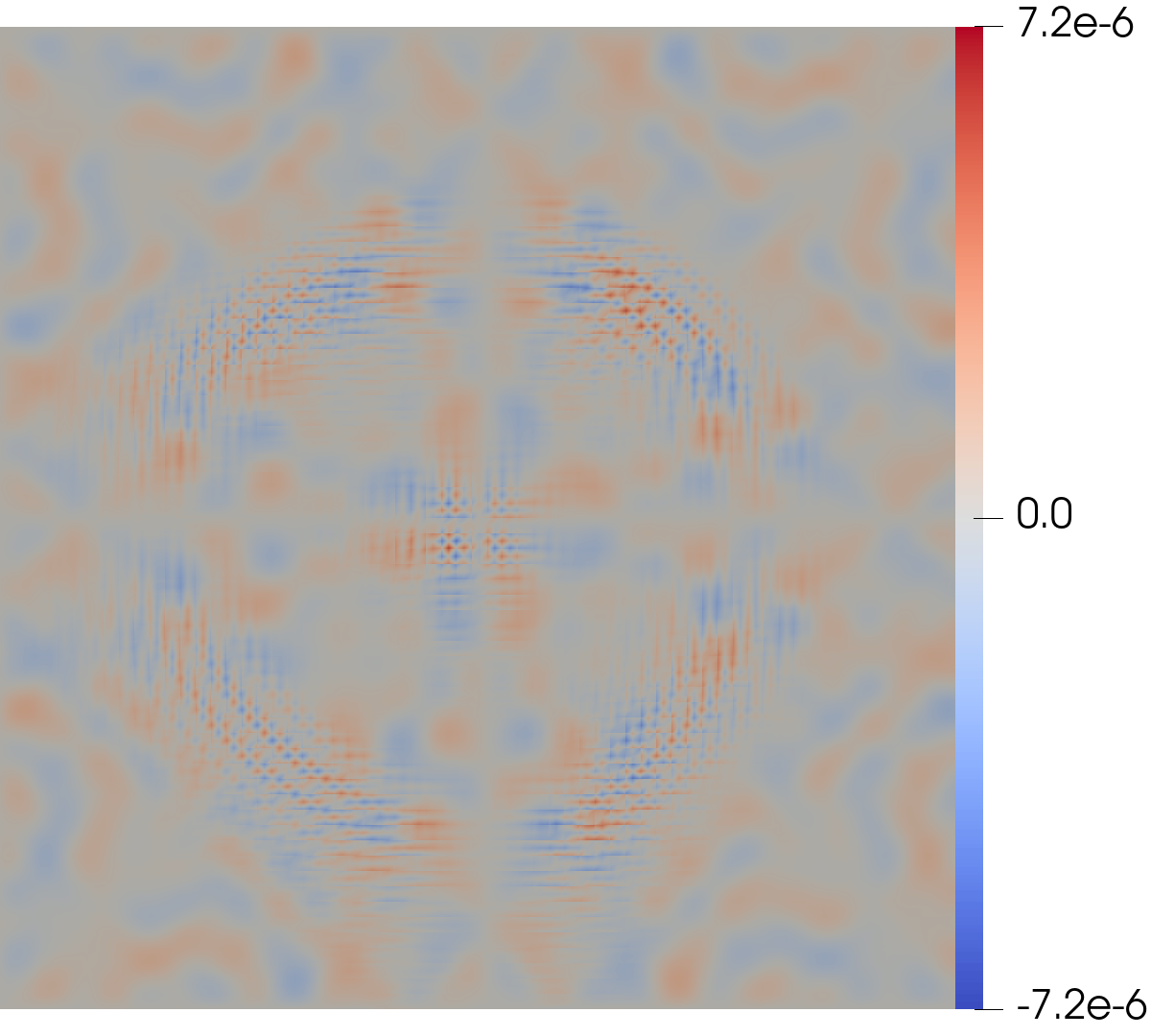}}
    \\
  \end{tabular}
  \caption{\label{fig:MagneticLoopPhi}The divergence cleaning field $\Phi$ for
    the magnetic loop advection problem after one period ($t=2.4$) comparing the
    DG-FD hybrid scheme, the $\Lambda\Pi^N$, Krivodonova, simple WENO, and HWENO
    limiters using P$_2$ DG, as well as the DG-FD scheme using P$_5$ DG. There
    are 192 degrees of freedom per dimension, comparable to what is used when
    testing FD schemes. In the plots from the DG-FD hybrid scheme the regions
    surrounded by black squares have switched from DG to FD at the final time.}
\end{figure*}

Since we are using hyperbolic divergence cleaning, violations of the
$\partial_i B^i=0$ constraint occur. In Fig.~\ref{fig:MagneticLoopPhi} we plot
the divergence cleaning field $\Phi$ at the final time $t=2.4$. The simple WENO,
HWENO, and DG-FD hybrid schemes all have $|\Phi|\sim5\times10^{-6}$, while the
$\Lambda\Pi^N$ limiter has $\Phi$ approximately one order of magnitude
larger. For the magnetic loop advection problem we find that all classical
limiters perform comparably, except the Krivodonova limiter completely
destroys the solution and remains stable only because of our conservative
variable fixing scheme. Nevertheless, the DG-FD hybrid scheme is better than
the classical limiters, and we conclude that the DG-FD hybrid scheme is both the
most robust and accurate method/limiting strategy for solving the magnetic loop
advection problem.

\subsection{2d magnetized Kelvin-Helmholtz instability\label{sec:KhInstability}}

The last 2-dimensional test problem we study is the magnetized Kelvin-Helmholtz
(KH) instability, similar to~\cite{Beckwith:2011iy}. The domain is $[0,1]^3$ and
we use the following initial conditions~\cite{Schaal:2015ila}:
\begin{align}
  \rho&=\begin{cases}
    1, & \left|y - 0.5\right| < 0.25\\
    10^{-2}, & \text{otherwise},
  \end{cases} \\
  p &= 1.0, \\
  v^x &= \begin{cases}
    0.5, & \left|y - 0.5\right| < 0.25\\
    -0.5, & \text{otherwise},
  \end{cases} \\
  v^y &= 0.1\sin(4\pi x)
          \left[\exp\left(-\dfrac{(y - 0.75)^2}{0.0707^2}\right)\right. \notag
  \\
      &\left.+\exp\left(-\dfrac{(y - 0.25)^2}{0.0707^2}\right)\right], \\
  v^z &= 0.0, \\
  B^x &= 10^{-3}, \\
  B^y &= B^z = 0.0.
\end{align}
We use an ideal gas equation of state with $\Gamma=4/3$, a final time $t_f=1.6$,
a time step size of $\Delta t = 10^{-3}$, an SSP RK3 time integrator, and
$[64\times1\times64]$ P$_2$ elements for the classical limiters. For the DG-FD
hybrid method we use both $[64\times1\times64]$ P$_2$ elements and
$[32\times1\times32]$ P$_5$ elements. We use a TVB constant of $1$ for all the
limiters. Using the flattening algorithm is crucial for the results obtained
here, while for other test problems it is significantly less important.

\begin{figure*}
  \begin{tabular}{ccc}
    \subfloat[DG-FD, P$_2$, $64^2$ elements]
    {\label{fig:KhInstabilitySubcellP2}
    \includegraphics[width=0.3\textwidth]{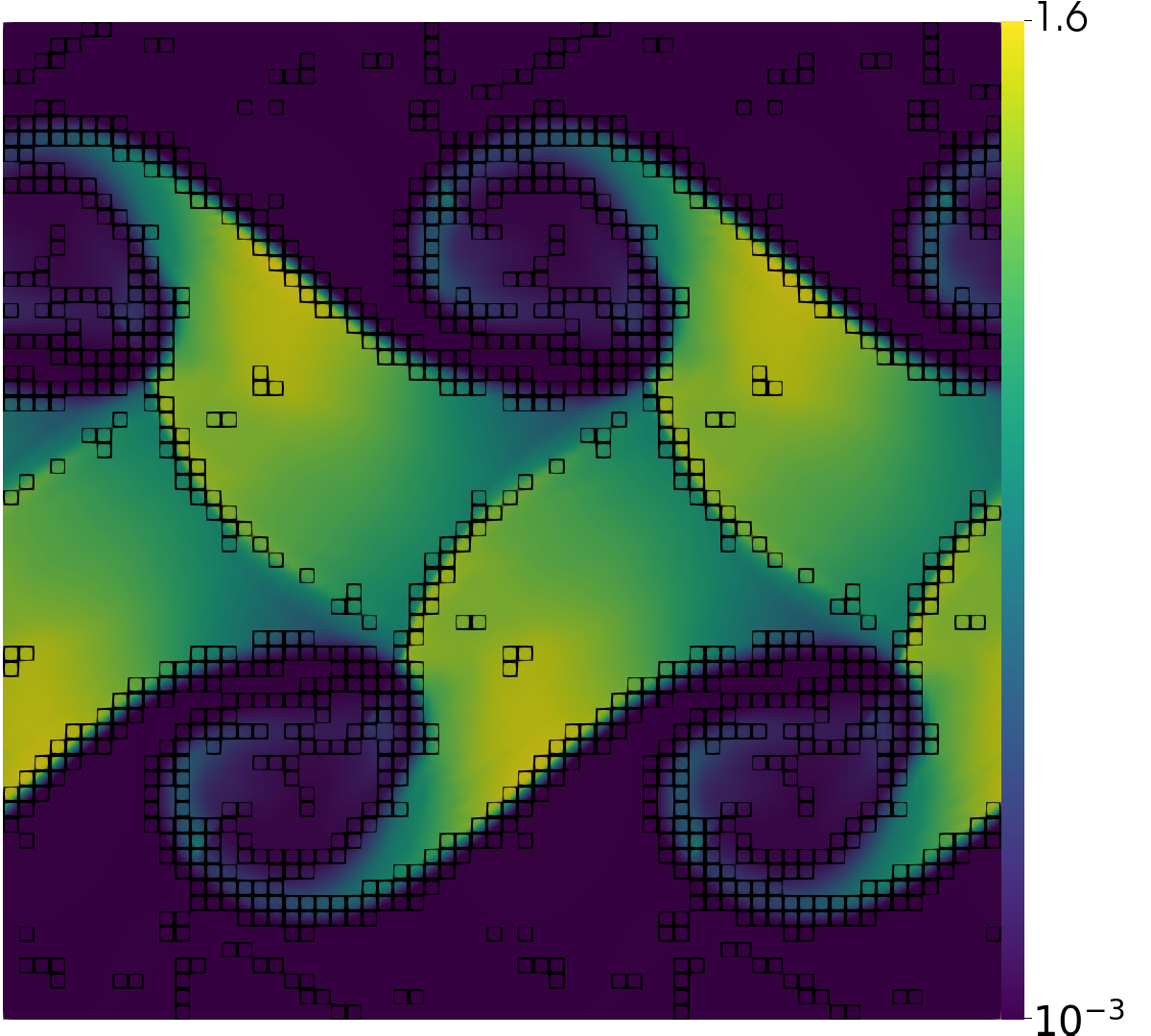}}
    &
      \subfloat[DG-FD, P$_5$, $32^2$ elements]
      {\label{fig:KhInstabilitySubcellP5}
      \includegraphics[width=0.3\textwidth]{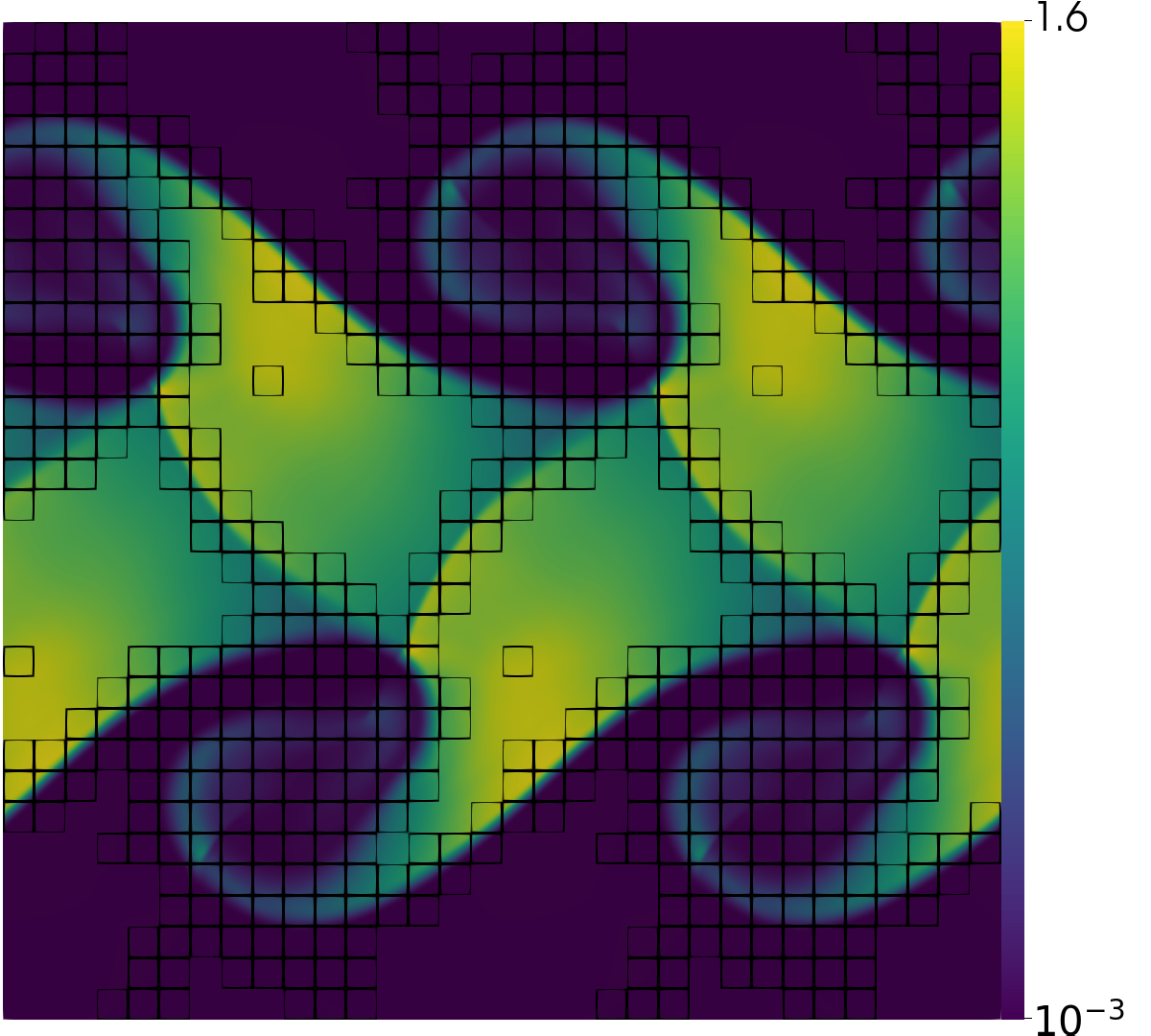}}
    &
      \subfloat[$\Lambda\Pi^N$, P$_2$, $64^2$ elements]
      {\label{fig:KhInstabilityLambdaPiN}
      \includegraphics[width=0.3\textwidth]{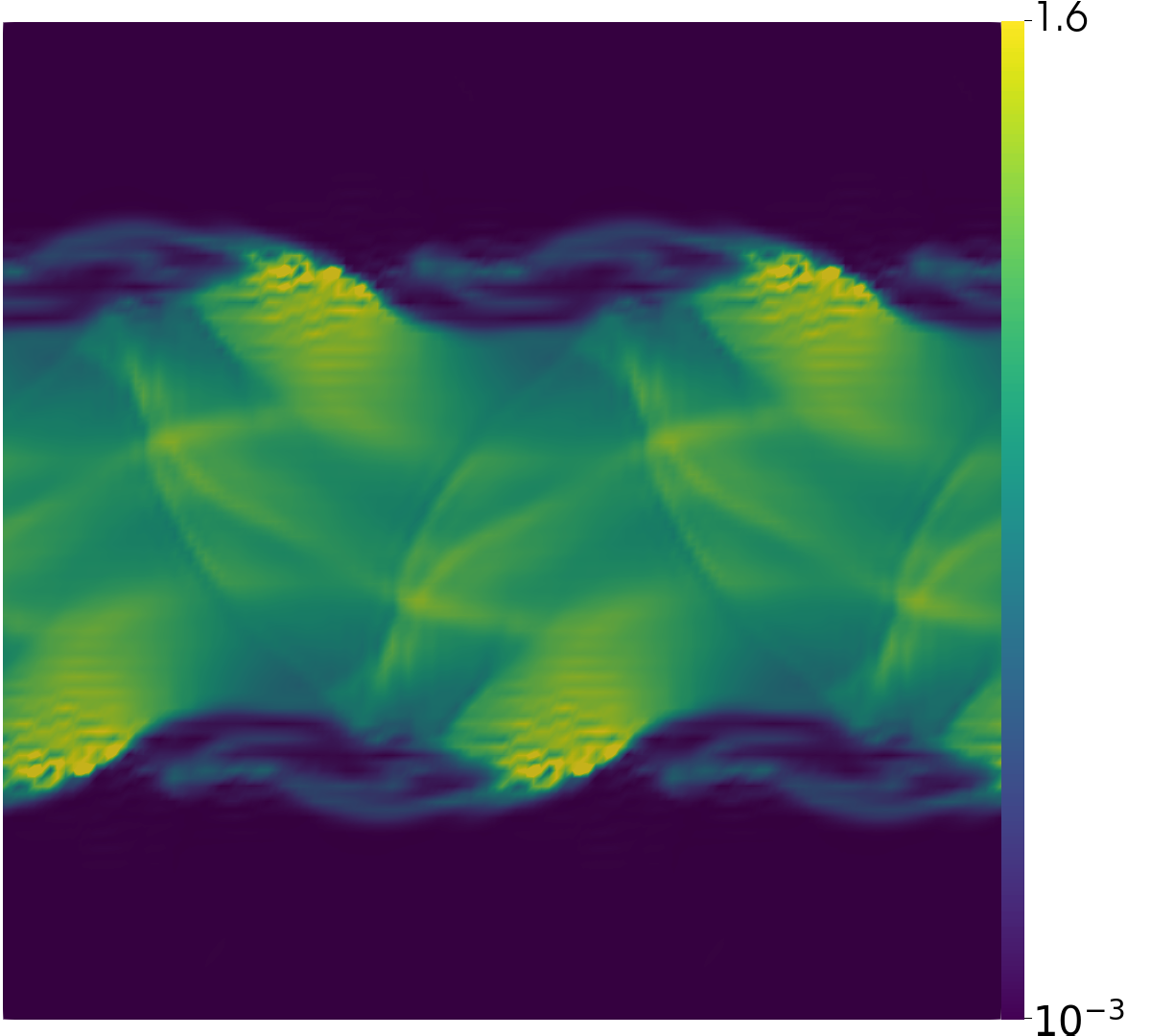}}
    \\
    \subfloat[Krivodonova, P$_2$, $64^2$ elements]
    {\label{fig:KhInstabilityKrivodonova}
    \includegraphics[width=0.3\textwidth]{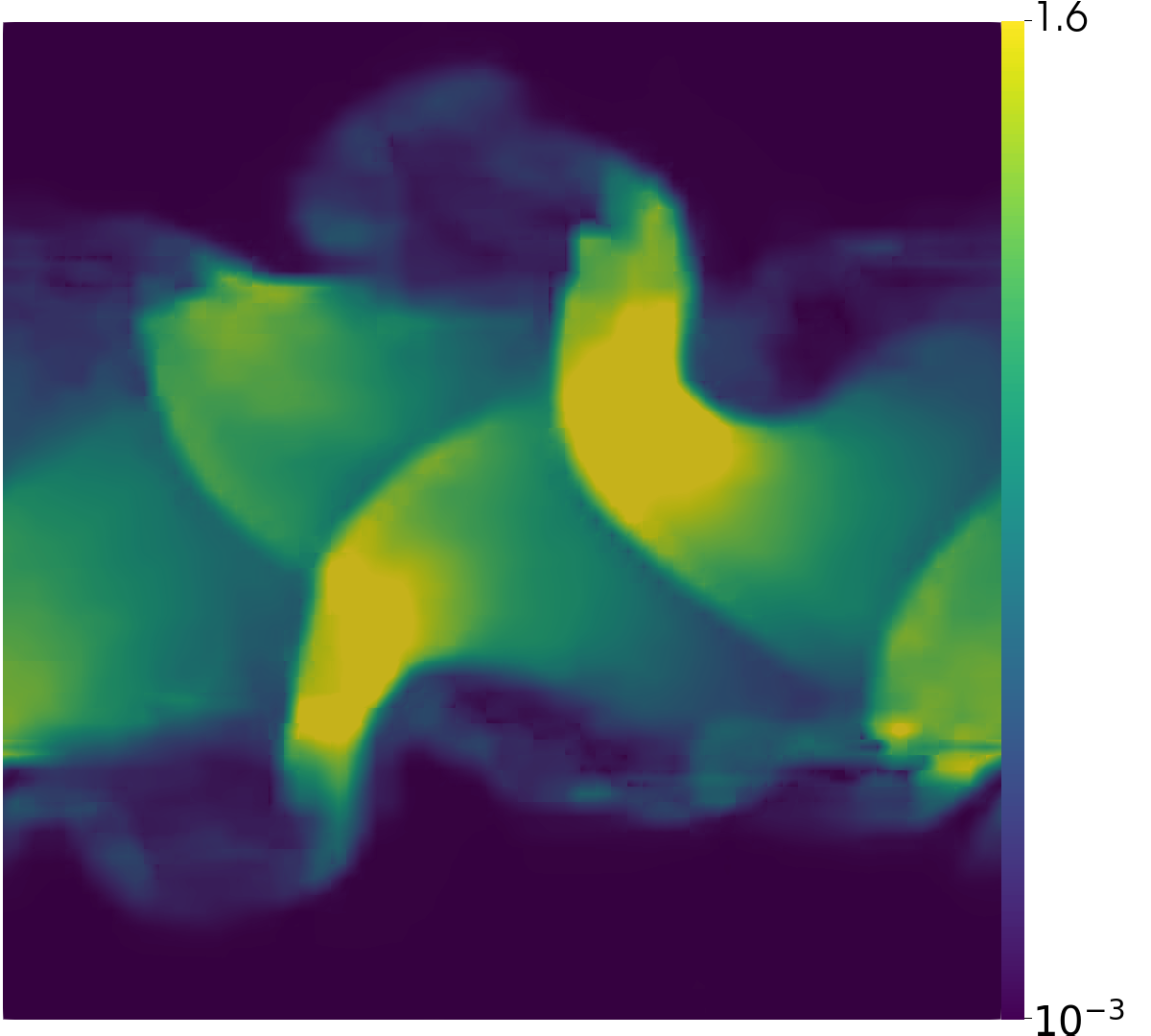}}
    &
      \subfloat[Simple WENO, P$_2$, $64^2$ elements]
      {\label{fig:KhInstabilitySimpleWeno}
      \includegraphics[width=0.3\textwidth]{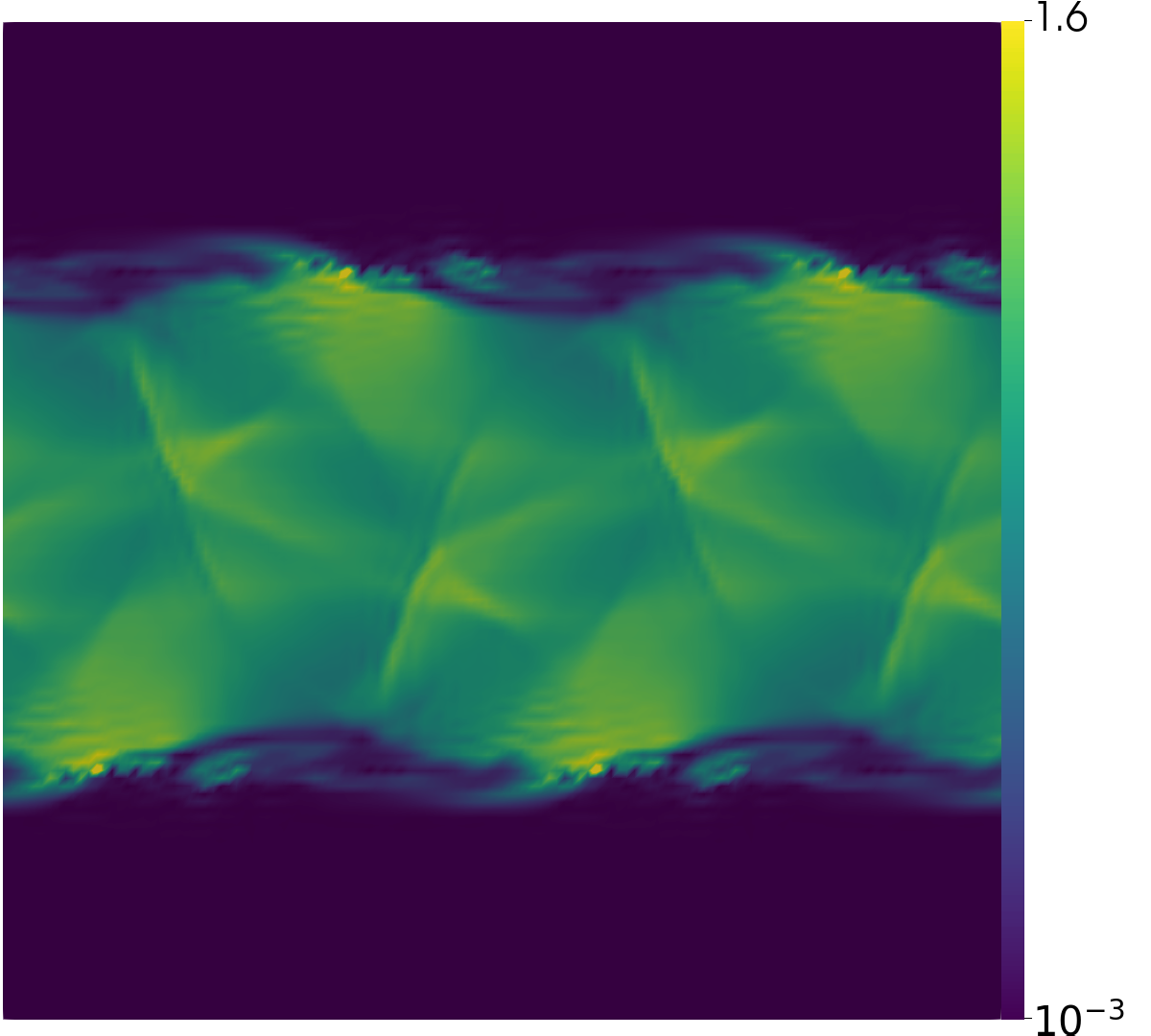}}
    &
      \subfloat[HWENO, P$_2$, $64^2$ elements]
      {\label{fig:KhInstabilityHweno}
      \includegraphics[width=0.3\textwidth]{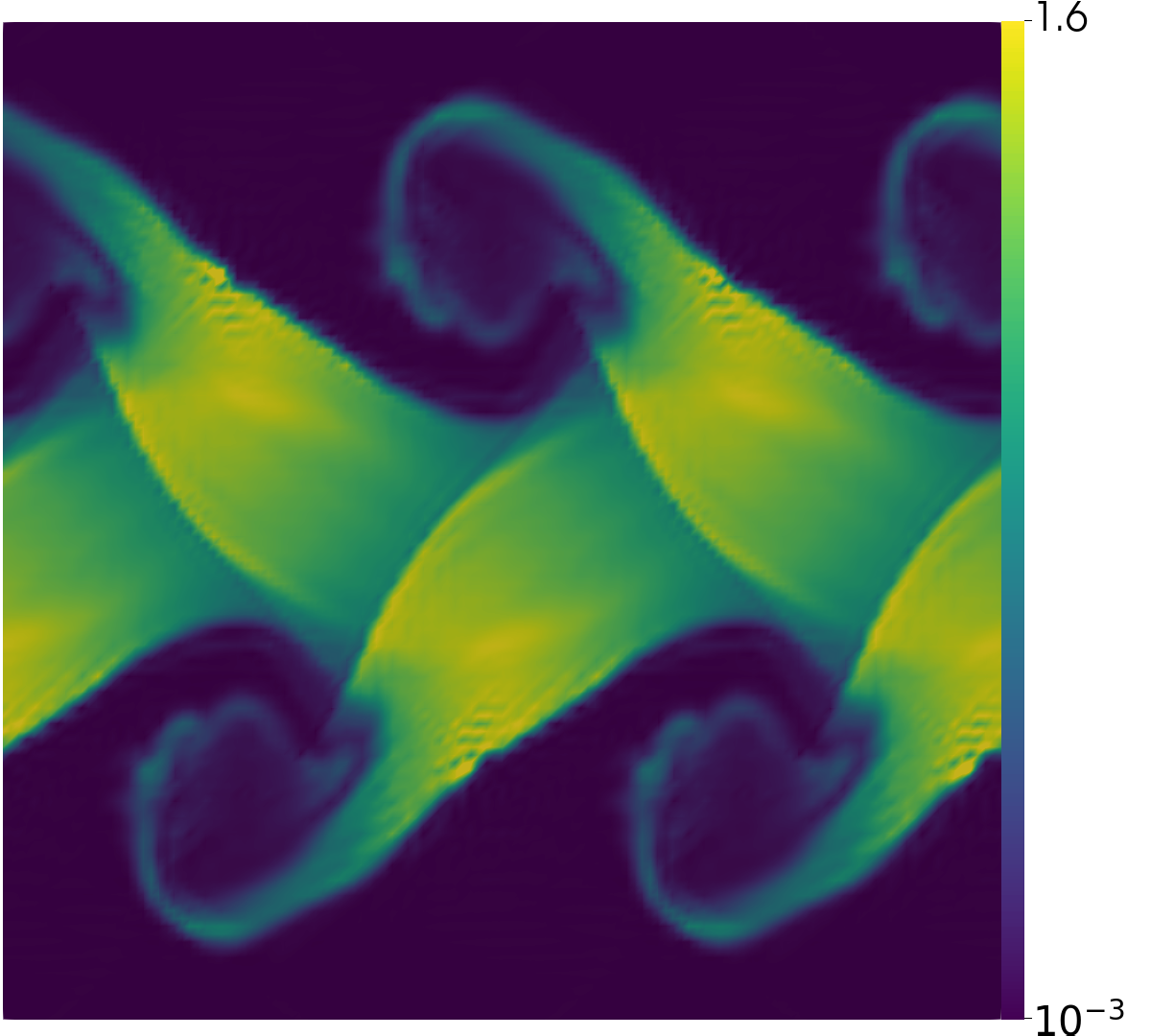}}
    \\
  \end{tabular}
  \caption{\label{fig:KhInstability}Magnetized Kelvin-Helmholtz instability
    $\rho$ at $t=1.6$ comparing the DG-FD hybrid scheme, the $\Lambda\Pi^N$,
    Krivodonova, simple WENO, and HWENO limiters using P$_2$ DG, as well as the
    DG-FD scheme using P$_5$ DG. There are 192 degrees of freedom per dimension,
    comparable to what is used when testing FD schemes. Only the DG-FD hybrid
    scheme and the HWENO limiter produce reasonable results, while the
    $\Lambda\Pi^N$ limiter has very low effective resolution, and the
    Krivodonova and Simple WENO limiters smear out the solution almost
    completely. In the plots of the DG-FD hybrid scheme the regions surrounded
    by black squares have switched from DG to FD at the final time.}
\end{figure*}

In Fig.~\ref{fig:KhInstability} we plot the density at the final time comparing
the different limiting strategies. From Fig.~\ref{fig:KhInstabilitySimpleWeno}
and~\ref{fig:KhInstabilityKrivodonova} we see that the Simple WENO and
Krivodonova limiters destroy the solution almost completely. The $\Lambda\Pi^N$
limiter (Fig.~\ref{fig:KhInstabilityLambdaPiN}) retains some hints of the
expected flow pattern, but also nearly completely destroys the solution. The
HWENO limiter is plotted in Fig.~\ref{fig:KhInstabilityHweno} and does by far
the best of the classical limiters. Ultimately, only the DG-FD hybrid method
(Fig.~\ref{fig:KhInstabilitySubcellP2} for P$_2$ and
Fig.~\ref{fig:KhInstabilitySubcellP5} for P$_5$) is able to produce the expected
vortices and flow patterns.

\subsection{TOV star\label{sec:TOV star}}

A rigorous 3d test case in general relativity is the evolution of a static,
spherically symmetric star. The Tolman-Oppenheimer-Volkoff (TOV)
solution~\cite{Tolman:1939jz, Oppenheimer:1939ne} describes such a setup. In
this section we study evolutions of both non-magnetized and magnetized TOV
stars. We adopt the same configuration as
in~\cite{Cipolletta:2019geh}. Specifically, we use a polytropic equation of
state,
\begin{align}
  \label{eq:polytropic EOS}
  p(\rho)=K \rho^\Gamma
\end{align}
with the polytropic exponent $\Gamma=2$, polytropic constant $K=100$, and a
central density $\rho_c=1.28\times10^{-3}$. When considering a magnetized star
we choose a magnetic field given by the vector potential
\begin{align}
  \label{eq:TOV vector potential}
  A_\phi=A_b(x^2+y^2)\max(p-p_{\mathrm{cut}},0)^{n_s},
\end{align}
with $A_b=2500$, $p_{\mathrm{cut}}=0.04p_{\max}$, and $n_s=2$. This
configuration yields a magnetic field strength in CGS units
\begin{align}
  \label{eq:2}
  |B_{\mathrm{CGS}}|=\sqrt{b^2}\times8.352\times10^{19}\,\mathrm{G},
\end{align}
of $|B_{\mathrm{CGS}}|=1.03\times10^{16}\,G$.  The magnetic field is only a
perturbation to the dynamics of the star, since
$(p_{\text{mag}}/p)(r=0)\sim5\times10^{-5}$. However, evolving the field stably
and accurately can be challenging.  The magnetic field corresponding to the
vector potential in Eq.~\eqref{eq:TOV vector potential} in the magnetized region
is given by
\begin{equation}
  \label{eq:TOV magnetic field}
  \begin{split}
    B^x&=\frac{1}{\sqrt{\gamma}}\frac{xz}{r}
    A_bn_s(p-p_{\mathrm{cut}})^{n_s-1}\partial_rp, \\
    B^y&=\frac{1}{\sqrt{\gamma}}\frac{yz}{r}
    A_bn_s(p-p_{\mathrm{cut}})^{n_s-1}\partial_rp, \\
    B^z&=-\frac{A_b}{\sqrt{\gamma}}\left[
      2(p-p_{\mathrm{cut}})^{n_s} \phantom{\frac{a}{b}}\right. \\
    &\left.+\frac{x^2+y^2}{r} n_s(p-p_{\mathrm{cut}})^{n_s-1}\partial_r p
    \right],
  \end{split}
\end{equation}
and at $r=0$ is
\begin{equation}
  \label{eq:TOV magnetic field origin}
  \begin{split}
    B^x&=0, \\
    B^y&=0, \\
    B^z&=-\frac{A_b}{\sqrt{\gamma}} 2(p-p_{\mathrm{cut}})^{n_s}.
  \end{split}
\end{equation}

We perform all evolutions in full 3d with no symmetry assumptions and in the
Cowling approximation, i.e., we do not evolve the spacetime. To match the
resolution usually used in FD/FV numerical relativity codes we use a domain
$[-20,20]^3$ with a base resolution of 6 P$_5$ DG elements and 12 P$_2$ DG
elements. This choice means we have approximately 32 FD grid points covering the
star's diameter at the lowest resolution, 64 when using 12 P$_5$ elements, and
128 grid points when using 24 P$_5$ elements. In all cases we set
$\rho_{\mathrm{atm}}=10^{-15}$ and $\rho_{\mathrm{cutoff}}=1.01\times10^{-15}$.

\begin{figure}
  \includegraphics[width=0.45\textwidth]{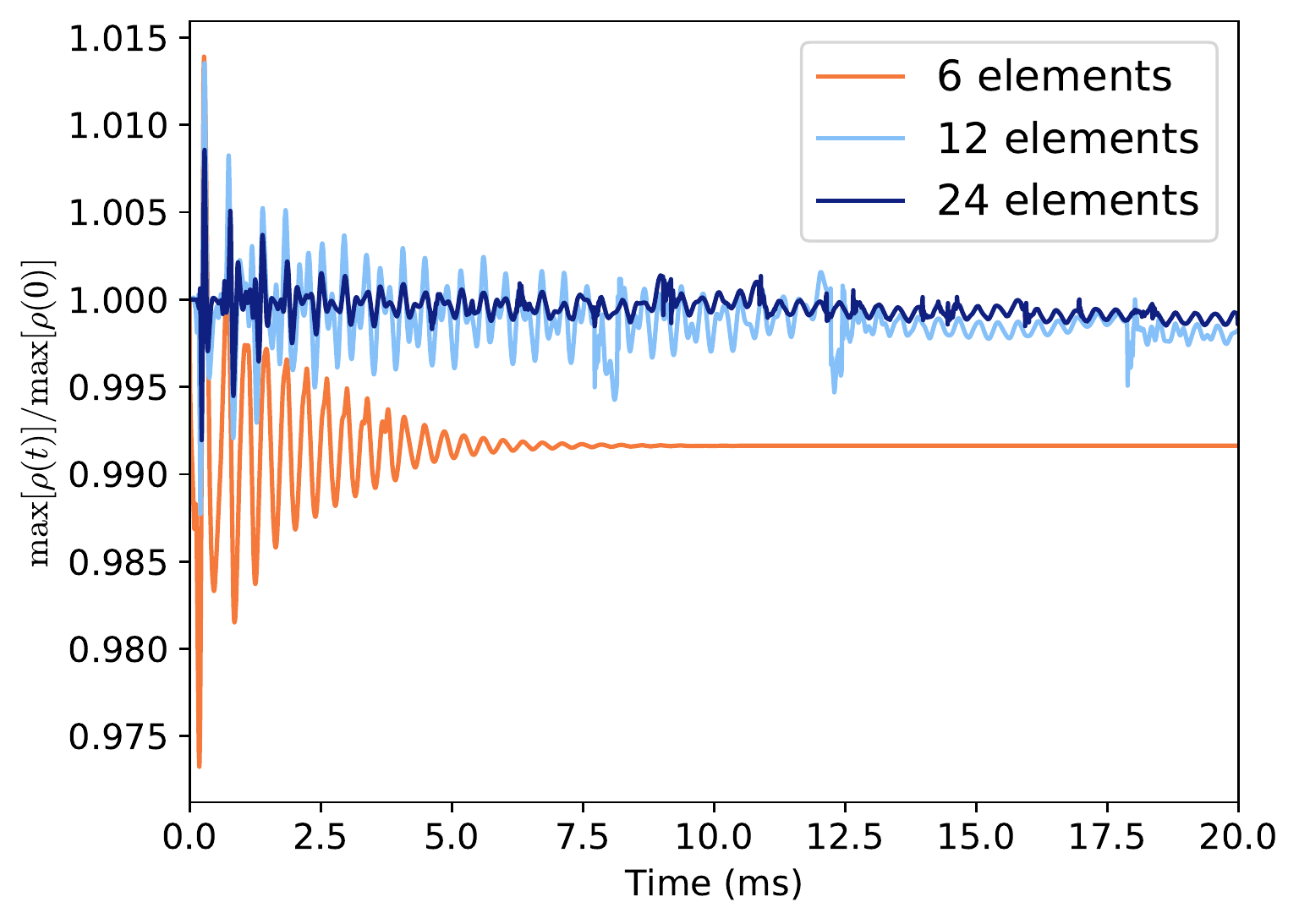}
  \caption{\label{fig:GrmhdTovStarDensities}The maximum density over the grid
    $\max(\rho)$ divided by the maximum density over the grid at $t=0$ for three
    different resolution for the non-magnetized TOV star simulations. The
    6-element simulation uses FD throughout the interior of the star, while 12-
    and 24-element simulations use DG. The increased high-frequency content in
    12- and 24-element simulations occurs because the high-order DG scheme is
    able to resolve higher oscillation modes in the star. The maximum density
    in the 6-element case drifts down at early times because of the low
    resolution and the relatively low accuracy of using FD at the center.}
\end{figure}

\begin{figure}
  \includegraphics[width=0.45\textwidth]{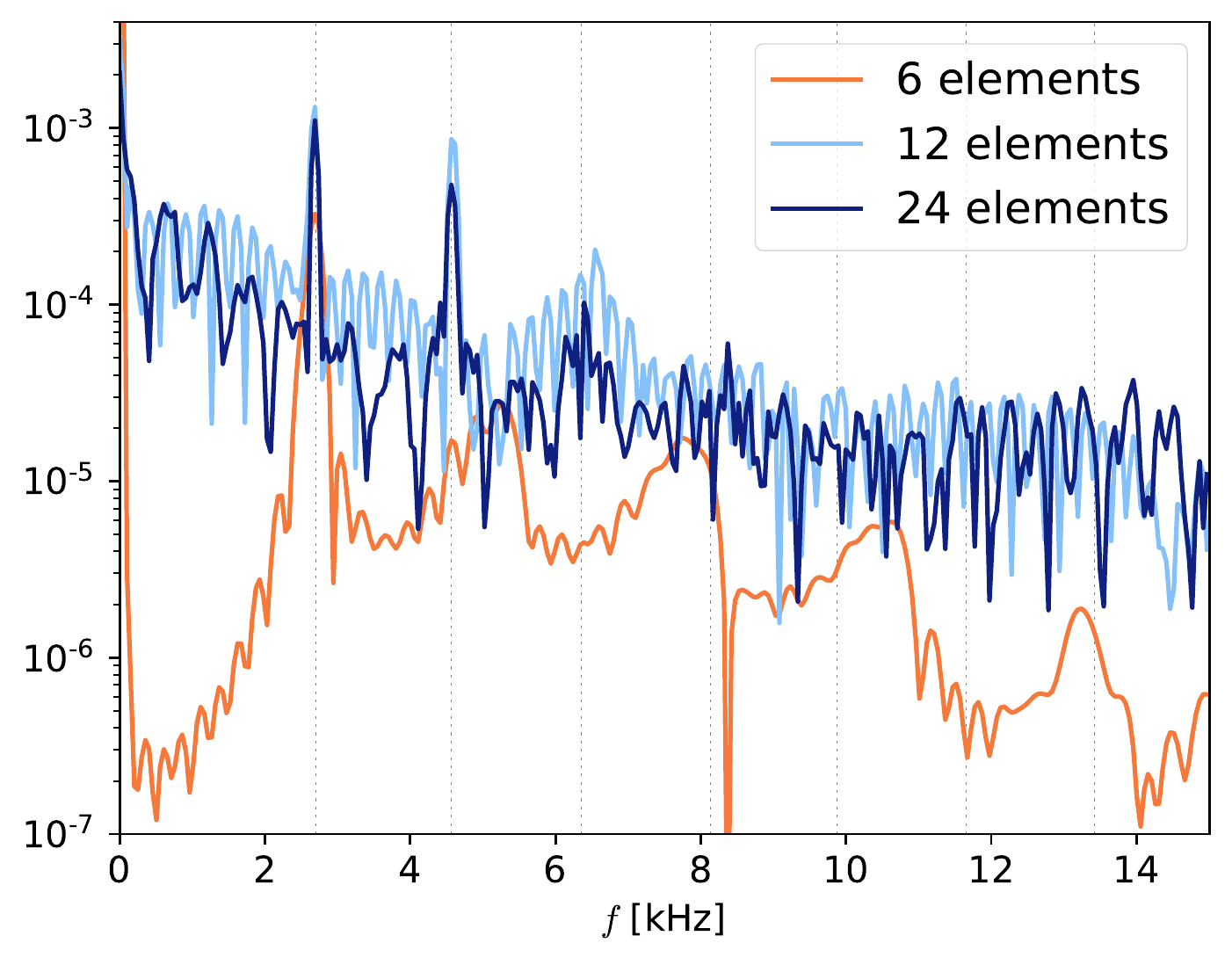}
  \caption{\label{fig:GrmhdTovStarSpectra}The power spectrum of the maximum
    density for three different resolutions for the non-magnetized TOV star
    simulations. The 6-element simulation uses FD throughout the interior of the
    star, while the 12- and 24-element simulations use DG. When the high-order
    DG scheme is used, more oscillation frequencies are resolved. The vertical
    dashed lines correspond to the known frequencies in the Cowling
    approximation~\cite{2002PhRvD..65h4024F}.}
\end{figure}

\begin{figure}
  \includegraphics[width=0.45\textwidth]{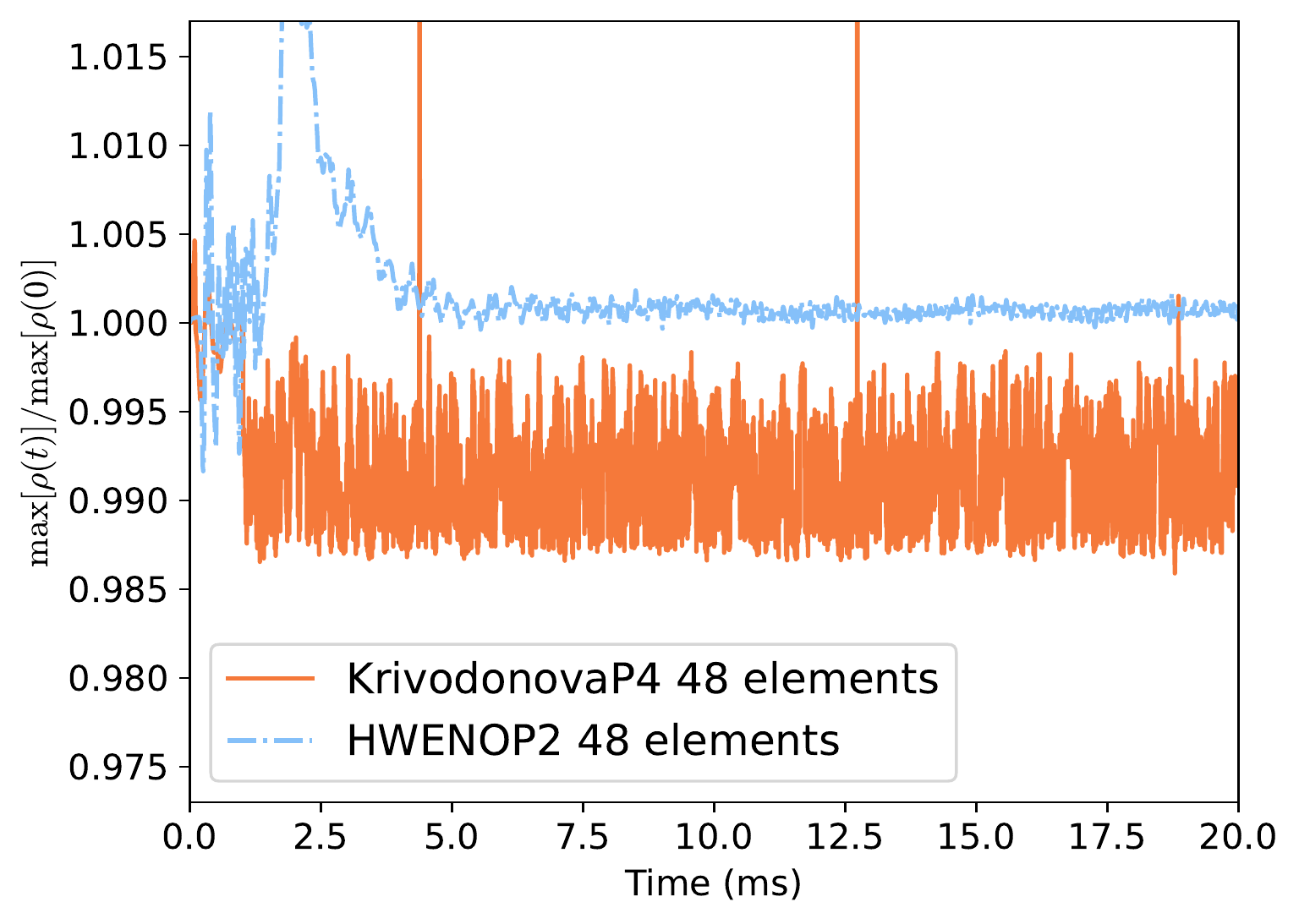}
  \caption{\label{fig:LimitersTOV}The maximum density over the grid
    $\max(\rho)$ divided by the maximum density over the grid at $t=0$
    for the best two cases using classical limiters for the
    non-magnetized TOV star simulations. The HWENO limiter is only
    stable for a P$_2$ DG solver.  Simple WENO (not plotted) gives
    similar results.  The Krivodonova limiter only succeeded at some
    resolutions (3 of the 16 attempted runs) and the shown best result
    is noticeably noisier than the subcell limiter.}
\end{figure}

In Fig.~\ref{fig:GrmhdTovStarDensities} we show the normalized maximum
rest mass density over the grid for the non-magnetized TOV star. The
6-element simulation uses FD throughout the interior of the star and
so there is no grid point at $r=0$.  This is the reason the data is
shifted compared to 12- and 24-element simulations, where the
unlimited P$_5$ DG solver is used throughout the star interior and so
there is a grid point at the center of the star. The increased
``noise'' in the 12- and 24-element data actually stems from the
higher oscillation modes in the star that are induced by numerical
error. In Fig.~\ref{fig:GrmhdTovStarSpectra} we plot the power
spectrum using data at the three different resolutions. The 6-element
simulation only has one mode resolved, while 12 elements resolve two
modes well, and the 24-element simulation resolves three modes well.
In Fig.~\ref{fig:LimitersTOV} we show the normalized maximum
rest mass density over the grid for the best two cases using the
classical limiters.  The simple WENO and HWENO limiters performed
similarly and were only stable for P$_2$.  The Krivodonova limiter
only succeeded at 3 of the 16 resolutions we attempted, and its best
result is noticeably noisier than the other limiters. Note that our experience
is consistent with that of reference~\cite{Bugner:2015gqa}, which was unable
to achieve stable evolutions of a 3d TOV star using the simple WENO limiter.

\begin{figure}
  \includegraphics[width=0.45\textwidth]{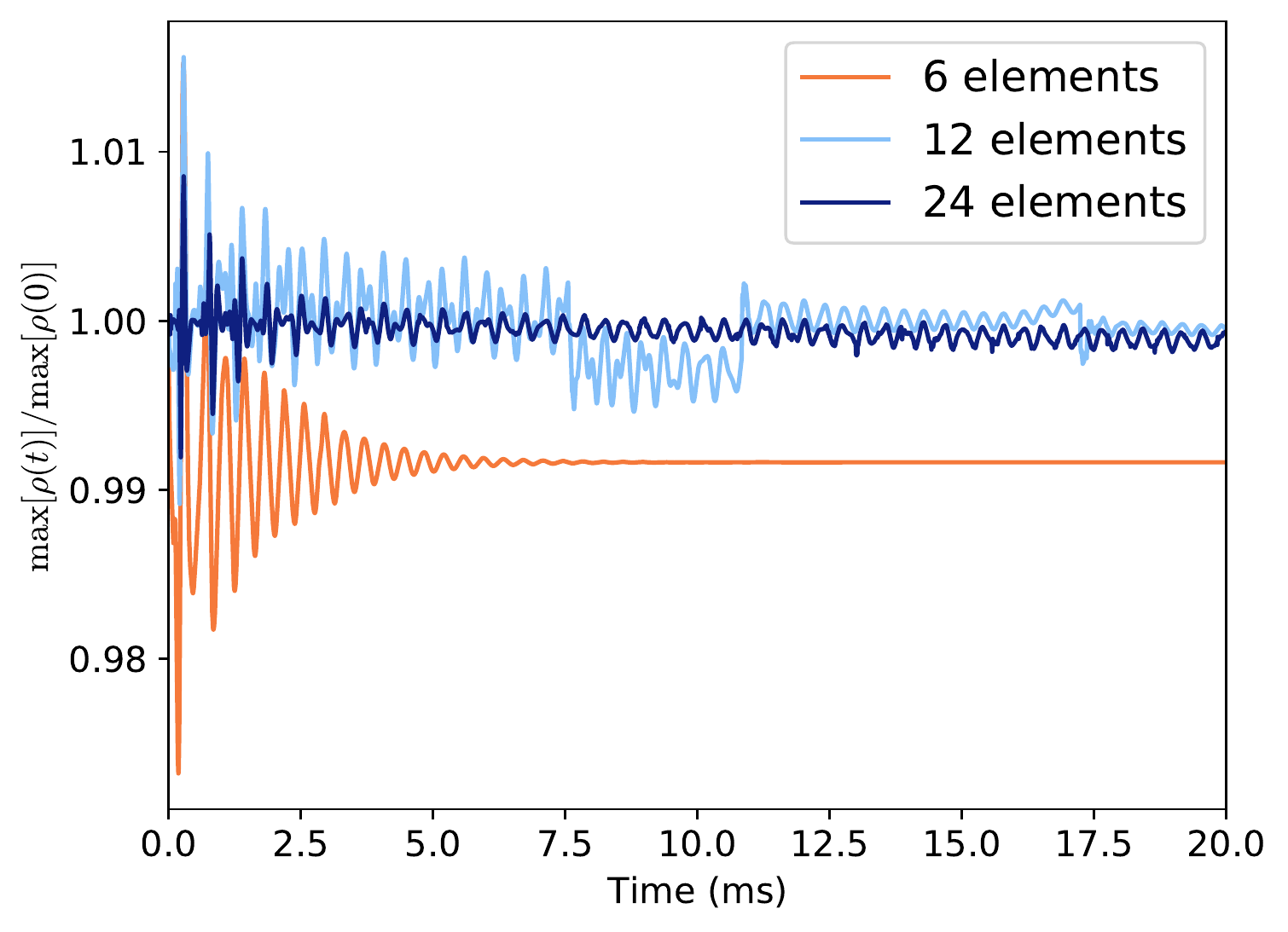}
  \caption{\label{fig:GrmhdMagnetizedTovStarDensities}The maximum density over
    the grid $\max(\rho)$ divided by the maximum density over the grid at $t=0$
    for three different resolution for the magnetized TOV star simulation. The
    6-element simulation uses FD throughout the interior of the star, while 12-
    and 24-element simulations use DG. The increased high-frequency content in
    12- and 24-element simulations occurs because the high-order DG scheme is
    able to resolve higher oscillation modes in the star. The maximum density
    in the 6-element case drifts down at early times because of the low
    resolution and the relatively low accuracy of using FD at the center.}
\end{figure}

\begin{figure}
  \includegraphics[width=0.45\textwidth]{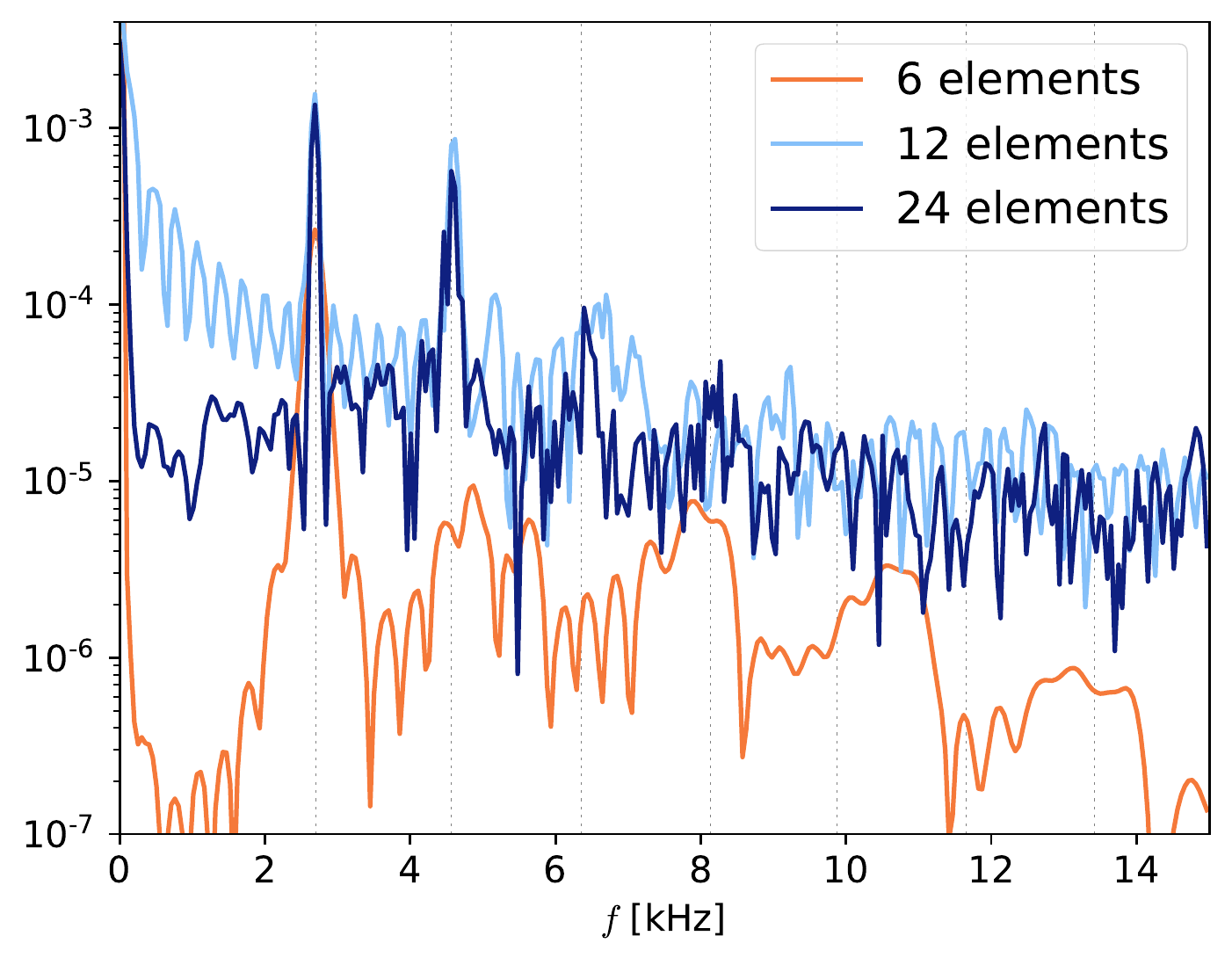}
  \caption{\label{fig:GrmhdMagnetizedTovStarSpectra}The power spectrum of the
    maximum density for three different resolutions of the magnetized TOV star
    simulations. The 6-element simulation uses FD throughout the interior of the
    star, while the 12- and 24-element simulations use DG. When the high-order
    DG scheme is used, more oscillation frequencies are resolved. The vertical
    dashed lines correspond to the known frequencies in the Cowling
    approximation.}
\end{figure}

We show the normalized maximum rest mass density over the grid for the
magnetized TOV star in Fig.~\ref{fig:GrmhdMagnetizedTovStarDensities}. Overall
the results are nearly identical to the non-magnetized case. One notable
difference is the decrease in the 12-element simulation between 7.5ms and 11ms,
which occurs because the code switches from DG to FD at the center of the star
at 7.5ms and back to DG at 11ms. Nevertheless, the frequencies are resolved just
as well for the magnetized star as for the non-magnetized case, as can be seen
in Fig.~\ref{fig:GrmhdMagnetizedTovStarSpectra} where we plot the power
spectrum. Specifically, we are able to resolve the three largest modes with our
P$_5$ DG-FD hybrid scheme. To the best of our knowledge, these are the first
simulations of a magnetized neutron star using high-order DG methods.

\section{Conclusions}

We compare various shock capturing strategies to stabilize the DG method applied
to the equations of general relativistic magnetohydrodynamics in the presence of
discontinuities and shocks. We use the open source numerical relativity code
\texttt{SpECTRE}~\cite{spectrecode} to perform the simulations. We
compare the classic $\Lambda\Pi^N$ method~\cite{Cockburn1999}, the hierarchical
limiter of Krivodonova~\cite{2007JCoPh.226..879K}, the simple WENO
limiter~\cite{2013JCoPh.232..397Z}, the HWENO
limiter~\cite{2016CCoPh..19..944Z}, and a DG-FD hybrid approach that uses DG
where the solution is smooth and HRSC FD methods where the solution contains
discontinuities\cite{Deppe:2021ada}. While many of the limiting strategies
appear promising in the Newtonion hydrodynamics case, we have found stable and
accurate simulations of GRMHD to be a much more challenging problem. This is in
part because limiting the characteristic variables is difficult since the
characteristic variables are not known analytically for the GRMHD system.

In the Newtonian hydrodynamics case, the literature advocates for using the
classical limiters ($\Lambda\Pi^N$, Krivodonova, simple WENO, HWENO) on the
characteristic variables of the evolution system to reduce oscillations, for
using more detailed troubled-cell indicators like that
of~\cite{Krivodonova2004}, and for supplementing the limiting with flattening
schemes to further correct any unphysical values remaining after limiting.  We
have found these techniques do somewhat improve the accuracy and robustness of
the limiters in the Newtonian case, but not enough to avoid the need for
problem-dependent tuning of parameters, or to obtain truly robust behavior.
Since these techniques do not all easily generalize to the relativistic
magnetohydrodynamics case we consider here, we use the classical limiters in
their simplest configuration. Our experience with limiters in Newtonian
hydrodynamics suggests that limiting characteristic variables with specialized
troubled-cell indicators and flatteners will likely still be problematic in the
more complicated GRMHD case.

A further challenge with the classical limiters lies in extending the DG method
to higher orders. With all these limiters, we consistently find large
oscillations and a corresponding loss of accuracy with $P_4$ or higher-order DG
schemes, both in Newtonian and relativistic hydrodynamics evolutions. The
difficulty in robustly applying these limiters to higher-order DG schemes gives
further motivation to favor the DG-FD hydrid method for scientific applications.

We find that only the DG-FD hybrid method is able to maintain stability when
using a sixth-order DG scheme. The other methods are unstable or in the case of
the $\Lambda\Pi^N$ limiter fall back to a linear approximation everywhere. The
classical limiters all work on only some subset of the test problems, and even
there some tuning of parameters is required. While it is certainly conceivable
that with enough fine-tuning each limiter could simulate most or all of the test
problems, this does not make the limiter useful in scientific applications where
a wide variety of different types of shock interactions and wave patterns
appear.  A realistic limiting strategy cannot require any fine-tuning for
different problems. The only method that presents such a level of robustness is
the DG-FD hybrid scheme. As a result, the DG-FD hybrid method is the only method
with which we are able to simulate both magnetized and non-magnetized TOV
stars. To the best of our knowledge this paper presents the first simulations of
a magnetized TOV star where DG is used.

While the DG-FD hybrid scheme is certainly the most complicated approach for
shock capturing in a DG code, our results demonstrate that such complexity
unfortunately seems to be necessary. We are not optimistic that any classical
limiting strategy can be competitive with the DG-FD hybrid scheme since none of
the methods presented in the literature are able to resolve discontinuities
within a DG element. This means that discontinuities are at best only able to be
resolved at the level of an entire DG element. Thus, at discontinuities the
classical limiting strategies effectively turn DG into a finite volume scheme
with an extremely stringent time step restriction. Switching the DG scheme to a
classical WENO finite-volume-type scheme was actually the only way
Ref.~\cite{Bugner:2015gqa} was able to evolve a non-magnetized TOV star.

It is
unclear to us how discontinuities could be resolved inside a DG element since
the basis functions are polynomials. By switching to FD, the hybrid scheme
increases resolution and is able to resolve discontinuities inside an
element. This can also be thought of as instead of solving the partial
differential equations governing the fluid dynamics, we want to solve as many
Rankine-Hugoniot conditions as possible to resolve the discontinuities as
cleanly as possible.

Alternatively, we can view the hybrid scheme as a FD method where in smooth
regions the solution is compressed to a high-order spectral representation to
increase efficiency.  The DG-FD hybrid scheme reduces the number of grid points
per dimension roughly in half, and so in theory a speedup of approximately eight
is expected in 3d. With the current code, we see more moderate speedups of
approximately two, so there is certainly room for optimizations in
\texttt{SpECTRE}.

In the future we plan to evolve the coupled generalized harmonic and GRMHD
system together as one monolithic coupled system, generalize the DG-FD hybrid
scheme to curved meshes, and use more robust positivity-preserving
adaptive-order FD schemes to achieve high-order accuracy even in regions where
the FD scheme is being used.

\section{Acknowledgements}

Charm++/Converse~\cite{laxmikant_kale_2020_3972617} was developed by the
Parallel Programming Laboratory in the Department of Computer Science at the
University of Illinois at Urbana-Champaign.  The figures in this article were
produced with \texttt{matplotlib}~\cite{Hunter:2007,
  thomas_a_caswell_2020_3948793}, \texttt{TikZ}~\cite{tikz} and
\texttt{ParaView}~\cite{paraview, paraview2}. Computations were performed with
the Wheeler cluster at Caltech. This work was supported in part by the Sherman
Fairchild Foundation and by NSF Grants No.~PHY-2011961, No.~PHY-2011968, and
No.~OAC-1931266 at Caltech, and NSF Grants No.~PHY- 1912081 and No.~OAC-1931280
at Cornell. P.K. acknowledges support of the Department of Atomic Energy,
Government of India, under project no. RTI4001, and by the Ashok and Gita Vaish
Early Career Faculty Fellowship at the International Centre for Theoretical
Sciences. M.D. acknowledges support from the NSF through grant PHY-2110287.
F.F. acknowledges support from the DOE through grant DE-SC0020435,
from NASA through grant 80NSSC18K0565 and from the NSF through grant
PHY-1806278. G.L. is pleased to acknowledge support from the NSF through grants
PHY-1654359 and AST-1559694 and from Nicholas and Lee Begovich and the Dan Black
Family Trust. H.R.R. acknowledges support from the Funda\c c\~ao para a
Ci\^encia e Tecnologia (FCT) within the projects UID/04564/2021,
UIDB/04564/2020, UIDP/04564/2020 and EXPL/FIS-AST/0735/2021.

% IAU suggested abbreviations
\newcommand\aap{Astron.\ Astrophys.}  \newcommand\mnras{Mon.\ Not.\ R.\ Astron.\
  Soc.}

\bibliography{refs}
\end{document}